\documentclass[11pt]{article}
\usepackage{geometry}                % See geometry.pdf to learn the layout options. There are lots.
\pdfoutput=1

\usepackage{setspace}
\usepackage{graphicx}
\usepackage{amssymb}
\usepackage{bm}
\usepackage{epstopdf}
\newcommand{\ra}[1]{\renewcommand{\arraystretch}{#1}}

\usepackage[square,authoryear]{natbib}
\usepackage[titletoc,title]{appendix}
\usepackage[font=small,labelfont=bf]{caption}
\usepackage{placeins}

\title{Enceladus's crust as a non-uniform thin shell:\\ I Tidal deformations}
\author{Mikael Beuthe\\
\it Royal Observatory of Belgium,
\it Avenue Circulaire 3, 1180 Brussels, Belgium\\
\it E-mail: mikael.beuthe@observatoire.be}      
\date{}

\begin{document}
\maketitle

\begin{abstract}

The geologic activity at Enceladus's south pole remains unexplained, though tidal deformations are probably the ultimate cause.
Recent gravity and libration data indicate that Enceladus's icy crust floats on a global ocean, is rather thin, and has a strongly non-uniform thickness.
Tidal effects are enhanced by crustal thinning at the south pole, so that realistic models of tidal tectonics and dissipation should take into account the lateral variations of shell structure.
I construct here the theory of non-uniform viscoelastic thin shells, allowing for depth-dependent rheology and large lateral variations of shell thickness and rheology.
Coupling to tides yields two 2D linear partial differential equations of the fourth order on the sphere which take into account self-gravity, density stratification below the shell, and core viscoelasticity.
If the shell is laterally uniform, the solution agrees with analytical formulas for tidal Love numbers; errors on displacements and stresses are less than 5\% and 15\%, respectively, if the thickness is less than 10\% of the radius.
If the shell is non-uniform, the tidal thin shell equations are solved as a system of coupled linear equations in a spherical harmonic basis.
Compared to finite element models, thin shell predictions are similar for the deformations due to Enceladus's pressurized ocean, but differ for the tides of Ganymede.
If Enceladus's shell is conductive with isostatic thickness variations,
surface stresses are approximately inversely proportional to the local shell thickness.
The radial tide is only moderately enhanced at the south pole.
The combination of crustal thinning and convection below the poles can amplify south polar stresses by a factor of 10, but it cannot explain the apparent time lag between the maximum plume brightness and the opening of tiger stripes.
In a second paper, I will study the impact of a non-uniform crust on tidal dissipation.

\end{abstract}

\vspace{\stretch{1}}
\newpage

{\small
\tableofcontents
\scriptsize
\listoffigures
\listoftables
}
\newpage

%%%%%%%%%%%%%%%%%%%%%%%%%%%%%%%%%%%%%%%%%%%%%%%%%%%%%%%%%%%
%%%%%%%%%%%%%%%%%%%%%%%%%%%%%%%%%%%%%%%%%%%%%%%%%%%%%%%%%%%
%%%%%%%%%%%%%%%%%%%%%%%%%%%%%%%%%%%%%%%%%%%%%%%%%%%%%%%%%%%
\section{Introduction}
\label{Introduction}

In 2005, the Cassini spacecraft discovered that the icy moon Enceladus is, next to Io, the most geologically active satellite in the Solar System \citep{porco2006}.
Surprisingly, this activity is concentrated in the south polar terrain where four long parallel cracks (nicknamed `tiger stripes') emit jets of water vapor and icy particles, together with gigawatts of heat \citep{spencer2006}.
These phenomena, together with the analysis of the plume composition, point to the existence of a subsurface reservoir of liquid salt water (e.g.\ \citet{postberg2011}).
Tides due to orbital eccentricity were immediately thought to be the culprit but, twelve years and more than twenty flybys later, we still don't know why geysers only occur at the south pole, and how a so small body
can lose all that heat into space without freezing and thus ending tectonic activity.

The tidal hypothesis concerns both tectonics and the heat budget.
The role of tides in tectonics is supported by the observed correlation between plume brightness and orbital position \citep{hedman2013,nimmo2014,ingersoll2017}, most easily explained by tidal influence on the opening/closing of south polar fractures \citep{hurford2007}.
However, one must invoke the unknown viscosity structure of the icy shell in order to explain the several hours' delay between apocentre and peak brightness \citep{behounkova2015}.
Regarding the heat budget, tidal dissipation seems to be the only source strong enough to power the thermal anomaly, but existing models predict that heat is lost to space faster than it is produced by tidal friction \citep{roberts2008}.

Modelling of Enceladus's tides started with spherically symmetric interior structures \citep{hurford2007,nimmo2007,roberts2008,smithkonter2008}, which are obviously incompatible with the north-south asymmetry.
Besides, it is well known that tidal dissipation is extremely non-uniform both radially and laterally (e.g.\ \citet{beuthe2013}), with concurrent effects on the shell structure.
Therefore, \citet{tobie2008} and \citet{behounkova2012} developed 3D models allowing for a local ocean and lateral viscosity variations, but in which the icy shell is of constant thickness and, incidentally, rather thick ($80\rm\,km$).
The local ocean model was introduced in order to minimize energy requirements and to explain the localization of the plumes at the south pole.
Tidal effects, however, are much smaller in that configuration, leading to the partial suppression of tidal tectonics and leaving the freezing conundrum unsolved.

Our picture of the interior has much evolved in the last three years, thanks to geodesy.
Gravity data clearly show that Enceladus is divided, at about three-fourths of its radius, into a hydrated silicate core surrounded by a $\rm{}H_2O$ mantle \citep{iess2014,mckinnon2015,beuthe2016b}.
The partition of the mantle into liquids and solids is more ambiguous.
Using classical isostasy and hydrostatic equilibrium for slow rotators, \citet{iess2014} interpreted gravity and topography data in terms of a local south polar sea under a 30 to 40~km-thick icy crust.
\citet{mckinnon2015} then pointed out two problems: large-scale isostasy is only possible if the ocean is global, and high-spin corrections to the figure of equilibrium imply a very thick icy shell ($50\rm\,km$ thick) probably grounded at the equator.
The detection of large librations, however, proved that the crust is decoupled from the core by a global ocean and that it is quite thin, between 21 and 26 km thick on average \citep{thomas2016}.
The conflict between librations and gravity-topography data actually arises from the breakdown of classical isostasy at the largest scales.
Using instead minimum stress isostasy and hydrostatic equilibrium for fast rotators, \citet{beuthe2016b} showed that gravity-topography data predict the same thin crust as the libration models ($23\pm4\rm\,km$ on average).
Once the mean shell thickness is known, the local thickness can be inferred from gravity data: it varies from $29\pm4\rm\,km$ at the equator to $7\pm4\rm\,km$ at the south pole \citep{beuthe2016b};
the south polar thickness estimate may be further reduced by a few km if one assumes that topography beyond degree~3 is isostatically compensated.
At the end of the Cassini mission, we can conclude that Enceladus is made of a large core of low density, a thick global ocean, and a thin icy shell (on average less than 10\% of the radius) with mostly latitudinal thickness variations resulting in a very thin crust at the south pole.

Tidal effects should thus be computed in a model allowing for radial and lateral variations of the icy shell structure.
Various methods have been designed to take into account the effect of lateral heterogeneities on the body tides of the Earth \citep{metivier2008b,latychev2009,lau2015}, the Moon \citep{zhong2012,qin2014,qin2016}, Europa and Ganymede \citep{a2014}, and Enceladus \citep{tobie2008,behounkova2012,soucek2016,behounkova2017}.
Many of these approaches do not seem to be well suited to an icy moon with a shell of variable thickness floating on a subsurface ocean.
A first exception is the work of \citet{a2014}, who use a finite-element approach to show that lateral variations in shell thickness (or rheology) have small effects ($<1\%$) on the tides of Europa and Ganymede.
The tides of these large satellites, however, are very different from those of Enceladus: the former nearly follow the response of the ocean, whereas the latter are mainly controlled by the elasticity of the icy shell (see Section~\ref{SoftHardShells}).
Furthermore, \citet{a2014} are mainly interested in checking the accuracy of future geodetic measurements (radial tide and geoid perturbation).
For Enceladus, tidal geodetic measurements seem extremely remote; the motivation is rather to compute the effect of crustal structure on tidal stresses and tidal dissipation.
For these different reasons, the results of \citet{a2014} are not relevant to Enceladus, although their method should be applicable.
\citet{soucek2016} attack the problem from another angle, by computing tidal deformations and stresses in an elastic shell of constant thickness locally broken through by the south polar faults.
They conclude that the breaking of the shell can increase deformations and stresses around the faults by up to a factor of 3, which should be considered as an upper bound since faults are not frictionless and might not extend through the whole crust.
Their approach was recently extended to elastic shells of variable thickness by \citet{behounkova2017}.
Finally, \citet{johnston2017} computed the effect of ocean overpressure on the stresses in a shell of variable thickness; their finite element method can probably be extended to tidal forcing.
I will review in Section~\ref{Why} in which respect these methods differ from thin shell theory regarding their assumptions.

In this paper, the first of a series of two, I set out the theory of the tidal deformations, stresses included, of a non-uniform thin shell floating on a global ocean.
The second paper will deal with tidal dissipation.
The rest of the paper is made of four parts: (1) non-uniform thin shell theory, (2) coupling to tides, (3) benchmarking against a laterally uniform shell, and (4) predicting tidal deformations of a fully non-uniform shell.
I will address right away the question of the adequacy of the thin shell assumption.

%%%%%%%%%%%%%%%%%%%%%%%%%%%%%%%%%%%%%%%%%%%%%%%%%%%%%%%%%%%
%%%%%%%%%%%%%%%%%%%%%%%%%%%%%%%%%%%%%%%%%%%%%%%%%%%%%%%%%%%
%%%%%%%%%%%%%%%%%%%%%%%%%%%%%%%%%%%%%%%%%%%%%%%%%%%%%%%%%%%
\section{Non-uniform thin shell theory}

First, a point of terminology: `uniform' refers to shell thickness and rheology, whereas `homogeneous' refers only to rheology.
The shell density is supposed to be homogeneous.
I will consider four types of shells listed from the simplest to the most complex: uniform (thickness and rheology are constant), laterally uniform (rheology can vary radially), radially homogeneous (thickness can vary), and non-uniform (thickness and rheology vary arbitrarily).

%=========================================================================================
%=========================================================================================
\subsection{Why thin shell theory?}
\label{Why}

In comparison with 3D methods, thin shell theory has the great advantage of being two-dimensional, and thus simpler and faster to solve.
After integrating all quantities over the thickness of the shell, one is left with 2D equations depending on scalar variables, instead of 3D equations depending on tensors.
The theory of thin elastic shells with variable thickness was worked out in \citet{beuthe2008} and applied to contraction and despinning tectonics in \citet{beuthe2010}, while tidal coupling of thin shells with constant thickness, but depth-dependent rheology, was studied in the membrane limit by \citet{beuthe2014,beuthe2015} for quasi-static tides, and by \citet{beuthe2016a} for dynamic tides.
Table~\ref{TableRef} gives a short summary of these papers, from which I will draw elements in order to build the theory of fully non-uniform thin shells coupled to quasi-static tides.

%TABLE 1
\begin{table}[ht]\centering
\ra{1.2}
\footnotesize
\caption[Papers about thin shell theory]{
Papers about thin shell theory relevant to this work.
The papers \textit{Be15a} and \textit{Be15b} are illustrated by the Mathematica notebook \textit{MembraneWorlds.nb} which is available at http://library.wolfram.com, subject class Science/Geology and Geophysics.
}
\begin{tabular}{@{}lllccc@{}}
\hline
Reference & Noted & Subject & Variable   & Depth-dependent & Tidal \\
                 &            &              & thickness & rheology & coupling \\
\hline
\citet{beuthe2008} & \textit{Be08} & Flexure of thin elastic shells  & \checkmark & - & - \\
\citet{beuthe2010} & \textit{Be10} & Contraction and despinning & \checkmark & - & - \\
\citet{beuthe2014} & \textit{Be15a} & Massless membrane approach & - & \checkmark & \checkmark \\
\citet{beuthe2015} & \textit{Be15b} & Massive membrane approach & - & \checkmark & \checkmark \\
\citet{beuthe2016a} & \textit{Be16} & Dynamic ocean tides & - & \checkmark & \checkmark \\
\hline
\end{tabular}
\label{TableRef}
\end{table}%

Membranes are thin shells in which bending effects are negligible, an approximation which is well suited to long-wavelength deformations such as tides.
What are the wavelength ranges for which the membrane and thin shell approximations are valid?
At long wavelength, the shell mainly deforms by lateral extension/compression so that the strains are nearly constant with depth: the shell is said to be in the \textit{membrane regime}.
At short wavelength, bending causes a variation of strains with depth: the shell is in the \textit{bending regime}.
Wavelengths $\lambda$ are approximately related to harmonic degrees $n$ by $\lambda\approx2\pi{}R/n$ ($R$ being the surface radius).
The transition (\textit{mixed regime}) between membrane and bending regimes occurs at wavelengths such that (\textit{Be08}, Eq.~(83))
\begin{equation}
n \approx \left(12\left(1-\nu^2\right)\right)^{1/4} \sqrt{R/d} \, .
\label{MembraneBendingTransition}
\end{equation}
The factor in front of the squared root is approximately equal to $1.8$ (see Table~\ref{TableParam} for $\nu$, $R$, and $d$).
At sufficiently short wavelengths, thin shell theory breaks down because the third and fourth thin shell assumptions (see Appendix~\ref{StrainThinShell}) are violated.
The breakdown threshold is approximately given by (see \textit{Be08}, Eq.~(95))
\begin{equation}
n \approx 2 \, (R/d) \, .
\label{ThinShellBreakdown}
\end{equation}
While it would be a bit easier to build a fully non-uniform membrane theory, there are three reasons to do the extra work and include bending corrections.
First, these corrections include a part of the next-to-leading thick shell corrections (see Section~\ref{UniformThickShell}).
Second, the non-uniformity of the shell couples the tidal degree~2 to higher harmonic degrees, so that the membrane approximation is not as good as if only degree~2 is involved.
Third, non-uniform thin shell theory is useful for other kinds of loading problems, for example topographic loading at short and long wavelengths.

As a common rule, thin shell theory is applied to shells with a thickness less than 5 to 10\% of the surface radius.
However, this is not a hard-and-fast rule because it depends on the wavelength of the deformation, as shown for example by Eq.~(\ref{ThinShellBreakdown}).
According to recent geodesy data (see Section~\ref{Introduction}), Enceladus's shell can be treated as a thin shell, though just barely.
It is thus preferable to estimate the error due to this approximation.
Suppose that Enceladus is made of three uniform layers: an incompressible elastic core, an incompressible ocean, and a compressible elastic shell having the same density as the ocean (Table~\ref{TableParam}).
I will compute below the radial displacement of Enceladus's surface, due to tides, assuming that the shell is either thick or thin.
The thick shell solution is computed with the program \textit{love.f} included in the software SatStress (http://code.google.com/p/satstress/) \citep{wahr2009}.
The thin shell solution is computed either in the membrane limit with the formulas of \textit{Be15a} (or the notebook \textit{MembraneWorlds.nb} mentioned in Table~\ref{TableRef}), or with bending corrections for a homogeneous shell (Eq.~(B.9) of \textit{Be16}).

\begin{figure}
   \centering
   \includegraphics[width=7.3cm]{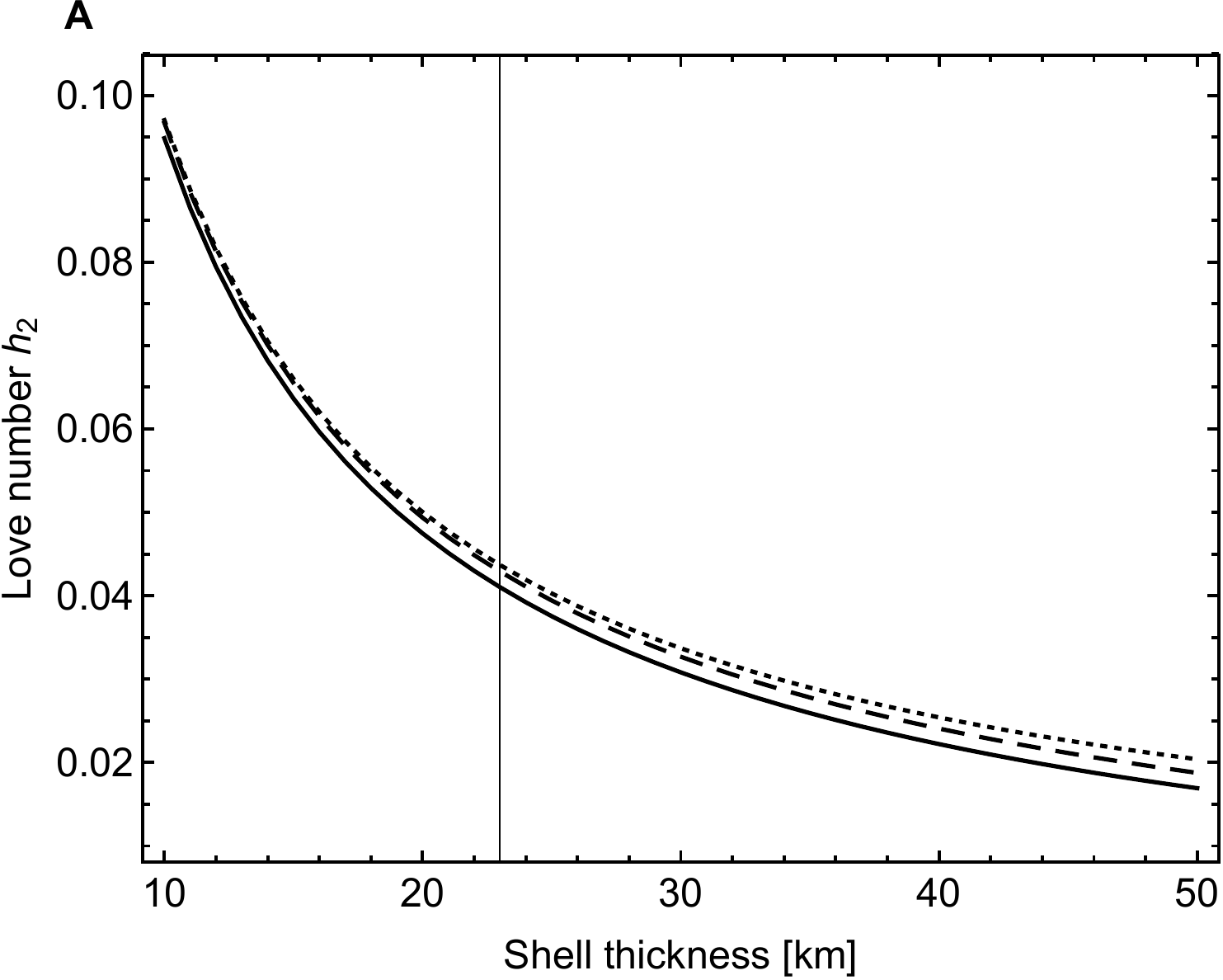}
    \includegraphics[width=7.3cm]{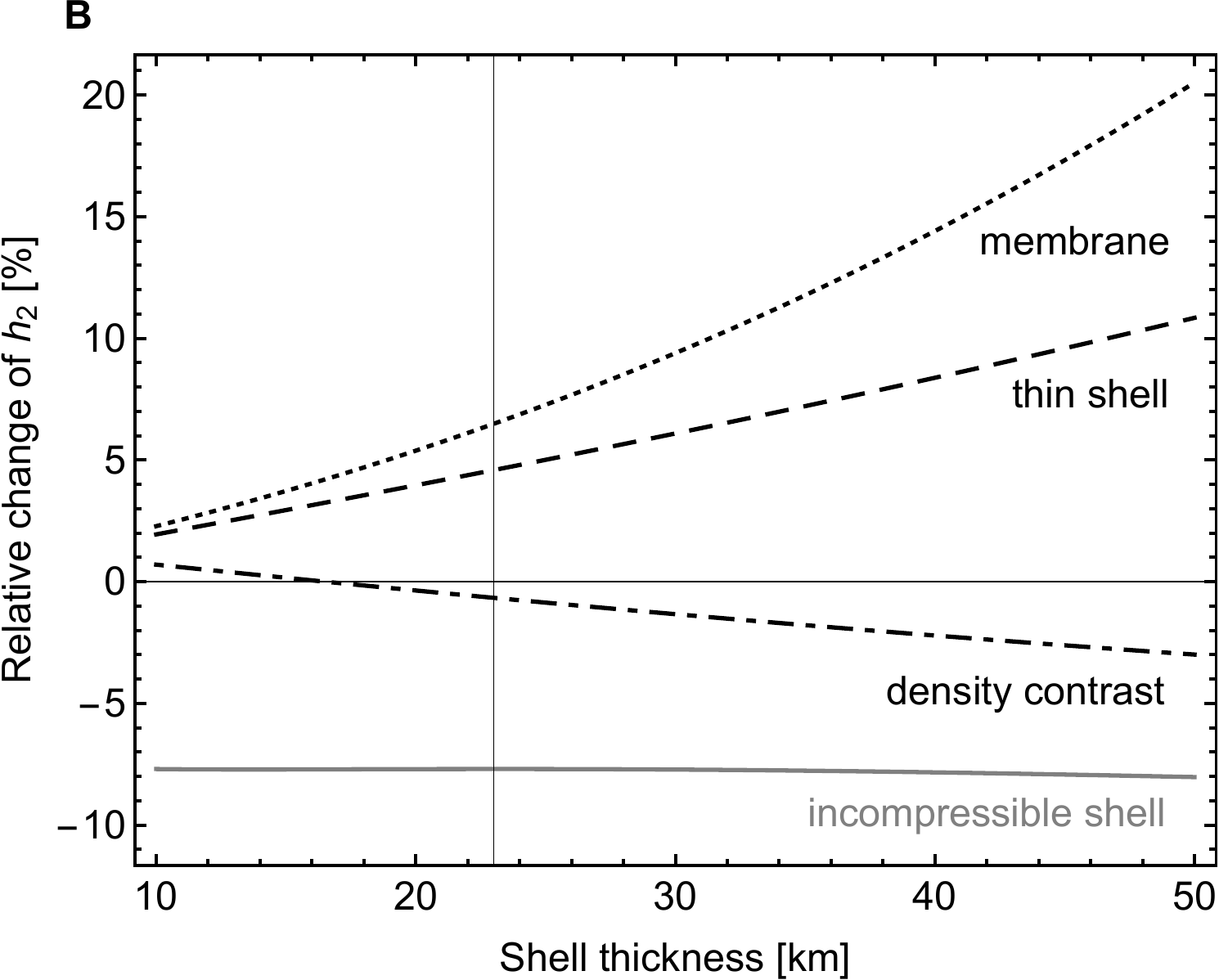}
   \caption[Tidal deformations of a uniform elastic shell]
   {Tidal deformations of a uniform elastic shell: various approximations.
   (A) Love number $h_2$ in the thick shell approach (solid), with the thin shell approximation (dashed), and in the membrane limit (dotted),
   for a compressible elastic shell having the same density as the ocean.
   (B) Relative change of $h_2$ with respect to the solid curve of panel~(A):
   in the membrane limit (dotted black), with the thin shell approximation (dashed black), if the thick shell and the ocean have different densities (dash-dotted black), or if the thick shell is incompressible (solid gray).
   The vertical line indicates the most likely thickness $d=23\rm\,km$.
   See Section~\ref{Why} for details.
}
   \label{FigLoveElastic}
\end{figure}

Since the internal structure is spherically symmetric, the radial displacement can be described by one number, called the Love number $h_2$, which is the proportionality factor between the radial displacement and the tidal potential.
Fig.~\ref{FigLoveElastic}A shows $h_2$ as a function of shell thickness for a thick shell, for a thin shell, and in the membrane limit.
The error due to the thin shell (resp.\ membrane) approximation is less than 5\% (resp.\ 7\%) for the most likely thickness $d=23\rm\,km$ (see Fig.~\ref{FigLoveElastic}B).
This error is not large in comparison with the effects due to the non-uniformity of the shell, as will be seen in Section~\ref{Results}.
Fig.~\ref{FigLoveElastic}B also shows that the effect of the shell-ocean density contrast is less than $1\%$ for $d=23\rm\,km$, if shell and ocean densities are equal to 930 and $1020\rm\,kg/m^3$ (the error is about 2\% if the ocean density remains at $1000\rm\,kg/m^3$).
This already provides a justification for the massless shell assumption of Section~\ref{MasslessShell}.
Neglecting self-gravity (the gravitational perturbation due to the tidal bulge) leads to an error of a few percents, depending on the shell thickness and viscoelastic properties (see Eq.~(\ref{LoveThickRigid})), whereas assuming an incompressible shell leads to an 8\% error on $h_2$.
Finally, the shell cannot be purely elastic: $h_2$ can be 5\% larger if the shell is conductive (see Section~\ref{ConductiveShell}).

As a common objection, one often hears that the thin shell approach is only a 2D approximation, whereas 3D methods are, well, 3D and thus supposedly exact.
3D methods, however,  have their own set of assumptions which are not always properly quantified.
As a first example, the 3D spectral approach of \citet{behounkova2012,behounkova2015} assumes that the shell is of uniform thickness and incompressible.
Overcoming the former assumption appears to be difficult in their framework.
Regarding the latter, neglecting compressibility has a larger effect on $h_2$ than the thin shell approximation if the shell thickness is less than $38\rm\,km$, or 15\% of the radius (see Fig.~\ref{FigLoveElastic}B).
The correct inclusion of compressibility can be a delicate issue for finite element models \citep{bangtsson2008,spada2011}.
As a second example, the finite-element method of \citet{a2014} is biased at long wavelength and affected by short-wavelength noise; moreover, problems of convergence arise if the shell is very thin, as it is likely on Europa (see further discussion in Section~\ref{TidalResponseGanymede}).
As a third example, the finite-element method of \citet{soucek2016} and \citet{behounkova2017} neglects self-gravity, density changes induced by volumetric deformation, and viscoelasticity (see the Supporting Information of \citet{soucek2016}).
The finite element method of \citet{johnston2017} makes similar assumptions.
These various approximations are not necessarily problematic, but illustrate the fact that each method has a limited domain of applicability defined by the underlying assumptions.
It can be difficult to quantify the impact of a given approximation if no complete theory is available.
I believe that thin shell theory and 3D methods should be better seen as complementary tools.

%TABLE 2
\begin{table}[h]\centering
\ra{1.2}
\footnotesize
\caption[Physical parameters]{
Physical parameters used in this paper.
}
\begin{tabular}{@{}llll@{}}
\hline
Parameter &  Symbol & Value & Unit \\
\hline
Mean eccentricity${}^a$ & $e$ & $0.0047$ & - \\
Rotation rate${}^a$    & $\omega$ & $5.307\times10^{-5}$ & $\rm\,s^{-1}$ \\
Surface radius${}^a$  & $R$ & $252.1$  & km \\
Bulk density${}^a$ & $\rho_b$ & $1610$ & $\rm kg \, m^{-3}$ \\
Surface gravity${}^a$ & $g$ & $0.1135$ & $\rm m \, s^{-2}$ \\
Mean shell thickness${}^b$ & $d$ & 23 & km \\
Core radius${}^b$ & $R_c$ & 192 & km \\
Density of ice and ocean & $\rho$ & 1000 & $\rm kg/m^3$ \\
Shear modulus of elastic ice & $\mu_e$ & 3.5 & GPa \\
Poisson's ratio of elastic ice & $\nu_{\rm e}$ & 0.33 & - \\
Shear modulus of elastic core & $\mu_m$ & 40 & GPa \\
Bulk modulus of the ocean${}^c$ & $K_o$ & 2 & GPa \\
\hline
\multicolumn{4}{l}{\scriptsize ${}^a$ See Table~1 of \citet{beuthe2016a}}
\vspace{-1mm}\\
\multicolumn{4}{l}{\scriptsize ${}^b$ \citet{beuthe2016b}}
\vspace{-1mm}\\
\multicolumn{4}{l}{\scriptsize ${}^c$ \citet{fine1973}}
\end{tabular}
\label{TableParam}
\end{table}%

%=========================================================================================
%=========================================================================================
\subsection{Thin elastic shell}
\label{ThinElasticShell}

In this section, I briefly review the theory of thin elastic shells with variable thickness.
Consider a spherical thin shell which is elastic and homogeneous, but of variable thickness.
The governing equations for the deflection of the shell submitted to a transverse load are (Eqs.~(58) and (66) of \textit{Be08}):
\begin{eqnarray}
{\cal C}( \chi_{\rm e} D_{\rm e} \,; w ) - (1-\nu_{\rm e}) \, {\cal A}( \chi_{\rm e} D_{\rm e} \,; w) + R_0^{\,3} \, {\cal A}(\chi_{\rm e} \,; F) &=& R_0^4 \, q \, ,
\nonumber \\
{\cal C}( \chi_{\rm e} \alpha_{\rm e}  \,;  F ) - (1+\nu_{\rm e}) \, {\cal A}( \chi_{\rm e} \alpha_{\rm e} \,; F) - R_0^{-1} \, {\cal A}(\chi_{\rm e} \,; w) &=& 0 \, ,
\label{ElasticGoverningEq}
\end{eqnarray}
where $w$ is the radial deformation (positive outward), $F$ is an auxiliary function called the \textit{stress function}, and $R_0$ is the radius of the middle surface of the shell.
The transverse load $q$ is measured per unit area of the reference surface and is positive outward (contrary to the convention of \textit{Be08}).
The differential operators ${\cal C}$ and ${\cal A}$ are defined in Appendix~\ref{OperatorsOrder4}.

The elasticity of a thin shell is characterized by three parameters: \textit{Poisson's ratio} $\nu_{\rm e}$, the \textit{bending rigidity (or stiffness)} $D_{\rm e}$,  and the \textit{extensibility} $\alpha_{\rm e}$.
The last parameter plays the role of the inverse of the \textit{extensional rigidity (or stiffness)} $B_e$ \citep{kraus1967}, to which it is related by $1/\alpha_{\rm e}=(1-\nu_{\rm e}^2)B_e$.
If the shell is radially homogeneous, the extensibility and the bending rigidity are given by
\begin{equation}
\left(  \alpha_{\rm e} \, , \, D_{\rm e}\, \right) = \left( \frac{E_{\rm e} d^3}{12(1-\nu_{\rm e}^2)} \, , \, \frac{1}{E_{\rm e} d} \right) \, ,
\label{defalphaDH}
\end{equation}
where $d$ is the variable shell thickness and $E_{\rm e}$ is the elastic Young's modulus (related to the elastic shear modulus $\mu_e$ by $E_{\rm e}=2(1+\nu_{\rm e})\mu_e$).
The governing equations also depend on a purely geometric parameter $\chi_{\rm e}$ (denoted $\eta$ in \textit{Be08}) which is close to one: 
\begin{equation}
\chi_{\rm e} = \left(1+ \frac{d^2}{12 R_0^2} \right)^{-1} .
\label{defetaxiH}
\end{equation}

A few comments about Eq.~(\ref{ElasticGoverningEq}):
\begin{enumerate}
\item
The solution has been benchmarked, for surface loading,  against a finite-element method by \citet{kalousova2012}, who concluded that it was very accurate even in regions with high thickness variability.
\item
Poisson's ratio must be uniform  but $\alpha_{\rm e}$ and $D_{\rm e}$ can vary independently, so that the equations remain valid if Young's modulus $E_{\rm e}$ varies laterally (though not radially).
\item
It implicitly assumes that thickness variations are symmetric with respect to the middle surface of the shell.
\item
It is derived in the second-order approximation of thin shell theory \citep{kraus1967,wan1988}, which consists in delaying the application of the first assumption of thin shell theory (see Appendix~\ref{StrainThinShell}).
This method is practical when deriving the governing equations because it is hard to know in advance which terms are negligible.
The first-order approximation of Eq.~(\ref{ElasticGoverningEq}) is obtained by setting $\chi_{\rm e}=1$ ($\Delta'$ is defined by Eq.~(\ref{DeltaPrime})):
\begin{eqnarray}
{\cal C}( D_{\rm e} \,; w ) - (1-\nu_{\rm e}) \, {\cal A}( D_{\rm e} \,; w) + R_0^{\,\,3} \, \Delta' F &=& R_0^4 \, q \, ,
\nonumber \\
{\cal C}( \alpha_{\rm e}  \,;  F ) - (1+\nu_{\rm e}) \, {\cal A}(\alpha_{\rm e} \,; F) - R_0^{-1} \, \Delta' w &=& 0 \, .
\label{ElasticGoverningEqApprox}
\end{eqnarray}
\end{enumerate}
In this paper, I show that the governing equations keep the same form if the shell is viscoelastic and radially inhomogeneous, although the definitions of the shell parameters $(\alpha_{\rm e},D_{\rm e},\chi_{\rm e})$ must be modified to reflect the variation with depth of the shell rheology.
In this new formulation, thickness variations do not need to be symmetric with respect to a spherical (middle) surface.

%=========================================================================================
%=========================================================================================
\subsection{Thin viscoelastic shell}
\label{ViscoShellRheology}

Tidal forcing and tidal response are periodic phenomena which can be expanded as Fourier series.
If the satellite is rotating synchronously with its mean motion on an eccentric orbit (as Enceladus does), there is only one tidal frequency equal to the rotation rate $\omega$ (ignoring the permanent tide).
Linear viscoelasticity is introduced as usual, with a version of the correspondence principle suited to periodic forcing: variables in the frequency domain become complex and depend on frequency as well as on the parameters of the rheological model (e.g.\ \citet{tschoegl1989}).
The time-domain variable $\tilde V(t)$ and the frequency-domain variable $V(\omega)$ are related by
\begin{equation}
\tilde V(t) = {\rm Re} \! \left( V(\omega) \, e^{i \omega t } \right) ,
\label{fourier}
\end{equation}
where Re(.) denotes the real part.
If the rheology is of Maxwell type, the shear modulus depends on the angular frequency $\omega$, the elastic shear modulus $\mu_e$, and the viscosity $\eta$ as
\begin{equation}
\mu = \frac{\mu_e}{1-i\, \delta} \, ,
\label{maxwell}
\end{equation}
where $\delta=\mu_e/(\omega\eta)$ is nondimensional \citep{wahr2009}.

Bulk compression is often supposed to be elastic, in which case the viscoelastic  bulk modulus $K$ is equal to its elastic value $K_e$.
Accordingly, Poisson's ratio $\nu$ must depend on frequency and viscosity (see \textit{Be15a} and \textit{Be15b}) and thus differs from its elastic value $\nu_{\rm e}$.
Theoretically, the reverse is also possible  ($K\neq{}K_e$ and $\nu=\nu_{\rm e}$).
In general, a non-uniform rheology implies that $K$ and $\nu$ cannot both be independent of frequency and viscosity, unless the material is incompressible.
Assuming that $\nu$ is uniform simplifies a lot the theory of thin elastic shells (Eq.~(\ref{ElasticGoverningEq})).
I will keep this assumption for a viscoelastic shell in order to simplify the integration of the stress-strain relation (Appendix~\ref{StressThinShell}).
A uniform Poisson's ratio is a good approximation because ${\rm Re}(\nu)$ varies between the elastic value $\nu_{\rm e}$ ($0.33$ for ice) and the fluid limit (equal to 0.5) whereas ${\rm Im}(\nu)$ is much smaller than ${\rm Re}(\nu)$ (Figs.~1 and 2 of \textit{Be15a}; Fig.~3 of \textit{Be15b}).
One way of implementing the uniform $\nu$ condition consists in working in the incompressible limit, because the viscoelastic Poisson's ratio of an incompressible material is always equal to 0.5.

When deriving thin elastic shell equations, one starts by linearly expanding strains and stresses around the middle surface of the shell (Appendices~\ref{StrainThinShell} and \ref{StressThinShell}).
Next, the stress and moment resultants are evaluated by computing the zeroth and first moments of the linearized stress over the shell thickness, which are then used in the equations of equilibrium (Appendix~\ref{StressThinShell}).
This procedure accounts for the appearance in the governing equations of the macroscopic shell parameters $1/\alpha\sim{}\mu{}d$ and $D\sim{}\mu{}d^3$.
But why is there no parameter proportional to $\mu{}d^2$?
Let us define the normalized moments of the shear modulus by
\begin{equation}
\mu_p = \frac{1}{d^{p+1}} \int_d \mu \, \zeta^p \, d\zeta \, ,
\label{momentsmu}
\end{equation}
where $\zeta$ is the radial coordinate measured with respect to the reference surface of radius $R_0$.
The zeroth moment $\mu_0$ (called \textit{effective shear modulus} in \textit{Be15a}) and the second moment $\mu_2$ give rise to the parameters $1/\alpha$ and $D$, respectively. 
If the shell is radially homogeneous, the first moment $\mu_1$ does not give rise to any macroscopic shell parameter (proportional to $\mu{}d^2$) because choosing the middle surface as the reference surface implies that the first moment vanishes.
If the shell thickness varies laterally, $\mu_1 = 0$ if thickness variations are symmetric with respect to the middle surface (an assumption implicitly made in \textit{Be08}).

If thickness variations are not symmetric about the middle surface or if the shell is radially inhomogeneous, the first moment $\mu_1$ with respect to the middle surface does not vanish.
In special cases (elastic and laterally uniform shells), one can find another reference surface so that $\mu_1=0$ everywhere (\citet{axelrad1987}, Eq.~(1.70)).
This is not possible here for two reasons.
First, such a reference surface is not spherical if the properties of the shell vary laterally.
Second, a viscoelastic shell has a complex shear modulus, so that the condition $\mu_1=0$ imposes two constraints on the location of the reference surface.
I will thus formulate the thin shell equations for a reference surface which is arbitrary except that it is spherical and located (if possible) within the shell boundaries.
An important observation in that regard is that the effective shear modulus $\mu_0$ is invariant under a change of the reference surface, and this is also true of the following combination (see Appendix~\ref{Covariance}):
\begin{equation}
\mu_{\rm inv} = \mu_2 - \frac{(\mu_1)^2}{\mu_0} \, .
\label{muinv}
\end{equation}

%=========================================================================================
%=========================================================================================
\subsection{Governing equations}
\label{Governing}

Consider a spherical shell which is thin, viscoelastic and fully non-uniform, in the sense that the shell thickness is variable and the rheology is radially and laterally variable, although Poisson's ratio must be uniform.
The shell is submitted to a transverse load $q$, measured per unit area of the spherical reference surface (of radius $R_0$) and positive outward.
Following closely the method of \textit{Be08}, I derive the governing equations for the radial displacement $w$ and the auxiliary stress function $F$ (see Appendices~\ref{FirstGoverningEquation} and \ref{SecondGoverningEquation}). 
Since there is no tangential load, the resulting equations (Eqs.~(\ref{master1}) and (\ref{master2})) can be written as
\begin{eqnarray}
{\cal C}( D \,; w ) - (1-\nu) \, {\cal A}( D \,; w) + R_0^{\,\,3} \, {\cal A}(\chi \,; F) 
&=& R_0^4 \, q \, ,
\nonumber \\
{\cal C}( \alpha  \,;  F ) - (1+\nu) \, {\cal A}( \alpha \,; F) - R_0^{-1} \, {\cal A}(\chi \,; w)
&=& 0 \, .
\label{GoverningEq}
\end{eqnarray}
The \textit{extensibility} $\alpha$ and \textit{bending rigidity} $D$ of the shell are defined by
\begin{equation}
\left(  \alpha\, , \, D \right)
= \chi\psi \left( \alpha_{\rm inv} \, , \, D_{\rm inv} \right) ,
\label{defalphaD}
\end{equation}
where
\begin{eqnarray}
\left( \alpha_{\rm inv} \, , \, D_{\rm inv} \right) &=& \left( \frac{1}{2\left(1+\nu\right) \mu_0 \, d} \, , \, \frac{2 \, \mu_{\rm inv} \, d^3}{1-\nu} \right) .
\label{defalphaDinv}
\end{eqnarray}
As their subscript indicates, $\alpha_{\rm inv}$ and $D_{\rm inv}$ are invariant under a change of the reference surface. 
Recall that the moments $(\mu_0,\mu_1,\mu_2)$ and $\mu_{\rm inv}$ are defined by Eqs.~(\ref{momentsmu}) and (\ref{muinv}).
The parameters $\chi$ and $\psi$ are given by
\begin{eqnarray}
\left( \chi \, , \, \psi \right) &=& \left( \frac{\mu_0 + \varepsilon \mu_1}{\mu_0 + 2 \varepsilon \mu_1 + \varepsilon^2 \mu_2} \, , \, \frac{\mu_0}{\mu_0+\varepsilon \mu_1} \right) ,
\label{defchiparam}
\end{eqnarray}
where $\varepsilon$ is the relative shell thickness (the expansion parameter of thin shell theory):
\begin{equation}
\varepsilon=d/R_0 \, .
\label{defepsilon}
\end{equation}
The parameters $(\chi,\psi)$ are complex, nondimensional, close to one and coincide to ${\cal O}(\varepsilon)$:
\begin{equation}
\chi \approx \psi \approx  1 - (\mu_1/\mu_0) \, \varepsilon \, .
\end{equation}
To sum up, the governing equations explicitly depend on two 2D variables $(w,F)$ and four viscoelastic 2D parameters $(\nu,\alpha,D,\chi)$, all complex and varying laterally (except $\nu$) on the reference surface of the shell. Note that the parameter $\psi$ is absorbed into $\alpha$ and $D$.

The governing equations satisfy the following properties, which are non-trivial and can be seen as additional checks on the validity of the governing equations:
\begin{enumerate}
\item
They reduce to the elastic governing equations (Eq.~(\ref{ElasticGoverningEq})) if the shell is elastic and homogeneous, if the middle surface is chosen as reference surface, and if thickness variations are symmetric with respect to the middle surface of the shell.
If these three conditions are met, the various parameters become:
$\nu=\nu_e$, $\mu_0=\mu_e$, $\mu_1=0$, $\mu_2=\mu_{\rm inv}=\mu_e/12$, $\chi=\chi_e$, $\psi=1$, $\alpha=\chi_e\alpha_e$, and $D=\chi_eD_e$.
\item
They are \textit{scalar}, i.e.\ they are invariant under 2D coordinate transformations on the reference surface.
\item
They are \textit{covariant under a change of the reference surface} (see Appendix~\ref{Covariance}).
Covariance means that the equations do not change their form - even though parameters and variables are transformed - if the radius $R_0$ of the reference surface is redefined.
It ensures that changing the reference surface does not affect observable quantities (such as radial displacement, strains, and stresses).
\item
They are \textit{invariant under degree-1 transformations of the radial displacement (rigid displacements) and of the stress function (gauge freedom)}.
In other words, the degree-1 harmonic components of $(F,w)$ drop out of the equations (Eq.~(\ref{nodeg1})).
\item
Their LHS are zero if projected on spherical harmonics of degree~1 (Eq.~(\ref{nodeg1image})). This requires that \textit{the sum of the external forces vanishes} (Eq.~(\ref{qOmega1})).
\item
The two equations have a strikingly similar form.
This reflects the \textit{static-geometric duality} \citep{goldenveizer1961,axelrad1987,wan1988} which exchanges their LHS: 
\begin{eqnarray}
\left( \, D \, , \, \nu \, , \, \chi \, \right) &\leftrightarrow& \left( -\alpha \, , -\nu \, , \, \chi \, \right) ,
\nonumber \\
w &\leftrightarrow& R_0^2 \, F \, .
\label{duality}
\end{eqnarray}
\end{enumerate}

The governing equations result from the second-order approximation of thin shell theory, meaning that they include small terms beyond the leading order (or first-order) approximation in the shell thickness.
The second-order approximation, however, is not fully satisfactory because it includes only a part of the next-to-leading corrections: for example, the radial displacement remains constant with depth in the second-order approximation.
However, we know from thick shell theory that the radial displacement within a thin shell is not constant but varies nearly linearly with depth (see Eq.~(50) of \textit{Be15b}).
Thus, the theory of thin elastic shells is usually restricted to the first-order approximation (see Eq.~(\ref{ElasticGoverningEqApprox})).
Should we do the same for the theory of non-uniform thin viscoelastic shells?
In Appendix~\ref{FirstOrderApprox}, I show that, for a laterally non-uniform shell, not only there is no computational advantage in doing the first-order approximation but that it would be an inconsistent procedure.
The task of classifying terms as being of first-order (leading order), second-order (next-to-leading order), or higher order is facilitated by the Thin Shell Expansion rule (or TSE rule) derived in Appendix~\ref{TSErule}.

As reference surface, I will choose the sphere with radius $R$ (the mean surface radius of the body) because the same radius appears in the tidal forcing term.
Another possibility consists in minimizing the influence of the spatially uniform part of $\chi$ by choosing a preferred reference surface such that
\begin{equation}
\langle \, {\rm Re}(\chi) \, \rangle = 1
\hspace{3mm} \leftrightarrow \hspace{3mm}
\delta{}R \approx \langle \, {\rm Re}(\mu_1/\mu_0) \, d \, \rangle \, ,
\label{etaAverage}
\end{equation}
where $\langle \, . \, \rangle$ denotes angular averaging over the whole sphere.
The second equation gives the radial coordinate shift to ${\cal O}(\epsilon)$ required to go from the original reference frame to the preferred reference frame.
If the shell is uniform, the latter coincides with the middle surface.
In any case, one should avoid the stronger constraint $\langle\chi\rangle=1$ because it does not correspond to any reference surface and leads to inconsistencies when computing the tidal heating rate (see second paper).

%%%%%%%%%%%%%%%%%%%%%%%%%%%%%%%%%%%%%%%%%%%%%%%%%%%%%%%%%%%
%%%%%%%%%%%%%%%%%%%%%%%%%%%%%%%%%%%%%%%%%%%%%%%%%%%%%%%%%%%
%%%%%%%%%%%%%%%%%%%%%%%%%%%%%%%%%%%%%%%%%%%%%%%%%%%%%%%%%%%
\section{Tidal coupling}

%=========================================================================================
%=========================================================================================
\subsection{Tidal forcing and massless shell approximation}
\label{MasslessShell}

For eccentricity tides, the Fourier-transformed tidal potential reads (e.g.\ \citet{beuthe2013}):
\begin{equation}
U = (\omega R)^2 \, e \Big(
- (3/2) \, P_{20}(\cos\theta) \, + \, (1/8) \, P_{22}(\cos\theta) \left( - e^{2i\varphi} + 7 e^{-2i\varphi} \right)
\Big) \, ,
\label{TidalPot}
\end{equation}
where $P_{nm}$ are the associated Legendre functions of degree $n$ and order $m$, $\theta$ is the colatitude, and $\varphi$ is the longitude.
In order to compute the tidal response of the shell, I need to specify how the transverse load $q$ is related to the tidal forcing and to the self-gravity of the whole body.

Thin shell theory typically deals with mechanical structures in a constant gravity field.
This theory is thus not well suited to describe a shell with gravitational couplings to direct and induced gravitational potentials.
Things become much easier, however, if the density contrast between shell and ocean is negligible with respect to the ocean density.
In that case, the mass of the shell can be included in the ocean so that the shell is treated as a viscoelastic massless layer.
This assumption was adopted in the \textit{massless membrane approach} which approximates very well the tidal Love numbers of a satellite with an ocean below a thin ice shell (see \textit{Be15a} for a detailed analysis).
It also fits neatly within the more general framework of the massive membrane approach (see \textit{Be15b}, Appendix~B).
For Enceladus, the error due to the membrane or thin shell approximation on the tidal Love number $h_2$ is less than $\epsilon=d/R$ if the shell is homogeneous and elastic (see Fig.~\ref{FigLoveElastic}); the error is smaller if part of the shell convects because the shell behaves as if it were thinner.

As the shell is massless and has a free outer surface, the transverse load acts on the bottom surface of the shell.
In the spherically symmetric case, it is equal to the water column pressure (of density $\rho$) resulting from the difference between the tidal bulge and the geoid $w_g$ (\textit{Be15a}, Eq.~(31)):
\begin{equation}
q = - \rho g \left( w - w_g \right) ,
\label{TransverseLoad1}
\end{equation}
where $(q,w,w_g)$ are all of harmonic degree~2 if the tidal forcing is given by Eq.~(\ref{TidalPot}).
To first order in the deformation and in the static approximation, $w_g=\Gamma/g$ where $\Gamma$ is the total perturbing potential, i.e.\ the sum of the tidal potential and the induced potential due to the deformation of the body.
If the shell is laterally non-uniform, the above relation can be extended to all degrees $n\geq2$ (see Section~\ref{Loads01} for $n=0$ and 1):
\begin{equation}
q_n = - \rho g  \left( w_n - \Gamma_n/g \right) .
\label{TransverseLoadn}
\end{equation}
Terms with $n>2$ can be interpreted as restoring forces since they do not include a direct contribution from the external forcing $U$.
I will use the massless shell approximation in the rest of the paper.
Its effect will be quantified in Section~\ref{LaterallyUniformThickShell}.

%=========================================================================================
%=========================================================================================
\subsection{Tidal load if rigid core and homogeneous ocean}

Computing the geoid for the transverse load (Eq.~(\ref{TransverseLoadn})) is simple if the ocean is homogeneous and the core (or mantle) below the ocean is infinitely rigid because the total perturbing potential $\Gamma$ depends only on the tidal potential $U$ and on the surface deformation $w$ (\textit{Be16}, Eq.~(10)),
\begin{equation}
\Gamma_n = U_n + g \, \xi_n \, w_n \, ,
\label{geoidR}
\end{equation}
where $\xi_n$ is the degree-$n$ ratio of the ocean density $\rho$ to the bulk density $\rho_b$:
\begin{equation}
\xi_n = \frac{3}{2n+1} \, \frac{\rho}{\rho_b} \, .
\label{xin}
\end{equation}
The second term in the RHS of Eq.~(\ref{geoidR}) represents the gravitational contribution of the ocean bulge (self-attraction or self-gravity).
If $n\geq2$, the transverse load thus reads
\begin{equation}
q_n = -\rho g \Big( ( 1-\xi_n ) \, w_n -U_n/g \Big) \, .
\label{qnrigid}
\end{equation}
For Enceladus, the error due to the approximation of an infinitely rigid core is small (see Fig.~3 of \textit{Be16}).
If $d/R<10\%$, the error on the radial displacement is less than 0.5\% if the core is silicate-rich ($\mu_m=40\rm\,GPa$) whereas the error is less than 5\% if the core has the rigidity of ice ($\mu_m=3.5\rm\,GPa$).

%=========================================================================================
%=========================================================================================
\subsection{Tidal load if viscoelastic core or inhomogeneous ocean}

A non-rigid core increases tidal deformations and generates additional tidal dissipation which could significantly contribute to the total heat budget.
In order to deal with this, I will couple the thin shell equations to a viscoelastic core (the case of a inhomogeneous ocean is solved at the same time).
If the whole body has a spherically symmetric structure, the spherical harmonics of the perturbing and tidal potentials are linearly related: $\Gamma_n=(1+k_n)U_n$ where $k_n$ is the degree-$n$ gravitational tidal Love number (see Eq.~(\ref{wnU}) below).
However, spherical symmetry is lost if the shell is laterally non-uniform.
A similar situation arises when studying dynamical tides in subsurface oceans (see \textit{Be16}).
In both cases, the massless shell does not contribute gravitationally to the perturbing potential: it only acts as a pressure load deforming the ocean and the core beneath it.
The perturbing potential $\Gamma$ can thus be computed as if the shell had no rigidity (`fluid-crust body') but the body must be submitted to an additional pressure load $U^P$ given by (\textit{Be16}, Eq.~(E.1))
\begin{equation}
U_n^P = \xi_n \, q_n/\rho \, .
\label{UnP}
\end{equation}
If the structure below the shell is spherically symmetric, the perturbing potential and the radial displacement can be expressed in terms of the tidal potential and the pressure load as (\textit{Be16}, Eqs.~(20)-(22)):
\begin{eqnarray}
\Gamma_n &=& \left(1+k_n^{\circ}\right) U_n + k_n^{\circ P} \, U_n^P \, ,
\label{GammaLove} \\
w_n &=& h_n^\circ \, U_n/g + h_n^{\circ P} \, U_n^P/g \, ,
\label{wn}
\end{eqnarray}
where $(k_n^{\circ},h_n^\circ)$ are the tidal Love numbers of the fluid-crust body, whereas $(k_n^{\circ P},h_n^{\circ P})$ are the pressure Love numbers of the fluid-crust body.
The former satisfy the well-known equipotential surface constraint,
\begin{equation}
k_n^\circ + 1 = h_n^\circ \, .
\label{knhn0}
\end{equation}
whereas the latter are given by
\begin{eqnarray}
k_n^{\circ P} &=& - h_n^\circ \, ,
\label{kPhn} \\
h_n^{\circ P} &=& -1/\xi_n - h_n^\circ  \, .
\label{hnPhn}
\end{eqnarray}
These two relations can be seen as the application to a fluid-crust body of the general relations between pressure, tidal, and load Love numbers: $(k_n^P,h_n^P)=(-h_n^T,h_n^L-h_n^T)$ (\textit{Be16}, Eq.~(E.5)).
Eq.~(\ref{hnPhn}) is also required for consistency between Eq.~(\ref{TransverseLoadn}) and Eqs.~(\ref{UnP}) to (\ref{kPhn}).
Thus all fluid-crust Love numbers are known once $h_n^\circ$ has been computed.

Solving Eqs.~(\ref{TransverseLoadn}) and (\ref{GammaLove}), I can write $\Gamma_n$ and $q_n$ in terms of $w_n$ and $U_n$ ($n\geq2$):
\begin{eqnarray}
\Gamma_n &=& \upsilon_n \left( g \, \xi_n \, w_n + U_n \right) ,
\label{GammaSol} \\
q_n &=& -\rho g \Big( (1 - \upsilon_n \, \xi_n ) \, w_n - \upsilon_n \, U_n/g \Big) \, ,
\label{qSol}
\end{eqnarray}
where all the Love numbers have been expressed in terms of $h_n^\circ$ with Eqs.~(\ref{knhn0}) to (\ref{hnPhn}).
The parameter $\upsilon_n$ ($n\geq2$) is defined by
\begin{equation}
\upsilon_n = \frac{h_n^\circ}{1+\xi_n h_n^\circ} \, .
\label{upsilon}
\end{equation}
If the core is infinitely rigid and the ocean is homogeneous, Eq.~(\ref{qSol}) should reduce to Eq.~(\ref{qnrigid}), which is the case if
\begin{equation}
\upsilon_n=1
\hspace{2mm} \Leftrightarrow \hspace{2mm}
h_n^\circ = \frac{1}{1- \xi_n} \, .
\label{upsi1}
\end{equation}
This is indeed the radial Love number of a two-layer body made of an infinitely rigid core and a homogeneous surface ocean (Eq.~(\ref{hn0})).
If the core is viscoelastic and the ocean is homogeneous, $h_n^\circ$ is given instead by Eq.~(\ref{hn0visco}).
The latter formula shows that the factor $\upsilon_n$ is close to one:
if the core is nearly fluid ($\hat\mu_m\ll1$), $h_2^\circ\approx1.97$ for Enceladus ($R_c/R=0.76$, $\rho/\rho_b=0.62$) so that $\upsilon_2\approx1.13$.
As the harmonic degree $n$ increases, $\upsilon_n$ decreases and tends to one.

%=========================================================================================
%=========================================================================================
\subsection{Loads of degrees 0 and 1}
\label{Loads01}

Loads of harmonic degree~1 vanish otherwise the shell would accelerate (Eq.~(\ref{qOmega1}); Section 4.3 of \textit{Be08}).
Deformations of degree 1 are allowed, but they do not affect the thin shell equations because they represent rigid displacements (Eq.~(\ref{S1w1}); Section 7.6 of \textit{Be08}).

Tides do not include a degree-0 term, though one can define a degree-0 tidal Love number for rotational deformations (see Section~2.5 of \citet{saito1974}).
Nevertheless, elastic couplings between degree~0 and higher degrees (in particular degree~2) can change the mean radius (degree-0 deformation) if the layers below the shell are compressible.
This does not lead to degree-0 surface loading as in Eq.~(\ref{TransverseLoadn}) because the total mass is conserved.
The compressed ocean, however, reacts by exerting a pressure on the shell $q_0$ (positive outwards).
The pressure on the ocean (equal to $q_0$ if taken to be positive inwards) and the radial displacement $w_0$ are related by Eqs.~(\ref{UnP})-(\ref{wn}) in which $U_0=0$:
\begin{equation}
q_0 = \frac{1}{\xi_0 \, h_0^{\circ P}} \, \rho g \, w_0 \, .
\label{Love0}
\end{equation}
The pressure Love number $h_0^{\circ P}$, a negative quantity, is computed in the same way as the load Love number of degree~0, by propagating the radial displacement $y_1(r)$ and the radial stress $y_2(r)$ from the centre to the ocean surface and imposing the boundary condition $y_2(R)=-\rho_b/3$ (e.g.\ Section 2.2 of \citet{saito1974} in which the parameter $k^2$ should read
$k^2=16\pi{}G\rho_b/3v_P^2$ with $v_P$ the compressional wave velocity).
The other boundary condition for pressure Love numbers, $y_6(R)=0$ (see Appendix~E of \textit{Be16}), yields $k_0^{\circ P}=0$ as expected from mass conservation.

Below the shell, it is likely that most of the compression occurs in the ocean.
For Enceladus, $K_o/(\rho_bgR)\approx43\gg1$ (Table~\ref{TableParam}), meaning that compression is a small effect.
Under the assumption of uniform density, Eq.~(\ref{h0Papprox}) yields $h_0^{\circ P}\approx-0.00143$.
This value differs by 2\% from the result of the numerical integration of a 2-layer model in which the core is compressible and of higher density.
The small value of $h_0^{\circ P}$ suggests that the compressibility of the interior (excluding the icy shell) has a small impact on tidal deformations.
As an alternative to Eq.~(\ref{Love0}), one can assume that all layers below the shell are incompressible ($h_0^{\circ P}=w_0=0$), in which case $q_0$ replaces $w_0$ as a free variable.

%=========================================================================================
%=========================================================================================
%\subsection{Tidal deformations if laterally varying thin shell}
\subsection{Tidal thin shell equations}

In the spatial domain, the tidal load (Eq.~(\ref{qSol})) can be compactly written as
\begin{equation}
q = -\rho g \left( w - {\cal G}(w) - \, U_\upsilon/g \right) ,
\label{TransverseLoad2}
\end{equation}
where
\begin{eqnarray}
{\cal G}(w) &=& \sum_{n\neq1} \upsilon_n \, \xi_n  \, w_{n} \, ,
\nonumber \\
U_\upsilon &=& \sum_{n\geq2} \upsilon_n \, U_n \, ,
\label{Gw}
\end{eqnarray}
in which $\xi_n$ is given by Eq.~(\ref{xin}), $\upsilon_n$ is given by Eq.~(\ref{upsilon}) if $n\geq2$, and $\upsilon_0$ reads
\begin{equation}
\upsilon_0 = \frac{1}{\xi_0} \left( 1 +  \frac{1}{\xi_0 h_0^{\circ P}} \right) .
\label{upsilon0}
\end{equation}
The tidal thin shell equations result from substituting the tidal load into the flexure equations in which $R_0=R$ (Eq.~(\ref{GoverningEq})):
\begin{eqnarray}
{\cal C} ( D \,; w ) - (1-\nu) \, {\cal A}( D \,; w) + R^{\,3} \, {\cal A}(\chi \,; F) 
&=& - \rho g R^4 \left(w - {\cal G}(w)  - U_\upsilon/g \right) ,
\nonumber \\
{\cal C} ( \alpha \,; F ) - (1+\nu) \, {\cal A}(\alpha \,; F) - R^{-1} \, {\cal A}(\chi \,; w)
&=& 0 \, .
\label{TidalThinShellEq}
\end{eqnarray}

%=========================================================================================
%=========================================================================================
\subsection{Tidal stress}
\label{TidalStressSec}

How does crustal non-uniformity affect tidal stresses?
I will not study the stress tensor component by component, because which ones matter depends on the type of tectonic analysis (fault initiation and propagation, faulting style, closing/opening of faults).
Instead, I will estimate the stress magnitude with the second invariant of the deviatoric stress, a positive definite scalar quantity appearing often in failure criteria \citep{jaeger2007}:
\begin{eqnarray}
J_2(t)
&=& \frac{1}{2} \left( {\rm Tr}(\tilde{\bm{\sigma}} \cdot \tilde{\bm{\sigma}}) - \frac{1}{3} \left( {\rm Tr}\,\tilde{\bm{\sigma}} \right)^2 \right) ,
\label{inv2time}
\end{eqnarray}
where the tilde denotes time-domain variables (as in Eq.~(\ref{fourier})).
Following usual conventions, $\tilde{\bm{\sigma}}$ is the $3\times3$ matrix associated with the full stress tensor (matrices are noted in boldface), $\bm{I}$ is the unit matrix, the dot denotes the matrix product, and Tr denotes the trace of the matrix.
In terms of Fourier components, the trace invariants read
\begin{eqnarray}
{\rm Tr}\,\tilde{\bm{\sigma}} &=&
 {\rm Re} \Big( {\rm Tr} \, \bm{\sigma} \, e^{i \omega t} \Big) \, ,
\nonumber \\
{\rm Tr}(\tilde{\bm{\sigma}} \cdot \tilde{\bm{\sigma}}) &=&
\frac{1}{2} \, {\rm Tr}(\bm{\sigma} \cdot \bm{\sigma}^*)
+ \frac{1}{2} \, {\rm Re} \Big( {\rm Tr}(\bm{\sigma} \cdot \bm{\sigma}) \, e^{2 i \omega t} \Big) \, .
\end{eqnarray}
The trace invariants in the frequency domain, ${\rm Tr}\,\bm{\sigma}$, ${\rm Tr}(\bm{\sigma} \cdot \bm{\sigma}^*)$, and ${\rm Tr}(\bm{\sigma} \cdot \bm{\sigma})$, can be expressed in terms of the scalar operators $\Delta'$ and ${\cal A}$ acting on $F$ and $w$ (Eq.~(\ref{inv2Fw})).
Such formulas are easy to handle because they can be evaluated with back-and-forth spectral transforms (\textit{transform method}): scalar differential operators (expressed in terms of Laplacians, see Eq.~(\ref{Aharmonics})) are evaluated in the spherical harmonic domain whereas products of non-uniform parameters and variables are computed on a grid in the spatial domain (the size of the grid is doubled in order to avoid aliasing).
With this method, it is not necessary to compute stress components by partial differentiation, a somewhat tricky procedure in spherical coordinates because of the apparent singularities at the poles.

In this paper, I will focus on the time average of $J_2(t)$ evaluated at the surface.
Crustal non-uniformity indeed affects in a similar way the time average and the time-dependent part of $J_2$. 
Moreover, surface stresses are of particular interest because faults form and are observed at the surface.
I thus quantify the time-averaged deviatoric stress at the surface with
\begin{equation}
\widehat\sigma_{\rm dev} = \sqrt{ \widehat J_2}\Big |_{r=R}  \, ,
\label{sigdev}
\end{equation}
where $\widehat J_2$ is the time average of $J_2(t)$:
\begin{eqnarray}
\widehat J_2
&=& \frac{1}{4} \left( {\rm Tr}(\bm{\sigma} \cdot \bm{\sigma}^*) - \frac{1}{3} \left| {\rm Tr}\,\bm{\sigma} \right|^2 \right) ,
\label{J2hat}
\end{eqnarray}
in which ${\rm Tr}(\bm{\sigma} \cdot \bm{\sigma}^*)$ and ${\rm Tr}\,\bm{\sigma}$ are computed with Eqs.~(\ref{inv2Fw})-(\ref{inv2Comp}).

In a similar fashion, I measure the magnitude of the resultant stress with $\widehat N_{\rm dev} = \sqrt{ \widehat N_2}$
where $\widehat N_2=({\rm Tr}\,(\bm{N} \cdot \bm{N}^*)-|{\rm Tr}\,\bm{N} |^2/3)/4$ is the time average of the second invariant of the deviatoric resultant stress.
$\bm{N}$ is the $2\times2$ matrix associated with the tensor $N_{\alpha\beta}$.
Similarly to $\widehat J_2$, it can be evaluated in terms of $F$ and $w$ by substituting $\bm{\sigma}\rightarrow\bm{N}$, $E\rightarrow1$, $\alpha\rightarrow\chi$, and $z\rightarrow-D(1-\nu^2)/R_0^2$ in Eqs.~(\ref{inv2Fw})-(\ref{inv2Comp}).

%%%%%%%%%%%%%%%%%%%%%%%%%%%%%%%%%%%%%%%%%%%%%%%%%%%%%%%%%%%
%%%%%%%%%%%%%%%%%%%%%%%%%%%%%%%%%%%%%%%%%%%%%%%%%%%%%%%%%%%
%%%%%%%%%%%%%%%%%%%%%%%%%%%%%%%%%%%%%%%%%%%%%%%%%%%%%%%%%%%
\section{Benchmarking against a laterally uniform thick shell}

In this section, I benchmark the solution of the tidal thin shell equations against the solution for a laterally uniform thick shell.
First, I recall how tidal deformations and tidal stresses at the surface are related to tidal Love numbers.
After deriving formulas for Love numbers in the thin shell approximation, I compare them to the exact results (`thick shell' Love numbers), either analytically if the shell is fully uniform, or numerically if the rheology varies radially.

%=========================================================================================
%=========================================================================================
\subsection{The central role of tidal Love numbers}
\label{CentralRoleLove}

If the whole body has a spherically symmetric structure, the problem can be solved degree by degree in terms of the degree-$n$ tidal potential $U_n$.
In that case, the surface values of the radial displacement $w_n$, the lateral displacement potential $S_n$, and the perturbing potential $\Gamma_n$ are proportional to $U_n$ (e.g.\ \citet{saito1974}):
\begin{equation}
\left( w_n \, , S_n \, , \frac{\Gamma_n}{g} \right) = \Big( h_n \, , l_n \, , k_n+1 \Big) \frac{U_n}{g}  ,
\label{wnU}
\end{equation}
where $h_n$, $l_n$, and $k_n$ are the radial, tangential, and gravitational tidal Love numbers, respectively.
These numbers encapsulate the effect of the internal structure on the surface tidal deformations.
In particular, surface tidal stresses can be written in terms of the viscoelastic moduli at the surface, the tidal Love numbers, and the tidal potential:
\begin{eqnarray}
\sigma_{\theta\theta} &=& \frac{2\mu}{1-\nu} \, \Big( h_n \, (1+\nu) + l_n \left( {\cal O}_1 -1 + \nu \, ( {\cal O}_2  - 1) \right) \Big) \, \bar U_n \, ,
\nonumber \\
\sigma_{\varphi\varphi} &=& \frac{2\mu}{1-\nu} \, \Big( h_n \, (1+\nu) + l_n \left( {\cal O}_2 - 1 + \nu \, ( {\cal O}_1 - 1 ) \right) \Big) \,  \bar U_n \, ,
\nonumber \\ 
\sigma_{\theta\varphi} &=& 2\mu \, l_n  \, {\cal O}_3  \,  \bar U_n \, ,
\label{tidalstress}
\end{eqnarray}
where $\bar U_n=U_n/(gR)$ is nondimensional and all quantities are defined at the surface (Eq.~(62) of \textit{Be15a} or equivalently Eqs.~(B.11)-(B.13) of \citet{wahr2009}).
Thus, the magnitude of the stresses and their  time lag with respect to the tidal potential can be studied in the frequency domain through the magnitudes and phases of the complex numbers $(h_n,l_n)$.

Furthermore, the time-averaged surface deviatoric stress $\widehat \sigma_{\rm dev}$ (Eq.~(\ref{sigdev})) can be directly evaluated in terms of $(h_2,l_2)$ and of $|U_2|^2$, the squared norm of the degree-2 tidal potential.
The deviatoric stress invariant $\widehat J_2$ can indeed be factorized in radial and angular parts if the internal structure is spherically symmetric, similarly to the tidal dissipation rate in \citet{beuthe2013}:
\begin{eqnarray}
\widehat J_2
&=&  |\mu|^2 \left( {\rm Tr}(\bm{\epsilon} \cdot \bm{\epsilon}^*) - \frac{1}{3} \left| {\rm Tr}\,\bm{\epsilon} \right|^2 \right) ,
\nonumber \\
&=&  |\mu|^2 \, \frac{e^2(\omega R)^4}{r^2} \, \left( f_A \Psi_A + f_B \Psi_B + f_C \Psi_C \right) ,
\label{J2hatABC}
\end{eqnarray}
where $\bm{\epsilon}$ is the $3\times3$ matrix associated with the full strain.
The radial functions $f_{A,B,C}$ depend on the interior structure (Eq.~(24) of \citet{beuthe2013}) whereas the angular functions $\Psi_{A,B,C}$ depend on the tidal potential.
The angular functions have a simple spherical harmonic expansion (Eq.~(36) and Table~1 of \citet{beuthe2013}, with the difference that the squared eccentricity $e^2$ has been factorized in Eq.~(\ref{J2hatABC}) above).
At the surface, the radial functions are related to $h_2$ and $l_2$ (Eq.~(29) of \citet{beuthe2013}):
\begin{equation}
\left( f_A \, , f_B \, , \, f_C \right)\Big|_{r=R} = \left( \frac{4}{3} \, \left| \frac{1+\nu}{1-\nu} \right|^2 \frac{ \left| h_2 - 3 l_2 \right|^2}{g^2} \, , \, 0  \, , \, 24 \, \frac{ \left| l_2 \right|^2}{g^2} \right) .
\label{fABC}
\end{equation}
Therefore, it is not necessary to compute the stress components: $\widehat \sigma_{\rm dev}$ can be directly evaluated once the tidal Love numbers are known.
Note that $f_A$ is much more sensitive than $f_C$ to errors on Love numbers because of the partial cancellation between $h_2$ and $3l_2$.

%=========================================================================================
%=========================================================================================
\subsection{Thin shell}
\label{LaterallyUniformShell}

%************************************************************************************************************************************
\subsubsection{The different forms of the flexure equation}
\label{DifferentForms}

If the shell is laterally uniform but with depth-dependent rheology, the governing equations (\ref{GoverningEq}) become (with $R_0=R$)
\begin{eqnarray}
D \left( \Delta' - 1 + \nu  \right) \Delta'  w + \chi R^3 \, \Delta' F &=& R^4 \, q  \, ,
\nonumber \\
\alpha \left( \Delta' - 1 - \nu \right) \Delta' F - \chi R^{-1} \, \Delta' w &=& 0 \, ,
\label{GoverningEqU}
\end{eqnarray}
where $(\alpha,D,\chi)$ are given by Eq.~(\ref{defalphaD}) and $\Delta'$ is given by Eq.~(\ref{DeltaPrime}).
I work in the second-order approximation so as to maintain numerical consistency between the cases of laterally varying and laterally uniform shells (this is convenient when benchmarking numerical codes).
Eliminating $\Delta'F$, I get a 6th-order equation for $w$:
\begin{equation}
D \left( \Delta \, \Delta' + 1 - \nu^2 \right) \Delta' w + \frac{\chi^2}{\alpha} \, R^2 \, \Delta' w = R^4\left( \Delta'- 1 - \nu \right) q \, .
\label{FlexConst1}
\end{equation}
The term proportional to $D(1 - \nu^2)\Delta'w$ is of next-to-leading order according to the TSE rule.
If I drop this term and impose $\chi\approx\psi\approx1$ (first-order approximation), Eq.~(\ref{FlexConst1}) becomes identical to the flexure equation for a homogeneous shell of constant thickness given by \citet{willemann1982} (their Eq.~(3)):
\begin{equation}
D_{\rm inv} \, \Delta \, \Delta' \Delta' w + \frac{R^2}{\alpha_{\rm inv}} \, \Delta' w = R^4\left( \Delta'- 1 - \nu \right) q \, .
\label{FlexConst2}
\end{equation}
As shown by Eqs.~(\ref{FlexConst1}) and (\ref{FlexConst2}), the standard flexure equation can take different forms because terms like $D\Delta'\Delta'w$ and $D\Delta'w$ can be added without affecting the leading order terms.
What really matters is that the operators acting on $w$ vanish if $w$ is of harmonic degree~1 (rigid displacement of the whole shell, see \textit{Be08}).
This is true of $\Delta'$ in Eq.~(\ref{FlexConst2}).
By contrast, the flexure equation of \citet{turcotte1981} does not satisfy this property because of a faulty derivation in \citet{kraus1967}.
In spite of being corrected by \citet{willemann1982}, this error is common in the literature.
It can lead to numerical problems if the properties of the shell vary laterally because of couplings between different harmonic degrees.

%************************************************************************************************************************************
\subsubsection{Thin shell spring constants}

The differential operators appearing in Eqs.~(\ref{GoverningEqU}) to (\ref{FlexConst2}) (i.e.\ $\Delta$ and $\Delta'$) have spherical harmonics as eigenfunctions.
It is thus convenient to expand variables in spherical harmonics denoted by their degree $n$: $F=\sum_n F_n$, $w=\sum_n w_n$, $q=\sum_n q_n$.
In this basis, Eq.~(\ref{FlexConst1}) takes the form of a Hooke's law:
\begin{equation}
q_{n}= \rho g \, \Lambda_n w_{n} \, ,
\label{hooke}
\end{equation}
where $\rho$ is the shell density, $g$ is the surface gravity, and $\Lambda_n$ is the \textit{thin shell spring constant} given by
\begin{equation}
\Lambda_n = \Lambda_n^M + \Lambda_n^B\, + \Lambda_n^{corr} \, ,
\label{LambdaTS}
\end{equation}
where the \textit{membrane spring constant} $\Lambda_n^M$ and \textit{bending spring constant} $\Lambda_n^B$ are defined by
\begin{eqnarray}
\Lambda^M_n &=& 2 \, \frac{ \delta'_n\left(1+\nu\right)}{\delta'_n-1-\nu} \, \frac{\mu_0}{\rho g R} \, \frac{d}{R} \, ,
\nonumber \\
\Lambda^B_n &=& \left( \delta'_n-1+\nu \right) \delta'_n \frac{D_{\rm inv}}{\rho{}gR^4} \, .
\label{LambdaMB}
\end{eqnarray}
The definition  of $\Lambda_n^M$ coincides with Eq.~(28) of \textit{Be15a} (degree~2) and with Eq.~(71) of \textit{Be15b} (degree $n$).
The definition of $\Lambda_n^B$ slightly differs from the one in \textit{Be16} where it is based on Eq.~(\ref{FlexConst2}) instead of Eq.~(\ref{FlexConst1}).
The third term (\textit{correction term}) in the RHS of Eq.~(\ref{LambdaTS}) is a next-to-leading correction (recall that $\varepsilon=d/R$):
\begin{eqnarray}
\Lambda^{corr}_n
&=& \left( \chi/\psi -1 \right) \Lambda_n^M + \left( \chi \psi -1 \right) \Lambda_n^B
\nonumber \\
&\approx& -  \varepsilon^2 \, \frac{\mu_{\rm inv}}{\mu_0}  \, \Lambda_n^M - 2 \varepsilon \, \frac{\mu_1}{\mu_0} \, \Lambda_n^B \, .
\label{Lambdacorr}
\end{eqnarray}

%************************************************************************************************************************************
\subsubsection{Example: conductive shell}
\label{ConductiveShell}

Suppose that the icy shell of Enceladus has a uniform thickness equal to $23\rm\,km$ (see physical parameters in Table~\ref{TableParam}).
I assume that heat is transported through the shell by radial conduction; conductivity varies with temperature as $k_s=(651{\rm\,W/m})/T$ \citep{petrenko1999}.
The surface temperature $T_s$ is approximated by the equilibrium temperature \citep{depater2001}:
\begin{equation}
T_{\rm eq} = \left( {\cal F}_{\rm sat} \, \frac{1-A}{4\epsilon\sigma} \right)^{1/4} \approx 59 \, \rm K \, ,
\end{equation}
where ${\cal F}_{\rm sat}=14.8\rm\,W/m^2$ is the solar irradiance at Saturn, $A=0.81$ is the Bond albedo \citep{spencer2006}, $\epsilon$ is the emissivity ($\epsilon\!=\!1$ for a blackbody), and $\sigma\!=\!5.67\times10^{-8}\rm\,Wm^{-2}K^{-4}$ is the Stefan-Boltzmann constant.
In fact, the surface temperature is lower at the poles than at the equator as solar insolation decreases at higher latitude.
The parametrization of \citet{roberts2008}, however, is not correct: first, the formula of \citet{ojakangas1989a} fails at large obliquities (the formula of \citet{nadeau2017} should be used instead) and second, the equatorial temperature of $80\rm\,K$ is much higher than the equatorial temperature associated with the mean annual insolation.
In any case, the temperature variation (about $11\rm\,K$) is much smaller than for Europa because of the large obliquity (it will be taken into account in the second paper of the series).
The temperature at the bottom of the shell (of radius $R_m$) is equal to the melting temperature $T_m=273\rm\,K$.
If the influence of dissipation and spherical corrections are neglected, the solution of the heat equation is
\begin{equation}
T(r) = T_m^{\,\,(r-R_s)/(R_m-R_s)} \, T_s^{\,\,(R_m-r)/(R_m-R_s)} \, .
\label{Tprofile}
\end{equation}
The viscosity of ice is related to temperature by an Arrhenius relation (e.g.\ \citet{barr2009})
\begin{equation}
\eta(T) = \eta_{\rm melt} \exp \left( \frac{E_a}{R_gT_m}\left(\frac{T_m}{T}-1\right)\right) \, ,
\label{arrhenius}
\end{equation}
where $\eta_{\rm melt}$ is the viscosity at the melting temperature, chosen to be either $10^{13}$ or $10^{14}\rm\,Pa.s$ ($E_a=59.4\rm\,kJ\,mol^{-1}$ is the activation energy for diffusion creep and $R_g$ is the gas constant).
Rheology is of Maxwell type (see Eq.~(\ref{maxwell})).
The corresponding viscoelastic parameters are given in Table~\ref{TableExample}.

%TABLE 3
\begin{table}[ht]\centering
\ra{1.3}
\footnotesize
\caption[Viscoelastic parameters if laterally uniform conductive shell]
{\small Viscoelastic parameters for Enceladus if the icy shell is $23\rm\,km$ thick, laterally uniform, and radially conductive (Eqs.~(\ref{Tprofile})-(\ref{arrhenius})).
Poisson's ratio is uniform ($\nu=0.33$).
The values of $(\mu_1,\mu_2)$, $(\chi,\psi)$, and $\Lambda^{corr}_2$ depend on the reference surface of the shell, which is here the sphere of radius $R$ (the mean surface radius of the body).
}
\begin{tabular}{@{}lrrc@{}}
\hline
&  $\eta_{\rm melt}=10^{13}$ Pa.s & $\eta_{\rm melt}=10^{14}$ Pa.s & Unit \\
\hline
$\mu_0$ & $3.341 + 0.115 \, i$ & $3.485 + 0.049 \, i$ & GPa \\
$\mu_1$ & $-1.595 - 0.109 \, i$ & $- 1.735 - 0.048 \, i $ & GPa \\
$\mu_2$ & $1.016 + 0.103 \, i$ & $1.152 + 0.047 \, i$ & GPa \\
$\mu_{\rm inv}$ & $0.255 + 0.026 \, i$ & $0.288 + 0.011 \, i$ & GPa \\
$\chi$ & $1.0449 + 0.0016 \, i$ & $1.0468 + 0.0006 \, i$ & -  \\
$\psi$ & $1.0456 + 0.0016 \, i$ & $1.0476 + 0.0007 \, i$ & -  \\
$\Lambda^M_2$ & $21.267 + 0.730 \, i$ & $22.179 + 0.311 \, i$ & - \\
$\Lambda^B_2$ & $0.378 + 0.038 \, i$ & $0.426 + 0.017 \, i$ & - \\
$\Lambda^{corr}_2$ & $0.020 + 0.003 \, i$ & $0.024 + 0.002 \, i$ & - \\
\hline
\end{tabular}
\label{TableExample}
\end{table}%

%=========================================================================================
%=========================================================================================
\subsubsection{Tidal Love numbers of a thin shell}
\label{LaterallyUniformThinShell}

If the shell is laterally uniform, the tidal thin shell equations become a differential equation of sixth order (Eq.~(\ref{FlexConst1})), whose solution in spherical harmonics takes the form of a Hooke's law (Eq.~(\ref{hooke})).
It is then possible to solve for $w_n$ by eliminating $q_n$ between Eq.~(\ref{qSol}) and Eq.~(\ref{hooke}).
Writing $w_n$ and $\Gamma_n$ in terms of the tidal potential, I obtain
the radial and gravitational tidal Love numbers (Eq.~(\ref{wnU})) in the thin shell approximation:
\begin{eqnarray}
h_n &=& \frac{ h_{n}^\circ }{1+ \left(  1 + \xi_n \, h_{n}^\circ  \right) \Lambda_n } \, ,
\nonumber \\
k_n + 1 &=& \left( 1 + \Lambda_n \right) h_n \, .
\label{LoveThin}
\end{eqnarray}
In the membrane limit ($\Lambda_n\rightarrow\Lambda_n^M$), these equations are identical to the formulas for Love numbers given in \textit{Be15a} (the present derivation differs in using pressure Love numbers instead of the gravity scaling method).
The stress function, decomposed into spherical harmonics, follows from the second line of Eq.~(\ref{GoverningEqU}):
\begin{equation}
F_{n} = \frac{\chi}{\alpha} \, \frac{1}{\delta_n'-1-\nu} \, \frac{w_{n}}{R} \, .
\label{Fn}
\end{equation}
Finally, the tangential Love number (Eq.~(\ref{wnU})) can be related to the radial Love number as follows.
As before, I assume that the reference surface of the shell coincides with the surface of the body ($R_0=R$).
The displacement potential $S$ is related to $F$ and $w$ by Eq.~(\ref{Spot}) whereas the toroidal potential $T$ is zero if the shell is laterally uniform.
Substituting Eq.~(\ref{Fn}) into Eq.~(\ref{Spot}) and applying the TSE rule, I obtain the ratio of $l_n$ to $h_n$ for a thin shell (recall that $\delta_n'<0$ if $n\geq2$):
\begin{equation}
\frac{l_n}{h_n} = \frac{1+\nu}{-\delta_n'+1+\nu} - \left(\chi -1 \right) \frac{-\delta_n'}{-\delta'_n+1+\nu} \, .
\label{lnhn}
\end{equation}
The first term of the RHS corresponds to the $l_n-h_n$ relation for a membrane (\textit{Be15b}, Eq.~(65)) while the second term is the new contribution from bending.
If the shell is homogeneous and $R_0=R$, one has to first order in $\varepsilon=d/R$
\begin{equation}
\chi -1 \approx \varepsilon/2 \, ,
\end{equation}
showing that the ratio $l_n/h_n$ decreases linearly with shell thickness.

%=========================================================================================
%=========================================================================================
\subsection{Thick shell}
\label{ComparisonThickShell}

%=========================================================================================
%=========================================================================================
\subsubsection{Fully uniform thick shell}
\label{UniformThickShell}

If the shell is fully uniform (laterally and radially), the thick and thin shell solutions can be analytically related.
This is useful in order to estimate the finite thickness corrections.
Moreover, it helps to understand the overlap between the thick shell correction and the bending contribution.

Consider the \textit{homogeneous thick shell model}: an incompressible three-layer body made of a viscoelastic core, a homogeneous ocean, and a viscoelastic shell having the same density as the ocean (Appendix~\ref{AnalyticalFormulas}).
It generalizes to a non-rigid core the solution given by \citet{love1909}, who was arguing against the existence of a fluid layer within the Earth (he was wrong about that, but his formula was correct).
The thin shell expansion to be done below requires working at degree $n$.
The tidal Love numbers of the homogeneous thick shell model are given by
\begin{eqnarray}
h_n &=& \frac{ h_n^\circ }{ 1 +  \left( 1 + \xi_n \, h_n^\circ \right) z_h \, \hat \mu } \, ,
\nonumber \\
k_n +1 &=& \left( 1 + z_h \, \hat \mu \right) h_n \, ,
\nonumber \\
l_n &=& z_l \, h_n \, ,
\label{LoveThick}
\end{eqnarray}
where the fluid-crust Love number $h_{n}^\circ$ is given by Eq.~(\ref{hn0visco}), while $\hat\mu=\mu/(\rho{}gR)$ is the nondimensional shear modulus of the shell..
The geometrical factors $z_h$ and $z_l$ (Eq.~(\ref{zh})) depend only on the relative radius of the shell-ocean boundary $x=1-d/R$ and on the harmonic degree $n$.

As regards the Love numbers $h_n$ and $k_n$, the above formulas have the same form as the thin shell formulas (Eq.~(\ref{LoveThin})) if I can establish the correspondence $z_h\hat\mu\leftrightarrow\Lambda_n$.
Expanding $z_h$ in $\varepsilon=1-x$ up to order ${\cal O}(\varepsilon^3)$, I get
\begin{equation}
z_h \, \hat\mu \approx
\Lambda_n^M \left( 1 + \frac{\delta_n'}{2\delta_n'-3} \, \varepsilon \right)
+ \Lambda_n^B \left( 1 + \frac{9\left(8\delta_n'-3\right)}{\left(2\delta_n'-1\right) \left(2\delta_n'-3\right)^3} \right) .
\label{zhExp}
\end{equation}
This expression coincides with the thin shell spring constant $\Lambda_n$ (Eq.~(\ref{LambdaTS}) with $\nu=1/2$) up to negligible terms according to the TSE rule.
The term proportional to $\varepsilon$ in the first brackets is the next-to-leading thick shell correction for an incompressible uniform thin shell, equal to $4\varepsilon/11$ if $n=2$ and tending to $\varepsilon/2$ for large $n$
(the term proportional to $\varepsilon$ in the second brackets is much smaller).
The degree-2 thick shell correction amounts to 3\% if the shell is $23\rm\,km$ thick, which is somewhat larger than the bending contribution ($\Lambda_2^B/\Lambda_2^M\approx2\%$).

As regards the Love number $l_n$, the expansion of $z_l$ up to order ${\cal O}(\varepsilon)$ reads
\begin{equation}
z_l = \frac{l_n}{h_n} \approx \frac{3}{-2\delta_n' + 3} - \frac{\varepsilon}{2} \, \left( \frac{2\delta_n'}{2\delta_n'-3} \right)^2 .
\label{zlExp}
\end{equation}
This expression coincides with the incompressible homogeneous thin shell $l_n/h_n$ ratio (Eq.~(\ref{lnhn}) with $\nu=1/2$ and $\chi-1=\epsilon/2$) up to negligible terms according to the TSE rule.
The term proportional to $\varepsilon$ is the next-to-leading thick shell correction to the $l_n-h_n$ membrane relation (first term in RHS of Eq.~(\ref{lnhn})).
In the large $n$ limit, it tends to the bending contribution (second term in RHS of Eq.~(\ref{lnhn})).
At degree~2, however, the thick shell correction is about 27\% smaller than the bending contribution.
If the shell is $23\rm\,km$ thick, the degree-2 thick shell correction is about 9\% whereas the bending contribution is about 12\%.

\begin{figure}
   \centering
   \includegraphics[width=7.3cm]{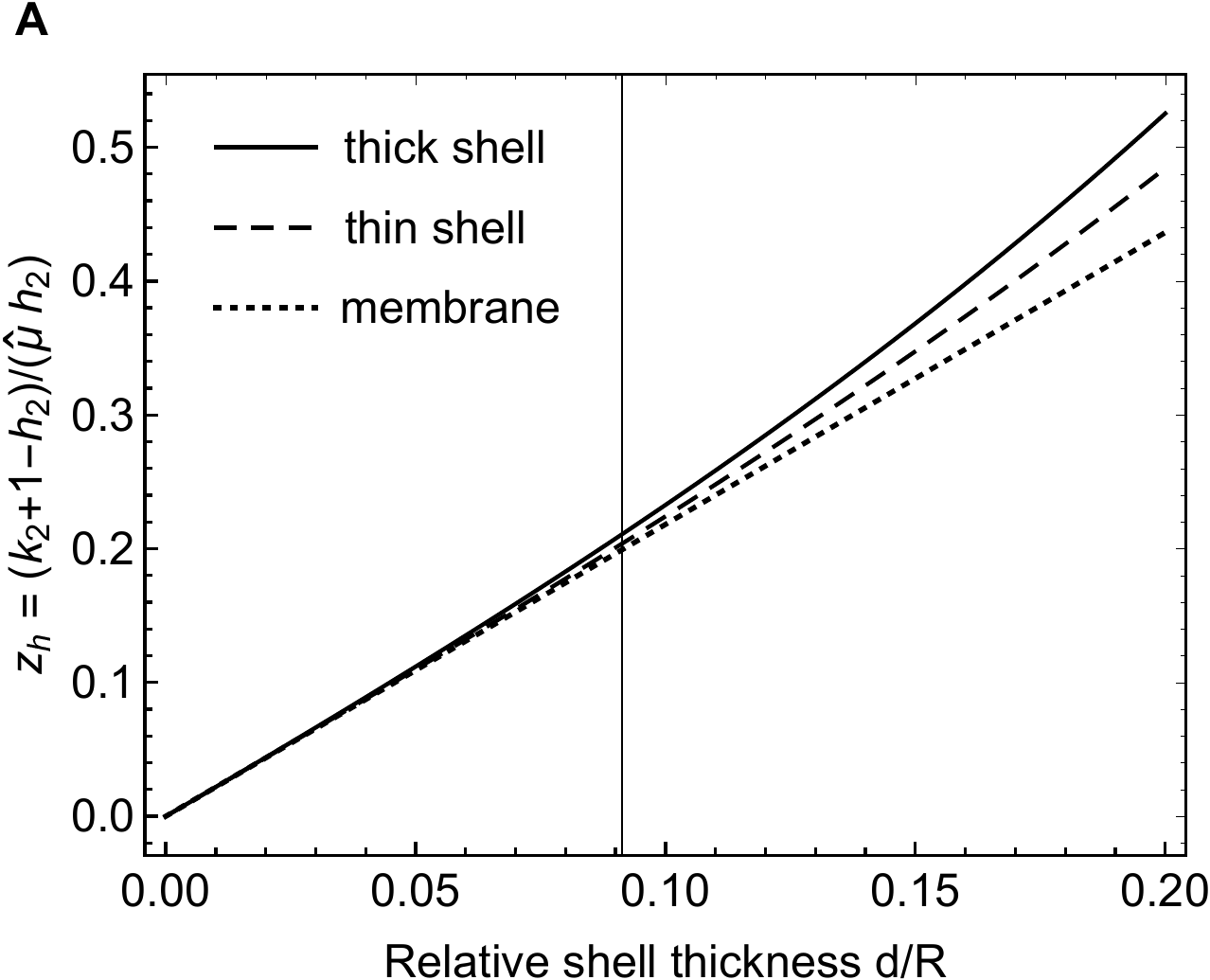}
    \includegraphics[width=7.3cm]{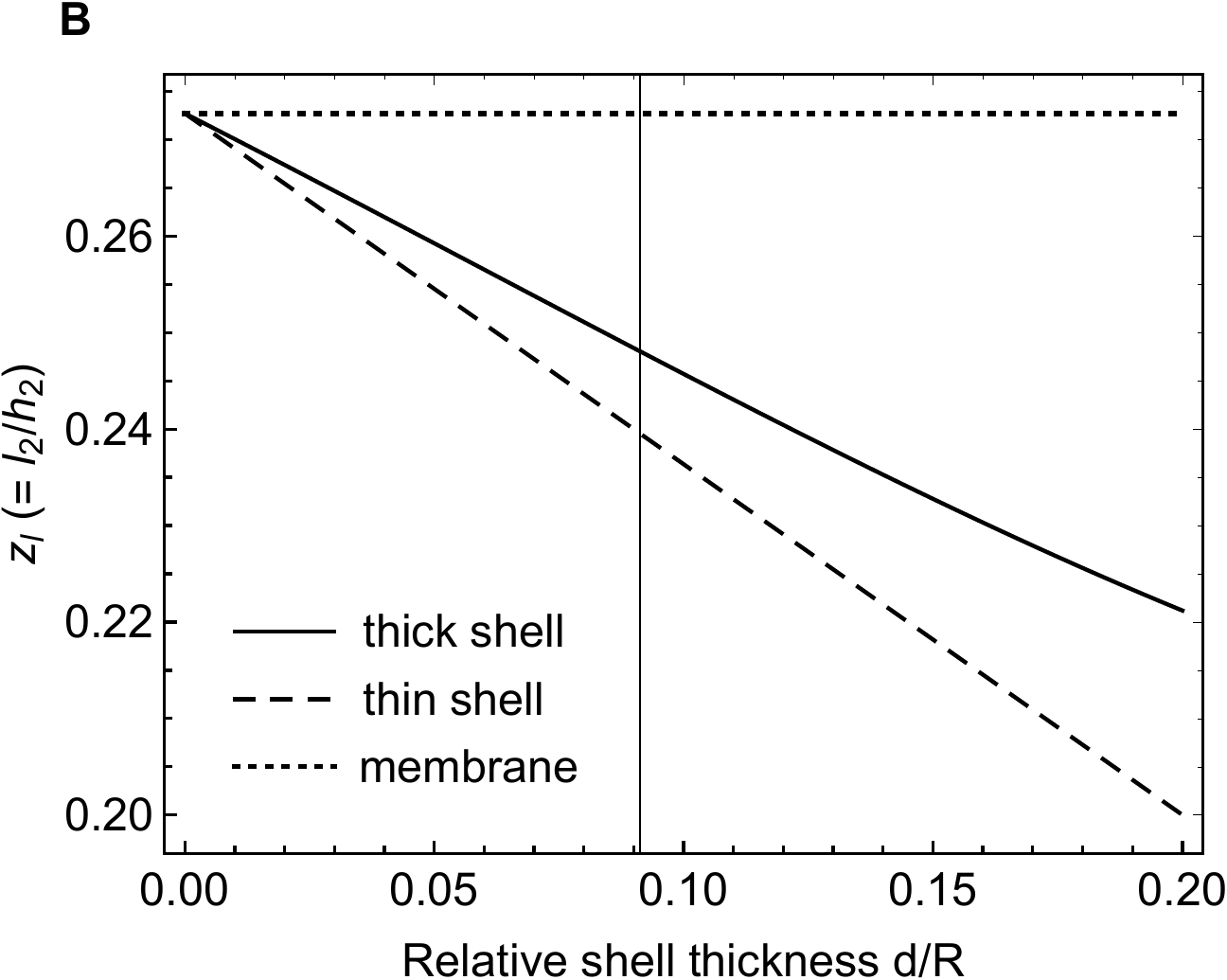}
   \caption[Tidal deformations of a uniform thick shell]
   {Tidal deformations of a uniform thick shell: geometrical factors (A) $z_h$ and (B) $z_l$ appearing in the degree-2 Love number formulas (Eq.~(\ref{LoveThick})).
   Solid/dashed/dotted curves show respectively the thick shell values (Eq.~(\ref{zh})), the thin shell approximation, and the membrane limit (Eqs.~(\ref{LoveThin})-(\ref{lnhn}) and (\ref{zhExp})-(\ref{zlExp})).
   The vertical line corresponds to the most likely thickness $d=23\rm\,km$.
   The shell is incompressible and homogeneous.
   See Section~\ref{UniformThickShell} for details.
}
   \label{FigGeomFactors}
\end{figure}

Fig.~\ref{FigGeomFactors} shows how the degree-2 factors $z_h$ and $z_l$ depend on the relative shell thickness.
If the shell is $23\rm\,km$ thick, the thin shell approximation underestimates the factors $z_h$ and $z_l$ by about 3\%.
For Enceladus, $\hat\mu\gg1$ so that $l_2$ approximately depends on the shell thickness through the ratio $z_h/z_l$.
The errors on $z_h$ and $z_l$ partly cancel in this ratio so that the error on $l_2$ is below 1\%.
To sum up, I showed that the bending corrections at low harmonic degree have a much larger effect on the tangential displacement than on the radial displacement.
While the tangential displacement is difficult to observe, it has an important impact on tectonics and tidal dissipation.

%=========================================================================================
%=========================================================================================
\subsubsection{Soft and hard shells}
\label{SoftHardShells}

The homogeneous thick shell model is a handy tool for back-of-the-envelope calculations.
I will use it now to explain the concept of soft and hard shells, which is needed for the membrane limit of non-uniform shells (Section~\ref{MembraneLimit}).
For periodic tides, it is generally a good approximation to work in the limit of an infinitely rigid core.
In that case, the fluid-crust Love number reads $h_n^\circ=1/(1-\xi_n)$ (Eq.~(\ref{hn0})), so that the degree-2 radial Love number (Eq.~(\ref{LoveThick})) becomes
\begin{equation}
h_2 = \frac{ 1 }{ 1 - \xi_2 + z_h \, \hat \mu } \, ,
\label{LoveThickRigid}
\end{equation}
where $\xi_2=3\rho/(5\rho_b)\approx0.37$ for Enceladus ($z_h$ is evaluated for $n=2$, see Eq.~(\ref{zh}) and Table~\ref{TablePoly}).
The tidal response of a floating shell can thus be characterized as follows:
\begin{itemize}
\item if $z_h\,\hat\mu\lesssim1$, $h_2$ is of order unity and close to the fluid-crust limit $h_2^\circ=1/(1-\xi_2)$: the shell is \textit{soft} and nearly follows the response of the ocean.
\item if $z_h\,\hat\mu\gg1$, $h_2$ is much smaller than 1 and close to $1/(z_h\,\hat\mu)$: the shell is \textit{hard} and the tidal deformation is controlled by the elastic properties of the shell.
\end{itemize}

If the shell is not extremely thick, the magnitude of $z_h\,\hat\mu$ can be estimated with the leading term of Eq.~(\ref{zhExp}): $\Lambda_2^M\approx2\hat\mu(d/R)$.
As $d/R$ is typically larger than a few percents, small moons like Enceladus ($\hat\mu\approx123$) and Dione ($\hat\mu\approx27$) have hard shells, whereas large moons like Europa ($\hat\mu\approx1.7$), Ganymede ($\hat\mu\approx0.9$) and Titan ($\hat\mu\approx1$) have soft shells.

If the shell is so thick as to fill the whole body, $\xi_2=3/5$ and $z_h=19/5$ (Eq.~(\ref{zhHomog})), so that Eq.~(\ref{LoveThickRigid}) reduces to the famous Love-Kelvin formula \citep{love1909}:
\begin{equation}
h_2 = \frac{5}{2} \, \frac{1}{ 1 +  (19/2)  \, \hat \mu } \, .
\end{equation}
Thus, the hard shell end-member is the generalization to floating shells of the uniform small body approximation, $h_2\approx5/(19\hat\mu)$ (e.g.\ \citet{peale1999,wisdom2004}).

Enceladus's hard shell accounts for the smallness of self-gravity corrections, which appear in Eq.~(\ref{LoveThickRigid}) through the factor $\xi_2$.
If the shell is incompressible and elastic, neglecting self-gravity leads to an error of 1, 2, and 3~\% if the shell thickness is 34, 17, and 11~km, respectively.
The error increases if the shell is viscoelastic.

%=========================================================================================
%=========================================================================================
\subsubsection{Laterally uniform thick shell}
\label{LaterallyUniformThickShell}

If the shell is laterally uniform but radially inhomogeneous, thin and thick shell solutions must be compared numerically.
I examine here the radial and tangential deformations of the surface and the time-averaged surface deviatoric stress.
Suppose that the shell is compressible, has the same density as the ocean, and is conductive with the rheology described in Section~\ref{ConductiveShell} ($\eta_{\rm melt}=10^{13}\rm\,Pa.s$).
Physical parameters are given in Table~\ref{TableParam}.
The `thin shell' Love numbers are computed with Eq.~(\ref{LoveThin}), in which $\Lambda_n$ and $h_n^\circ$ are given by Eq.~(\ref{LambdaTS}) and Eq.~(\ref{hn0visco}).
The `thick shell' Love numbers are computed numerically by integrating with Mathematica the standard elastic-gravitational equations in the static limit (Eq.~(82) of \citet{takeuchi1972} with $\omega=0$) and applying the boundary conditions for the tidal problem (the solution is propagated through the fluid layer as in \citet{saito1974}).

\begin{figure}
   \centering
      \includegraphics[width=15cm]{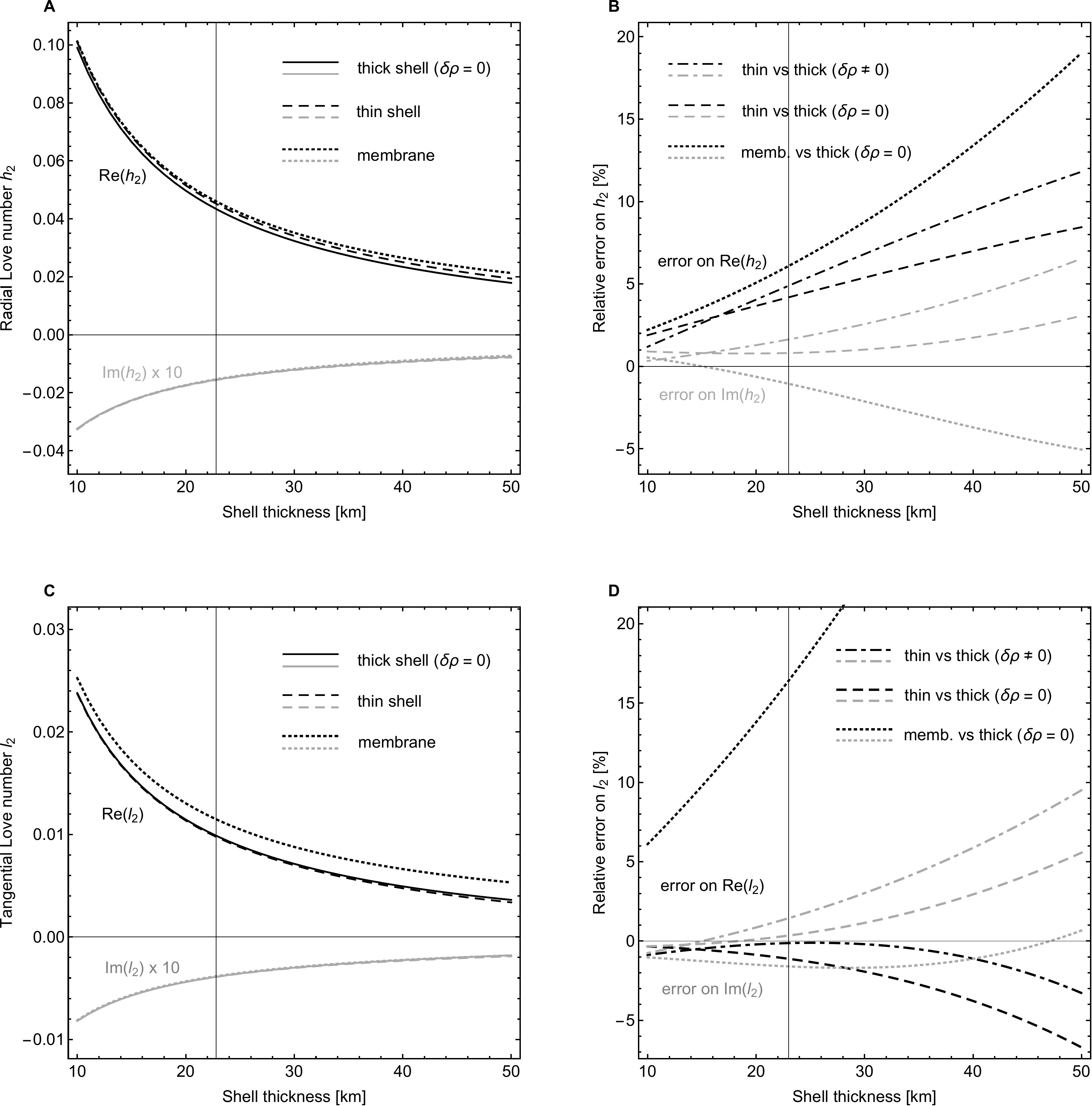}
   \caption[Tidal deformations of a laterally uniform shell]
   {Tidal deformations of a laterally uniform shell:
   (A) Love number $h_2$,
   (B) relative error on $h_2$, with respect to the thick shell approach,
    (C) Love number $l_2$,
    (D) relative error on $l_2$, with respect to the thick shell approach.
    Black (resp.\ gray) curves show the real (resp.\ imaginary) part of $h_2$ and $l_2$.
    The imaginary part is multiplied by a factor of 10.
    In panels A and C, solid/dashed/dotted curves show respectively the thick shell values, the thin shell approximation, and the membrane limit (the three models have no shell-ocean density contrast).
    In panels B and D, dashed and dotted curves show respectively the errors of the thin shell and membrane predictions, compared to the thick shell model without a shell-ocean density contrast ($\delta\rho=0$).
    The dash-dotted curves show the error of the thin shell predictions, compared to the thick shell model with a shell-ocean density contrast ($\delta\rho\neq0$).
   The vertical line indicates the most likely thickness $d=23\rm\,km$.
   The shell is compressible and conductive.
   See Section~\ref{LaterallyUniformThickShell} for details.
}
   \label{Figh2l2Conductive}
\end{figure}

Fig.~\ref{Figh2l2Conductive} compares the thin shell and membrane approximations with the thick shell solution for the real and imaginary parts of the Love numbers $h_2$ and $l_2$ (Table~\ref{TableLove} gives their values if $d=23\rm\,km$).
Suppose first that there is no shell-ocean density contrast.
If the shell is $23\rm\,km$ thick, the error on $h_2$ decreases from 6\% (membrane) to 4\% (thin shell) for the real part and decreases below 1\% for the imaginary part;
the error on $l_2$ decreases from 16\% (membrane) to 1\% (thin shell) for the real part and from 2\% to less than 1\% for the imaginary part.
The presence of a shell-ocean density contrast increases the error on the thin shell values by about $1\%$ if the shell is $23\rm\,km$ thick, assuming that the shell and ocean densities are equal to 930 and $1020\rm\,kg/m^3$ (dash-dotted curves in panels B and D).
The thin shell approach is a thus significant improvement on the membrane approximation, in particular for the tangential Love number $l_2$.
This confirms the analytical comparison of Section~\ref{UniformThickShell}.

\begin{figure}
   \centering
   \includegraphics[width=12.cm]{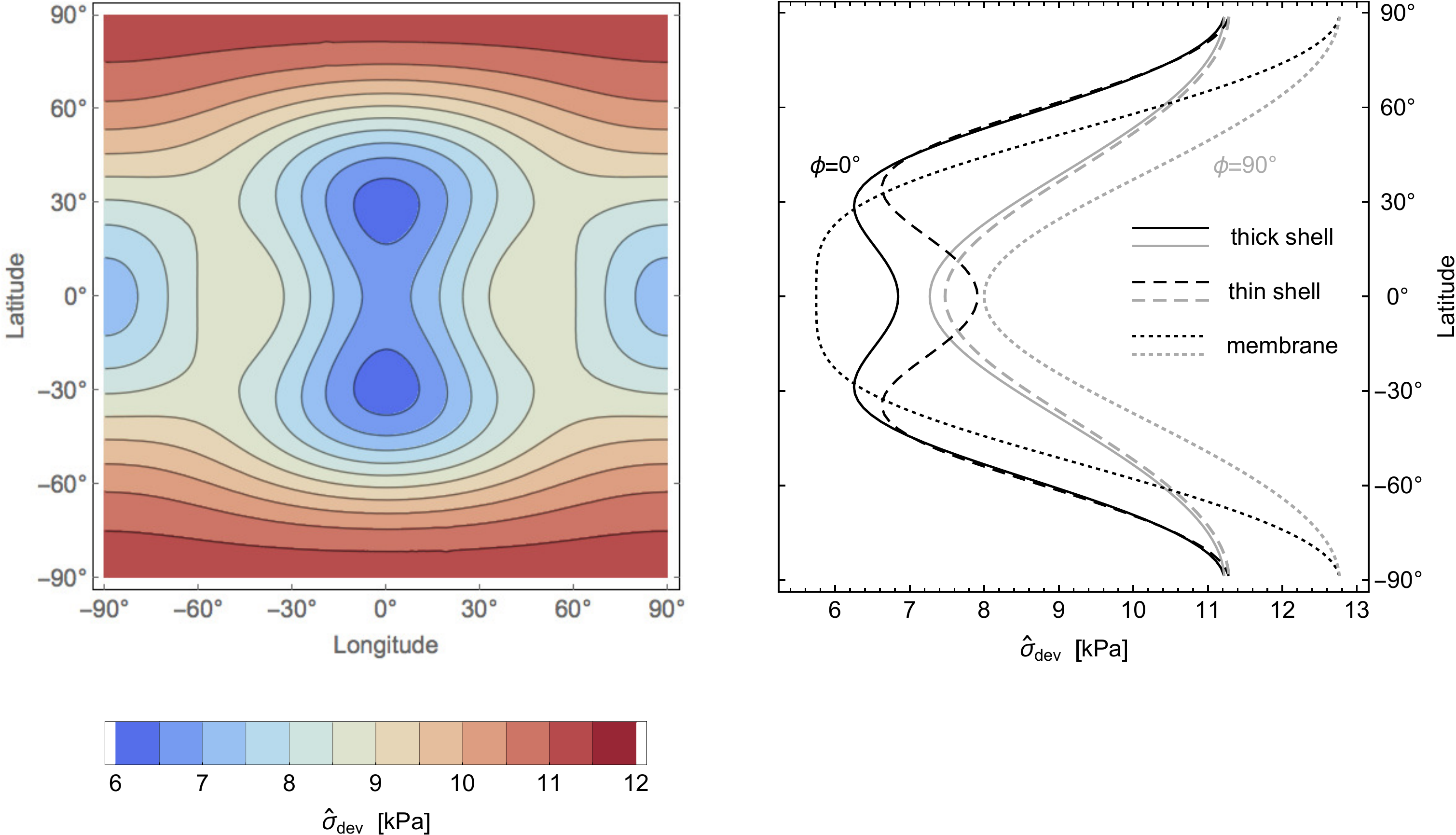}
   \caption[Time-averaged deviatoric stress at the surface]
   {Time-averaged surface deviatoric stress $\widehat\sigma_{\rm dev}$ for a laterally uniform shell: (A) spatial pattern for a thick shell; (B) crosscuts at longitudes $0^\circ$ (solid black) and $90^\circ$ (solid gray).
   For comparison, dashed and dotted curves show the thin shell approximation and the membrane limit, respectively (Eqs.~(\ref{sigdev}) and (\ref{J2hatABC})-(\ref{fABC})).
   The required Love numbers are given in Table~\ref{TableLove}.
   The shell is $23\rm\,km$ thick, compressible, and conductive.
   See Section~\ref{LaterallyUniformThickShell} for details.}
   \label{FigsigdevUNI}
\end{figure}

Fig.~\ref{FigsigdevUNI}A shows the spatial pattern of the time-averaged deviatoric stress at the surface of a thick shell ($\widehat\sigma_{\rm dev}$ given by Eq.~(\ref{sigdev}) with $\widehat J_2$ given by Eq.~(\ref{J2hatABC})).
Fig.~\ref{FigsigdevUNI}B compares meridional crosscuts of $\widehat\sigma_{\rm dev}$ at two different longitudes for a thick shell, a thin shell, and a membrane for the values of $h_2$ and $l_2$ given in Table~\ref{TableLove}.
From these figures, I conclude that the thin shell approximation gives a much better estimate - compared to the membrane limit - of the magnitude of the surface stress.
The improvement mainly results from a better estimate of the tangential strain.
The thin shell error is largest along the tidal axis (about 15\%) because $\hat\sigma_{\rm dev}$ along the tidal axis depends exclusively on the radial function $f_A$ which is very sensitive to errors on Love numbers (see Eq.~(\ref{fABC})). 

%TABLE 4
\begin{table}[ht]\centering
\ra{1.3}
\footnotesize
\caption[Degree-2 Love numbers if the shell is laterally uniform]
{\small
Degree-2 Love numbers for Enceladus if the shell is laterally uniform.
The $23\rm\,km$-thick shell is compressible and radially conductive with $\eta_{\rm melt}=10^{13}\rm\,Pa.s$.
See Section~\ref{LaterallyUniformThickShell} and Fig.~\ref{FigsigdevUNI}.
}
\begin{tabular}{@{}lcc@{}}
\hline
                   &  $h_2$                     & $l_2$ \\
\hline
Thick shell & $0.0430- 0.0015 \, i$  & $0.0098 - 0.0004 \, i$
\vspace{1mm}  \\
Thin shell   & $0.0448- 0.0015 \, i$ & $0.0097 - 0.0004 \, i$
\vspace{1mm}  \\
Membrane  & $0.0456- 0.0015 \, i$ & $0.0114 - 0.0004 \, i$
\vspace{1mm}  \\
\hline
\end{tabular}
\label{TableLove}
\end{table}%

\FloatBarrier

%%%%%%%%%%%%%%%%%%%%%%%%%%%%%%%%%%%%%%%%%%%%%%%%%%%%%%%%%%%
%%%%%%%%%%%%%%%%%%%%%%%%%%%%%%%%%%%%%%%%%%%%%%%%%%%%%%%%%%%
%%%%%%%%%%%%%%%%%%%%%%%%%%%%%%%%%%%%%%%%%%%%%%%%%%%%%%%%%%%
\section{Tidal deformations of a fully non-uniform shell}

%=========================================================================================
%=========================================================================================
\subsection{Shell structure}
\label{ShellStructure}

As far as we know, Enceladus can be modelled as a 3-layer body: a compressible shell floats on a global ocean surrounding an elastic core (physical parameters are given in Table~\ref{TableParam}).
In the thin shell approach, the shell has the same density as the ocean.
The ocean is homogeneous and compressible ($h_0^{\circ P}\approx-0.0014$) though it makes nearly no difference for degree~0 to treat it as incompressible (see Section~\ref{Loads01}).
For degrees larger than two, ocean compressibility is irrelevant in the static tide limit \citep{saito1974}.
The core is supposed to be homogeneous and incompressible: the fluid-crust Love numbers characterizing the structure below the shell are thus given by Eq.~(\ref{hn0visco}).

Improving on previous gravity-topography isostatic analyses (see Section~\ref{Introduction}), \citet{beuthe2016b} determined the average thickness of the shell and its lateral variations at low harmonic degrees:
\begin{equation}
d = d_{00} + d_{20} \, P_{20}(\cos\theta) + d_{22} \, P_{22}(\cos\theta) \cos2\varphi + d_{30} \, P_{30}(\cos\theta) \, ,
\label{ThicknessVarGRL}
\end{equation}
with $(d_{00},d_{20},d_{22},d_{30})=(22.8,-12.1,1.3,3.7)\pm(4,2.4,0.3,0.7)\rm\,km$; $P_{nm}$ are the unnormalized associated Legendre functions.
Thickness variations of higher degree can be estimated if topography beyond degree~3 is isostatically compensated, though there are no gravity data confirming this hypothesis.
With the harmonic degrees $n=4-8$ determined by \citet{nimmo2011}, the south polar thickness estimate is reduced by 3~km (from $7\pm4\rm\,km$ to $4\pm4\rm\,km$; the value at the north pole does not change much).
However, there is no reduction at the south pole if the estimate (\ref{ThicknessVarGRL}) is combined with the thickness variations inferred from the harmonic degrees $n=1$ and $n=4-16$ of the new shape determined by \citet{tajeddine2017}.
Therefore, it seems better to stick to the estimate (\ref{ThicknessVarGRL}) while waiting for a consistent inversion of the gravity field of \citet{iess2014} and the shape of \citet{tajeddine2017}.

\begin{figure}
   \centering
   \includegraphics[width=11cm]{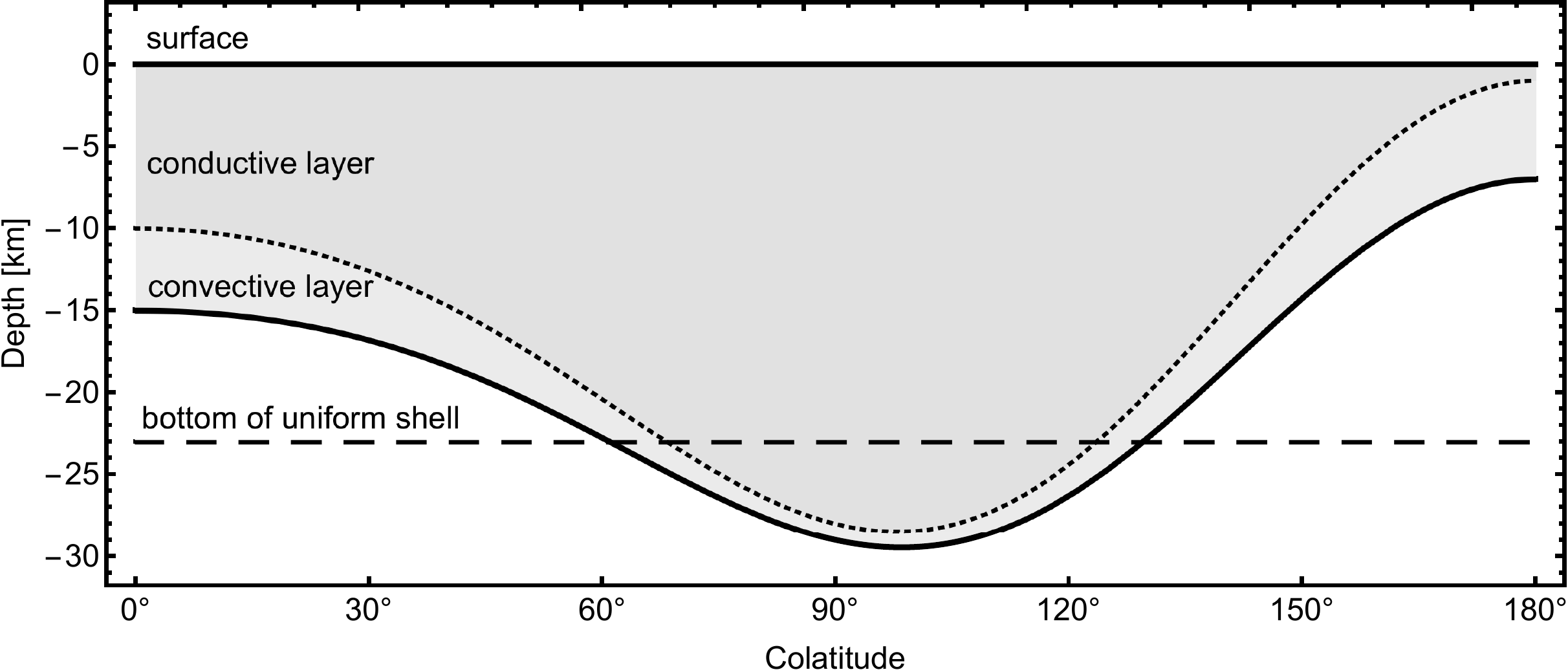}
   \caption[Shell structure]
   {Shell structure.
   The shaded area represents shell thickness variations, with darker and lighter shading indicating the conductive and convective parts (the latter is not present in the conductive model).
   The horizontal dashed line shows the bottom of the shell for the laterally uniform model.
   See Section~\ref{ShellStructure} for details.}
   \label{FigThickness}
\end{figure}

Since thickness variations are much smaller in longitude than in latitude, I choose to ignore them, which greatly accelerates the computation.
Rounding off the above coefficients, I assume that shell thickness variations are given (in km) by
\begin{equation}
d = 23 -12 \, P_{20}(\cos\theta) + 4 \, P_{30}(\cos\theta) \, .
\label{ThicknessVariation}
\end{equation}
These variations result from topography at the top and at the bottom of the shell.
The bottom topography, however, is larger than the top topography by a factor of about 10 (the ratio is close to $\rho/\Delta\rho$ with $\rho$ being the shell density and $\Delta\rho$ the shell-ocean density contrast).
As another simplification, I will assume that the surface of the shell is spherical: thickness variations exclusively result from bottom topography (Fig.~\ref{FigThickness}).
For comparison, I consider a laterally uniform model with a $23\rm\,km$-thick shell.

Given that the shell thickness is smaller than $40\rm\,km$, the shell is most likely in a conductive state (e.g.\ \citet{barrmckinnon2007}; \citet{mitri2008})), which also protects shell thickness variations against lateral flow.
My reference shell structure is therefore a conductive shell with the rheology described in Section~\ref{ConductiveShell} ($\eta_{\rm melt}=10^{13}\rm\,Pa.s$).
Nevertheless, it is interesting to apply the thin shell approach to a shell having strong lateral variations in rheology.
Moreover, radiometry data suggest the possibility that the conductive part of the shell could be as thin as $1-2\rm\,km$ in the South Polar Terrain \citep{legall2017}.
I will thus consider a model in which the lower part of the shell convects vigorously below the poles.
The total shell thickness varies as in the conductive model.
The upper conductive layer is given by Eq.~(\ref{ThicknessVarGRL}) with $(d_{00},d_{20},d_{22},d_{30})=(20.5,-15,0,4.5)\rm\,km$ so that its thickness is 10, 28, and $1\rm\,km$ at the north pole, equator, and south pole, respectively.
The viscosity of the convective layer is $\eta_{\rm melt}=10^{13}\rm\,Pa.s$ and the temperature of the conductive/convective boundary is 273~K.

%=========================================================================================
%=========================================================================================
\subsection{Numerical method}
\label{NumericalMethod}

%************************************************************************************************************************************
\subsubsection{Spherical harmonics and other methods}
\label{General}

If the shell is laterally non-uniform, the tidal thin shell equations must be solved numerically.
I describe below a method based on spherical harmonics to solve these equations whatever the variations of the shell structure with latitude, longitude, and depth.
The essential assumption of this method is that the shell is decoupled from the core by a global ocean.
If the ocean is local, the tidal thin shell equations should be solved instead in the spatial domain with appropriate boundary conditions at the edge of the shell portion covering the regional ocean.

The problem becomes simpler if the longitudinal variations of the shell structure are negligible, which seems to be the case for Enceladus (see Section~\ref{ShellStructure}).
In the spherical harmonic domain, it means that the viscoelastic parameters $(\alpha,D,\chi)$ are zonal (harmonic order 0), whereas the variables $(F,w)$ depend on the same harmonic orders as the tidal potential (0 and $\pm2$, see Eq.~(\ref{TidalPot})).
Indeed, the thin shell operators ${\cal C}(a;b)$ and ${\cal A}(a;b)$ do not change the harmonic order of $b$ if $a$ is zonal (Eq.~(\ref{Aharmonics})).
It is thus possible to decompose the problem into three decoupled sub-problems, each of given harmonic order.
This is what I will do in this paper.

Instead of going further with spherical harmonics, one could consider the above decomposition in the spatial domain, where $F$ and $w$ can each be split into three components (proportional to 1, $e^{2i\varphi}$, and $e^{-2i\varphi}$), which are eigenfunctions with respect to $\varphi$-derivatives.
Factoring out the longitude dependence, we are left with three systems of two differential equations depending only on latitude.
Is it possible to solve these three sub-problems in the spatial domain with standard numerical software?
This method worked well for planetary contraction and despinning in the membrane limit (see \textit{Be10}, Section~4.2).
A first complication is the presence of the self-attraction term which would take the form of a convolution integral in the spatial domain (Eq.~(\ref{Gw})).
For a small body like Enceladus, this difficulty could be bypassed by neglecting self-gravity because elastic effects are dominant.
A second complication is the presence of bending which increases the order of the differential equations and introduces coefficients of very different magnitudes ($\alpha$ and $D$).
I will not pursue this possibility further because solutions in spherical harmonics are well suited to a shell floating on a global ocean.
Nevertheless, the number of required spherical harmonics becomes large if the shell structure varies in longitude.
In that case, it could be a good idea to solve the equations on a grid in the spatial domain, for example with an iterative method starting from the membrane limit for a hard shell (see Section~\ref{MembraneLimit}).

%************************************************************************************************************************************
\subsubsection{Nondimensionalization}
\label{NonDim}

The tidal thin shell equations (\ref{TidalThinShellEq}) are nondimensionalized with
\begin{equation}
\left( \bar w ,  \bar F ,  \bar\alpha , \bar D , \bar U \right) = \left( \frac{w}{R} \, , \,  \alpha_{\rm ref} F \, , \,  \frac{\alpha}{\alpha_{\rm ref}} \, , \,  \frac{\alpha_{\rm ref}}{R^2} D \, , \,  \frac{U}{gR} \right) ,
\end{equation}
where $\alpha_{\rm ref}$ is a constant having the dimensions of $\alpha$ (possible choices are $1/(\rho{}gR^2)$ or a quantity closer in magnitude to $\alpha$ -- such as the surface average of $\alpha$).
The tidal thin shell equations become
\begin{eqnarray}
{\cal C}( \bar D \,; \bar w ) - (1-\nu) \, {\cal A}( \bar D \,; \bar w) + {\cal A}(\chi \,; \bar F) &=& - \gamma \left(\bar w - {\cal G}(\bar w)  -  \bar U_\upsilon \right) ,
\nonumber \\
{\cal C}( \bar\alpha \,\,; \bar F ) - (1+\nu) \, {\cal A}( \bar\alpha \,; \bar F) - {\cal A}(\chi \,; \bar w) &=& 0 \, ,
\label{TidalThinShellMat}
\end{eqnarray}
where the coupling $\gamma$ is given by
\begin{equation}
\gamma = \rho g R^2 \alpha_{\rm ref} \, .
\label{defgamma}
\end{equation}
Suppose that $\alpha_{\rm ref}=1/(E_{\rm e}d_{\rm ref})$, where $E_{\rm e}$ is Young's modulus for elastic ice (about $9.3\rm\,GPa$) and $d_{\rm ref}$ is the mean shell thickness.
Then
\begin{equation}
\gamma = \left( \frac{E_{\rm e}}{\rho{}gR} \, \frac{d_{\rm ref}}{R} \right)^{-1} \, .
\end{equation}
With this choice for $\alpha_{\rm ref}$, the coupling $\gamma$ is smaller than 1 by one order of magnitude for small icy satellites like Enceladus ($E_{\rm e}/(\rho{}gR)\approx330$).
Referring to Section~\ref{SoftHardShells}, the coupling $\gamma$ plays the role of the factor $1/(z_h\,\hat\mu)$ which served to distinguish hard shells from soft shells.
In that context, Enceladus's shell is hard.
By contrast, $\gamma$ is larger than 1 by one order of magnitude for large icy satellites like Europa ($E_{\rm e}/(\rho{}gR)\approx4.6$): their shell is soft.

%************************************************************************************************************************************
\subsubsection{Matrix solution for a thin shell}
\label{MatrixSolution}

The nondimensional tidal thin shell equations (\ref{TidalThinShellMat}) are solved by
\begin{enumerate}
\item
expanding the variables $(\bar w,\bar F)$, the viscoelastic parameters $(\bar\alpha,\bar{D},\chi)$ and the tidal potential $\bar U_\upsilon$ in spherical harmonics of degree $n$ and order $m$;
\item
expanding the product of spherical harmonics as a sum of spherical harmonics via Clebsch-Gordan coefficients (e.g.\ Appendix~A of \citet{kalousova2012});
\item
projecting the result on a spherical harmonics basis truncated to degree $n_{\rm max}$ and writing the result in matrix form.
\end{enumerate}
If the shell structure only varies in latitude, the projection results in decoupled systems of given harmonic order (the same orders that appear in the tidal forcing).

If the harmonic coefficients of $\bar w$ and $\bar F$ are ordered into vectors ${\mathbf {\bar w}}$ and ${\mathbf {\bar F}}$, operators become matrices (\textit{Be10}, Section~4.3):
\begin{eqnarray}
\left( \Delta' \bar w \, , \, \Delta' \bar F \right) &\rightarrow& \left( {\mathbf\Delta'} \, {\mathbf {\bar w}} \, , \, {\mathbf\Delta'} \, {\mathbf {\bar F}} \right) ,
\nonumber  \\
\left( \, {\cal C}( \bar{D} \,;  {\bar w}) \, , \, {\cal A}( \bar{D} \,;  {\bar w}) \, , \, {\cal A}(\chi \,;  {\bar w}) \, \right)
&\rightarrow&
\left( \, {\mathbf C}_{\bar{D}} \, , \, {\mathbf A}_{\bar{D}} \, , \, {\mathbf A}_{\chi} \, \right) {\mathbf {\bar w}} \, ,
\nonumber \\
\left( \, {\cal C}( \bar{\alpha} \,;  {\bar F}) \, , \, {\cal A}( \bar{\alpha} \,;  {\bar F}) \, , \, {\cal A}(\chi \,;  {\bar F}) \, \right)
&\rightarrow&
\left( \, {\mathbf C}_{\bar{\alpha}} \, , \, {\mathbf A}_{\bar{\alpha}} \, , \, {\mathbf A}_{\chi} \, \right) {\mathbf {\bar F}} \, ,
\nonumber  \\
\bar w - {\cal G}(\bar w) &\rightarrow& {\mathbf Q} \, {\mathbf {\bar w}} \, .
\end{eqnarray}
The vectors $\bar w$ and $\bar F$ are defined without degree-1 components while the matrices ${\mathbf C}_x$ and ${\mathbf A}_x$ ($x=\bar\alpha,\bar D,\chi$) have neither rows nor columns associated with degree-1 harmonics.
The justification is twofold.
First, degree-1 columns are removed because the operators ${\cal C}$ and ${\cal A}$ give zero when their second argument is a degree-1 harmonic (Eq.~(\ref{nodeg1})).
Second, degree-1 rows are removed because degree-1 harmonics do not belong to the images of ${\cal C}$ and ${\cal A}$ (Eq.~(\ref{nodeg1image})).
Consistency requires that the matrix ${\mathbf Q}$ has no rows or columns of degree~1.
Projecting out the degree~1 is important for numerical stability when coupling occurs between different harmonic degrees: degree-1 components of $\bar w$ and $\bar F$ are decoupled from the viscoelastic equations, even if the spatial variations of crust thickness and rheology have nonzero degree-1 components.
Although the thin shell solution is computed in a frame in which the degree-1 radial displacement vanishes, it can be rigidly translated (Eq.~(\ref{S1w1})) and rotated to another frame while leaving stresses and strains invariant.

The \textit{membrane matrix} ${\mathbf M}$ (\textit{Be10}, Eq.~(52)) and the \textit{bending matrix} ${\mathbf B}$ are defined by
\begin{eqnarray}
{\mathbf M} &=& {\mathbf C}_{\bar{\alpha}} - (1+\nu) \, {\mathbf A}_{\bar{\alpha}} \, ,
\nonumber \\
{\mathbf B} &=& {\mathbf C}_{\bar{D}} - (1-\nu) \, {\mathbf A}_{\bar{D}} \, .
\label{Bmatrix}
\end{eqnarray}
An explicit example of the membrane matrix can be found in Appendix~I of \textit{Be10}.
If the shell structure is zonal, different matrices are associated with the three possible harmonic orders (see Section~\ref{General}), with size $(n_{\rm max},n_{\rm max})$ if $m=0$ (no degree~1) and $(n_{\rm max}-1,n_{\rm max}-1)$ if $m=\pm2$ (since $n\geq{}m$).
If the shell structure also varies in longitude, the number of rows (or columns) scales with $n_{\rm max}^2$ though east-west symmetry with respect to the tidal axis reduces the size.

In matrix form, the tidal thin shell equations read
\begin{eqnarray}
\left( {\mathbf B} + \gamma \, {\mathbf Q} \right) {\mathbf {\bar w}} + {\mathbf A}_{\chi}  \, {\mathbf {\bar F}} &=& \gamma \, {\mathbf {\bar U}_\upsilon} \, ,
\nonumber \\
{\mathbf M} \, {\mathbf {\bar F}} - {\mathbf A}_{\chi}  \, {\mathbf {\bar w}} &=& 0 \, ,
\label{TidalThinShellM2}
\end{eqnarray}
where ${\mathbf {\bar U}_\upsilon}$ is the vector associated with the potential $\bar U_\upsilon$.
If the core and the ocean are incompressible, $w_0=0$ and $q_0$ replaces $w_0$ as a free variable (Eq.~(\ref{Love0}) becomes irrelevant).
In such a case, one can drop the first row of the inhomogeneous tidal thin shell equation in matrix form, as it only serves to determine $q_0$.

The solution of Eq.~(\ref{TidalThinShellM2}) is formally given by
\begin{eqnarray}
{\mathbf {\bar F}}
&=& \left( {\mathbf A}_{\chi} + \left( {\mathbf B} + \gamma {\mathbf Q} \right) {\mathbf A}_{\chi}^{-1} {\mathbf M} \right)^{-1} \gamma {\mathbf {\bar U}_\upsilon} \, ,
\nonumber \\
{\mathbf {\bar w}} &=& \left( {\mathbf B} + \gamma {\mathbf Q} + {\mathbf A}_{\chi} \, {\mathbf M}^{-1} {\mathbf A}_{\chi} \right)^{-1} \gamma {\mathbf {\bar U}_\upsilon} \, .
\label{TidalThinShellSol}
\end{eqnarray}
In practice, the linear system (\ref{TidalThinShellM2}) is solved with the linear algebra package LAPACK.
Spherical harmonic transformations and Clebsch-Gordan coefficients are computed with the library SHTOOLS \citep{wieczorek2016}.
Grid points are evenly spaced in longitude ($2n_{\rm max}+1$ points) and unevenly spaced in latitude ($n_{\rm max}+1$ Gauss-Legendre points).
The maximum harmonic degree is set to $n_{\rm max}=40$.

Once the solution $(F,w)$ is known, related scalar quantities (such as the time-averaged surface stress invariant $\widehat J_2$) can be computed with the transform method already mentioned in Section~\ref{TidalStressSec}: scalar derivatives are evaluated in the spherical harmonic domain whereas multiplication is done on a grid in the spatial domain.
Computing derivatives is all the more easy if the spherical differential operators are expressed in terms of $\Delta'$ (Eq.~(\ref{Aharmonics})).

If the shell is laterally uniform with viscoelastic constants $(\bar\alpha^{\rm uni},\bar D^{\rm uni},\chi^{\rm uni})$, the matrices $({\mathbf M},{\mathbf B},{\mathbf A}_{\chi})$ are diagonal (${\mathbf Q}$ being diagonal in any case) with non-zero elements given by
\begin{eqnarray}
({\mathbf M}^{\rm diag})_{nn} &=& \bar \alpha^{\rm uni} \left( \delta_n'- 1 - \nu \right) \delta_n' \, ,
\nonumber \\
({\mathbf B}^{\rm diag})_{nn} &=& \bar D^{\rm uni} \left( \delta_n'-1+\nu \right) \delta_n' \, ,
\nonumber \\
({\mathbf A}_{\chi}^{\rm diag})_{nn} &=& \chi^{\rm uni} \, \delta_n' \, ,
\nonumber \\
({\mathbf Q}^{\rm diag})_{nn} &=& 1- \upsilon_n \, \xi_n \, .
\label{matdiag}
\end{eqnarray}
In that case, the solution of Eq.~(\ref{TidalThinShellM2}) reduces to the solution for a laterally uniform shell (Eqs.~(\ref{LoveThin})-(\ref{Fn})).

%************************************************************************************************************************************
\subsubsection{Mode coupling and membrane limit}
\label{MembraneLimit}

Tidal deformations of Enceladus can (and will be) computed with the general matrix method of Section~\ref{MatrixSolution} for the isostatic shell structure shown in Fig.~\ref{FigThickness}.
Before doing that, I will discuss some features of the solution which can be inferred from the general structure of the equations.
Since tidal forcing of degree~3 (and higher) is negligible, non-degree-2 components of the displacement (or stress function) are not directly excited by the tidal potential but rather result from couplings with the laterally non-uniform shell structure.
The matrix solution in the spectral domain makes it very easy to compute the spectrum of the tidal response given the lateral variation of shell thickness or rheology.

Mode coupling analysis consists in studying the tidal response due to a harmonic component (of given degree and order) of the tidal potential if the laterally non-uniform shell structure is described by a specific spherical harmonic (of given degree and order).
One can start with two general observations.
First, the basic symmetries of the lateral variations in shell structure constrain the spectrum of the tidal response.
For example, a zonal variation implies that the tidal response is of the same harmonic order as the tidal forcing (see Section~\ref{General}).
Moreover, a variation symmetric with respect to the equator (i.e.\ of even harmonic degree) induces a similarly symmetric tidal response (i.e.\ including only even harmonic degrees).
Second, if the lateral variations of the shell structure are treated as perturbations, the non-degree-2 tidal response is mainly due to first-order coupling between the lateral variations and the tidal forcing.
Well-known selection rules then predict which are the largest spherical harmonics in the tidal response \citep{zhong2012}; this analysis can be extended to second perturbative order \citep{qin2014}.
Note that the tidal thin shell equations in matrix form are not perturbative in that sense, but they can be analytically solved with a perturbative series to arbitrary order (see Appendix~I of \textit{Be10}).
Mode coupling analysis is particularly relevant to tidal tomography, which consists in constraining the lateral variations of the shell structure with non-degree-2 tides \citep{zhong2012,a2014,qin2016}.
Unfortunately, tidal geodetic measurements will not available for a long time, not even for degree-2 tides..
It is actually the other way around: we constrain tides by using our knowledge of Enceladus's non-uniform shell based on gravity-topography data.
Therefore, the usefulness of mode coupling analysis is limited here to consistency checks (see Section~\ref{Results}) and benchmarking studies.

The membrane approximation throws additional light on some important features of the numerical solution.
If the lateral variations of the shell properties are of long wavelength, the components of the response get smaller as their harmonic degree increases.
It thus makes sense to examine the solution in the membrane limit:
\begin{equation}
{\mathbf B}\approx 0
\hspace{5mm} \mbox{and} \hspace{5mm}
{\mathbf A}_{\chi}\approx{\mathbf \Delta'} \, .
\end{equation}
In that regime, one can study the effect of the shell on tidal deformations by defining $\alpha_{\rm ref}$ as the spatial average of $\alpha$ and by varying the parameter $\gamma$ (Eq.~(\ref{defgamma})).
Recall that the magnitude of $\gamma$ indicates whether the shell is soft or hard (Section~\ref{NonDim}), and that this classification is valid whatever the shell thickness (Section~\ref{SoftHardShells}).

In the \textit{soft shell} limit ($\gamma\gg1$, relevant to Europa), the radial displacement tends to the surface ocean solution (of degree~2) whereas the stress function still depends on the non-uniform shell properties included in ${\mathbf M}$:
\begin{eqnarray}
{\mathbf {\bar w}} &\approx& {\mathbf Q}^{-1} \, {\mathbf {\bar U}_\upsilon} \, ,
\nonumber \\
{\mathbf {\bar F}} &\approx& {\mathbf M}^{-1} {\mathbf \Delta'} \, {\mathbf Q}^{-1} \, {\mathbf {\bar U}_\upsilon} \, .
\label{SoftShellApprox0}
\end{eqnarray}
The soft shell assumption was adopted in \textit{Be10} for the study of contraction and despinning tectonics.

In the \textit{hard shell} limit ($\gamma\ll1$, relevant to Enceladus), the stress function does not depend on the shell properties, whereas the radial displacement does:
\begin{eqnarray}
{\mathbf {\bar F}} &\approx& {\mathbf\Delta'}^{-1} \gamma  {\mathbf {\bar U}_\upsilon} \, ,
\nonumber \\
{\mathbf {\bar w}} &\approx& {\mathbf\Delta'}^{-1} {\mathbf M} \, {\mathbf\Delta'}^{-1} \gamma  {\mathbf {\bar U}_\upsilon} \, .
\label{HardShellApprox0}
\end{eqnarray}
One can improve on this estimate by keeping the diagonal part of the membrane correction:
\begin{equation}
{\mathbf {\bar F}} \approx \left( {\mathbf \Delta'} +  \gamma {\mathbf Q} \, {\mathbf \Delta'}^{-1} {\mathbf M}^{\rm diag} \right)^{-1} \gamma {\mathbf {\bar U}_\upsilon} \, ,
\label{Fmem}
\end{equation}
where ${\mathbf M}^{\rm diag}$ is defined by Eq.~(\ref{matdiag}) with $\bar\alpha^{\rm uni}=\langle\bar\alpha\rangle$ (spatial average of $\bar\alpha$).
This expression is equivalent to the uniform membrane solution, i.e.\ Eqs.~(\ref{LoveThin})-(\ref{Fn}) in which $\Lambda_n$ is replaced by $\langle\Lambda_n^M\rangle$.
Going back to spherical harmonics and dimensional quantities, I can write the hard-shell membrane approximation as
\begin{eqnarray}
F_{n} &\approx& \left( 1 - \frac{h_n}{h_n^\circ} \right) \left( \delta_n' \right)^{-1} \rho R \, \upsilon_n \, U_n \, ,
\nonumber \\
w_n &\approx& R \left( \delta_n' \right)^{-1} \Big[ {\cal C} ( \alpha \,; F ) - (1+\nu) \, {\cal A}(\alpha \,; F) \Big]_n \, .
\label{MemApprox}
\end{eqnarray}
This approximation does not require the explicit construction of the membrane matrix (with Clebsch-Gordan coefficients) because the terms ${\cal C}(\alpha \,; F)$ and ${\cal A}(\alpha \,; F)$ can be evaluated with the transform method (see Section~\ref{MatrixSolution}).

In the membrane limit, the stress resultants are given by the first term of Eq.~(\ref{eqN}),
\begin{equation}
\left( N_{\theta\theta} \, , N_{\varphi\varphi} \, , N_{\theta\varphi} \right) \approx \left(  {\cal O}_2 \, , {\cal O}_1 \, , -{\cal O}_3 \right) F \, ,
\label{NF}
\end{equation}
while the stresses (at the surface or within the crust) are given by the first term of Eq.~(\ref{stressesFw}):
\begin{equation}
\sigma_{ij} \approx E\alpha_{\rm inv} \,N_{ij}
\hspace{5mm} (i,j=\theta,\varphi) \, ,
\end{equation}
where $E$ is the only factor depending on depth.
For Enceladus, Eq.~(\ref{NF}) means that the stress resultants depend little on the lateral variations in shell structure.
By contrast, the surface stresses should be approximately proportional to the magnitude of the extensibility $\alpha_{\rm inv}$ ($E$ is considered to be uniform at the surface).
For a conductive shell, the depth-averaged shear modulus ($\mu_0$) on which the extensibility depends is nearly uniform (see Eq.~(\ref{defalphaDinv})) so that the surface stresses are approximately inversely proportional to the local thickness $d$.

%=========================================================================================
%=========================================================================================
\subsection{Comparison with finite element methods (FEM)}
\label{FEM}

%************************************************************************************************************************************
\subsubsection{Pressurized ocean in Enceladus}
\label{PressurizedOcean}

\citet{johnston2017} studied the effect of ocean overpressure on Enceladus's tectonics.
Their models include either a regional sea at the south pole or a global ocean under an axisymmetric shell of variable thickness.
The thickness variation is specified by spherical indentations at the bottom of the shell, either at the south pole only (`south polar indentation') or at both poles (`two polar indentations').
The ocean overpressure is uniform in magnitude and is everywhere normal to the bottom surface of the shell.
They compute shell stresses with the commercial FEM software \textit{COMSOL Multiphysics}.
The solved models are publicly available as \textit{COMSOL} \textit{mph} files \citep{johnston2017Z}.

With the thin shell approach, I can duplicate the results of their global ocean models, paying attention to the following points: the forcing is not tidal but due to static surface loading; self-gravity is not included; the surface load must be projected on the spherical reference surface of the shell (of radius $R$), on which it has both radial and tangential (here latitudinal) components.
This projection involves a rescaling of the load magnitude by the factor $(R-d)^2/R^2$ (Eq.~(\ref{defLoads})).
The tangential load and its associated moment are integrated into a load potential $\Omega$ and a moment potential $\Omega_M$, respectively (Eq.~(\ref{defOmega})).
The radial load $q$ and the tangential potential $\Omega$ both include spherical harmonic components of degree~1, denoted $q_1$ (which drops out of the thin shell equations) and $\Omega_1$, respectively.
The total force due to the overpressure integrated over the bottom surface of the shell is zero (Eq.~(\ref{qOmega1})).
I solve Eqs.~(\ref{master1}) and (\ref{master2}) with the matrix method of Section~\ref{MatrixSolution}.
Beware that ${\mathbf Q}=\mathbf{1}$ (no self-gravity) and that the RHS of Eq.~(\ref{TidalThinShellM2}) is modified as follows: $\gamma{\mathbf{\bar U}_\upsilon}$ and 0 are respectively replaced by ${\mathbf {\bar p}_1}$ and ${\mathbf {\bar p}_2}$, the vectors of spherical harmonics coefficients given by the correspondence
\begin{eqnarray}
\bar q + \Delta\bar\Omega-\Delta'\left(\chi\left(\bar\Omega+\bar\Omega_M\right)\right) &\rightarrow& {\mathbf {\bar p}_1} \, ,
\nonumber \\
- (1-\nu) \, \Delta' \left(\alpha\left(\bar\Omega+\bar\Omega_M \right) \right) &\rightarrow& {\mathbf {\bar p}_2} \, ,
\end{eqnarray}
where $\bar q=R\alpha_{\rm ref}q$ and $(\bar\Omega,\bar\Omega_M)=\alpha_{\rm ref}(\Omega,\Omega_M)$ are nondimensional.

Another detail is that the thin shell solution is computed in a frame in which the degree-1 radial displacement $w_1$ vanishes (it can be rigidly translated with Eq.~(\ref{S1w1})).
By contrast, \citet{johnston2017} constrain the FEM solution so that the bottom of the shell, at the equator, does not move along the $z$-axis.
In this frame, $w_1\neq0$.
Before comparing FEM and thin shell results, I must thus translate the FEM displacements to a frame in which $w_1=0$.
Note that \textit{COMSOL} automatically selects this frame if one gives up on the displacement constraint.

\begin{figure}
   \centering
   \includegraphics[width=15cm]{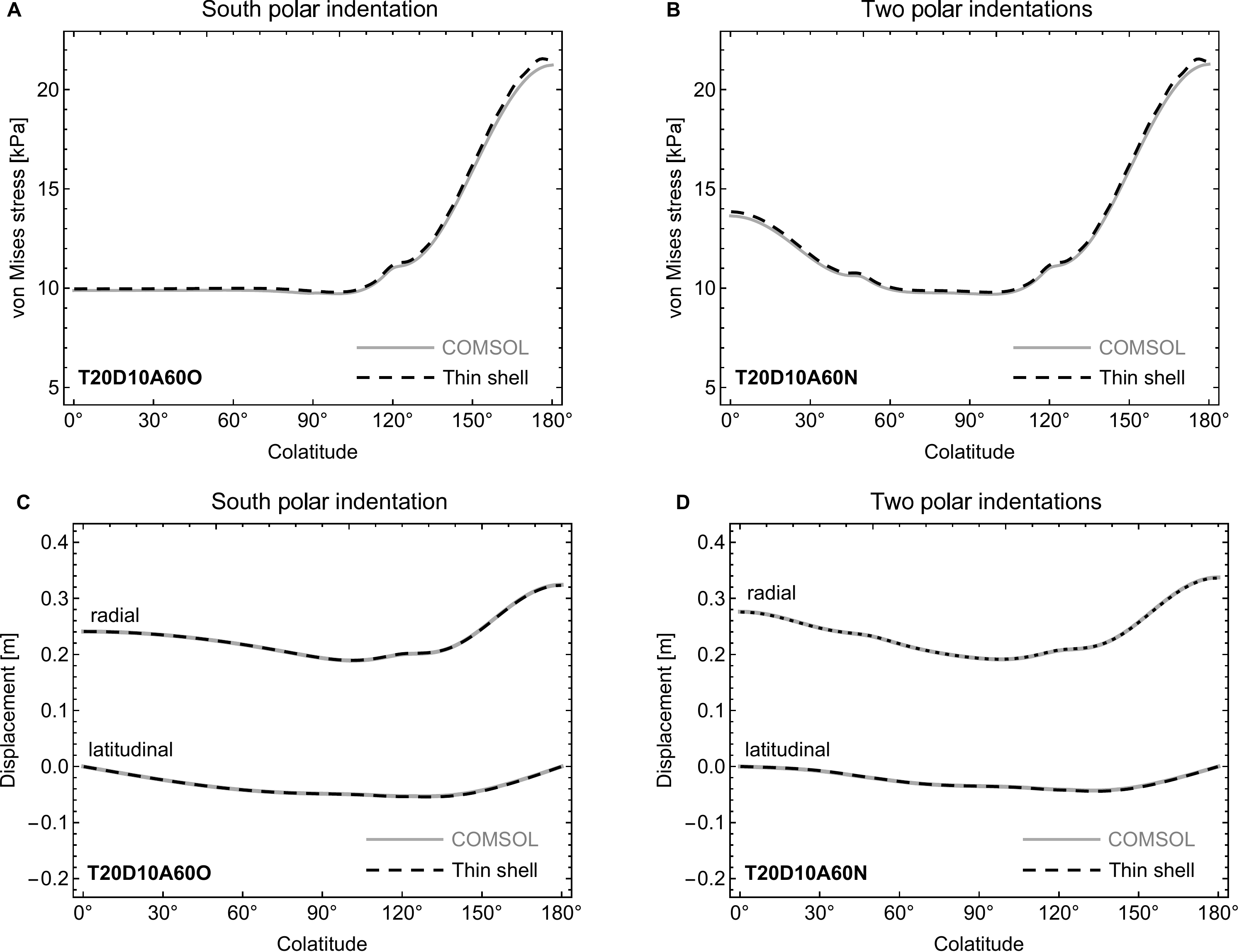}
   \caption[Effect of pressurized global ocean]
   {Impact of pressurized global ocean on Enceladus's surface: (A,B) von Mises surface stress $\sqrt{3J_2}$ and (C,D) surface displacement, as functions of the colatitude.
   In panels A and C, the crust is thinner at the south pole; in panels B and D, it is thinner at both poles. 
   Gray curves show the finite element results of \citet{johnston2017Z}, while dashed black curves show the thin shell results.
   Bold labels refer to the $20\rm\,km$ shell thickness outside of the indentations (\textbf{T20}), the $10\rm\,km$ depth of the south polar indentation (\textbf{D10}), the $60^\circ$ half-angle of the south polar indentation (\textbf{A60}), and the absence (\textbf{O}) or presence (\textbf{N}) of a $5\rm\,km$ deep indentation at the north pole.}
   \label{FigJM}
\end{figure}

For the comparison, I chose the FEM models having a shell structure similar to Fig.~\ref{FigThickness}: the shell thickness is $20\rm\,km$ outside of the indentations, the south polar indentation is $10\rm\,km$ deep, and the half-angle of the south polar indentation is $60^\circ$.
Fig.~\ref{FigJM} shows the von Mises stress $\sqrt{3J_2}$ (as in Fig.~5 of \citet{johnston2017}) and the surface displacements for the models with one or two polar indentations. 
The agreement between thin shell  and FEM results is excellent.

%************************************************************************************************************************************
\subsubsection{Tidal response of Ganymede}
\label{TidalResponseGanymede}

\citet{a2014} (hereafter \textit{A14}) studied how lateral variations of the icy shell thickness or shear modulus affect the tidal response of Ganymede and Europa.
They computed the radial displacement (or uplift) and the geoid perturbation with the FEM software CitcomSVE (not publicly available).
All layers, including the ocean, are treated as Maxwell viscoelastic solids.
The system is forced with a static tidal potential and subsequently relaxes viscously until the ocean reaches a fluid state.
Due to the limited mesh resolution, this approach does not converge well if the shell is very thin and of non-uniform thickness (as on Europa).
CitcomSVE was first applied to a tidal problem by \citet{zhong2012} who benchmarked it against spherically symmetric models of the Moon's interior.
\textit{A14} then applied the method to icy satellites with internal oceans, benchmarked it against spherically spherically models for Ganymede, and performed several convergence and resolution tests on 3D models (described in detail in their paper).
Next, \citet{qin2014,qin2016} benchmarked CitcomSVE against their perturbation method for 3D structures of the Moon's crust and mantle, but always for layers bounded by spherically symmetric interfaces.
Furthermore, \citet{qin2016} studied the effect of lateral variations in crustal thickness by modelling them as lateral density variations around the crust-mantle boundary.
However, this configuration is not analogous to the thickness variations of an icy shell floating on a global ocean because the Moon's crust does not freely slip on a fluid mantle.
To sum up, CitcomSVE has been shown to work well for lateral variations of the structure of spherical layers, but the code has not yet been benchmarked for shell thickness variations of icy satellites with global oceans.

I will first compare FEM and thin shell predictions for a spherically symmetric model of Ganymede, for which the Love numbers can be very computed accurately with other methods.
Model parameters are given in Table~1 of \textit{A14}; the shell is $108\rm\,km$ thick.
Using the algorithms of \citet{wahr2009}, \textit{A14} obtain the `semianalytic solution' of Table~\ref{TableLove2}, but their solution is slightly wrong.
I obtain lower values (`numerical integration' in Table~\ref{TableLove2}) with the numerical integration method used to compute thick shell Love numbers in Section~\ref{LaterallyUniformThickShell} (\textit{A.~Trinh} independently checked my solution with a code developed for dynamical Love numbers).
The thin shell values (`thin shell' in Table~\ref{TableLove2}) are computed with Eq.~(\ref{LoveThin}) in which the fluid-crust Love number $h_2^\circ=1.518$ results from numerical integration, as in Section~\ref{LaterallyUniformThickShell}.
Compared to the reference values $(h_2,k_2)=(1.432,0.487)$, $h_2$ is more accurately predicted by the thin shell approach ($0.5\%$ error) than by CitcomSVE ($1.6\%$ error), whereas the thin shell result for $k_2$ is nearly exact.
Errors at the 1\% level can seem small, but they are larger than the corrections due to non-uniform shell thickness.

The thin shell error on $h_2$ is mostly due to shell compressibility.
The latter effect is not fully taken into account in thin shell theory because of the assumption that the upper and lower boundaries of the shell deform in the same way (see Appendix~B of \textit{Be15b}).
The massive membrane approach of \textit{Be15b} does not make this assumption, so that the resulting Love numbers are nearly the same as the reference values (`massive membrane' in Table~\ref{TableLove2}).
The FEM error on $h_2$ and $k_2$ most likely results from discretization:
\textit{A14} observe that the error on $h_2$ decreases from 8\% to 2\% as the mesh resolution increases between their models A2-1D and A4-1D.
They could not check, however, whether the error continues to decrease as the mesh becomes finer.
Another explanation could be that viscoelastic relaxation has a significant effect (of the order of 1\%) beyond the end of the model run time (see their Fig.~A1).

%TABLE 5
\begin{table}[ht]\centering
\ra{1.3}
\footnotesize
\caption[Tidal Love numbers $h_2$ and $k_2$ for Ganymede]
{\small
Tidal Love numbers $h_2$ and $k_2$ for the spherically symmetric model of Ganymede.
See Section~\ref{TidalResponseGanymede} for details.
}
\begin{tabular}{@{}lcc@{}}
\hline
 & $h_2$ &  $k_2$ \\
 \hline
 \multicolumn{3}{l}{\citet{a2014} (Table A1):}\\
Semianalytical solution & $1.440$ & $0.495$ \\
FEM model A4-1D & $1.409$ & $0.481$ \\
\multicolumn{3}{l}{\textit{This paper:}}\\
Numerical integration & $1.432$ & $0.487$ \\
Thin shell & $1.425$ & $0.487$ \\
Massive membrane & $1.432$ & $0.487$ \\
\hline
\end{tabular}
\label{TableLove2}
\end{table}%

\begin{figure}
   \centering
      \includegraphics[width=15cm]{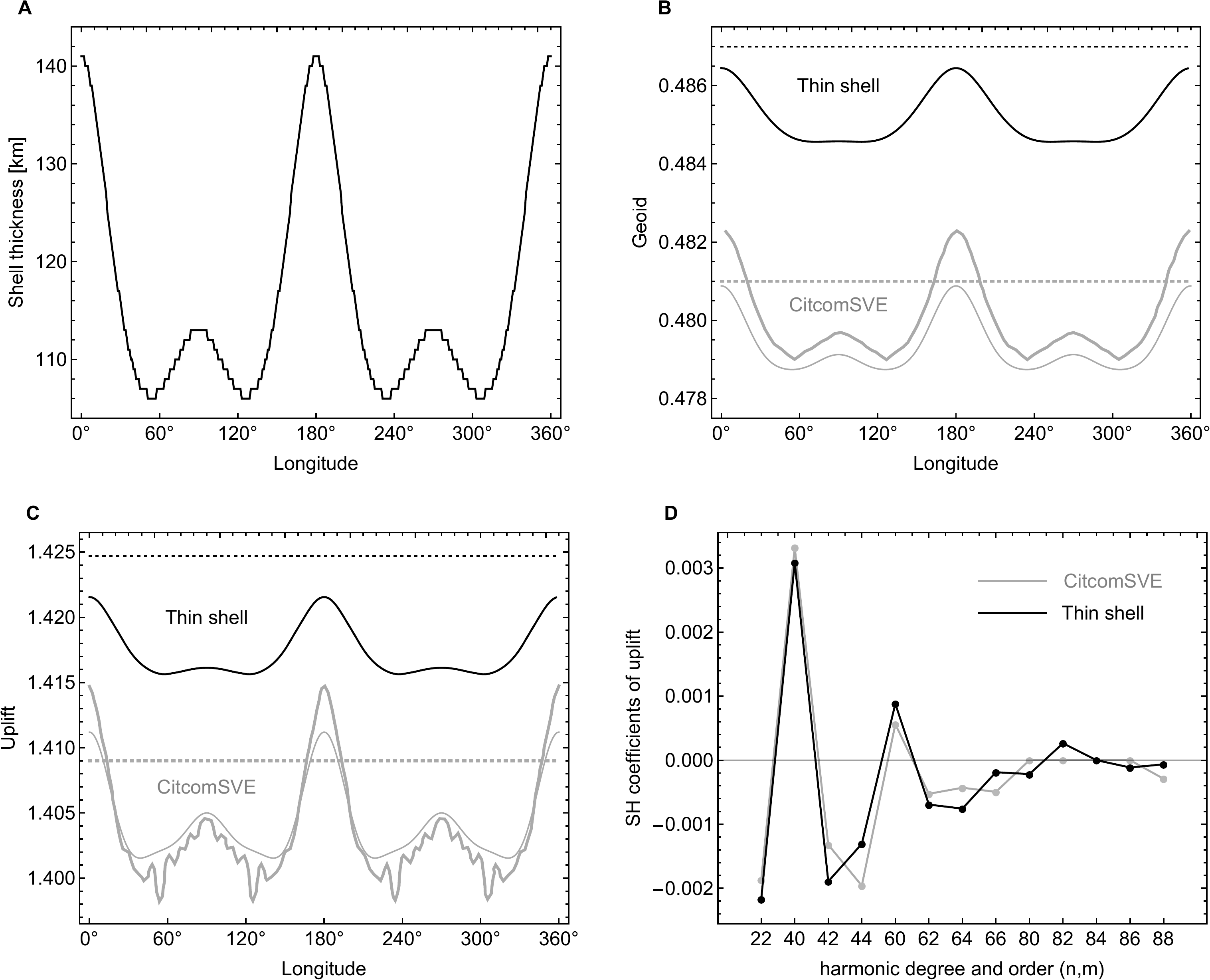}
   \caption[Tidal response of Ganymede with non-uniform shell thickness]
   {Tidal response of Ganymede with non-uniform shell thickness: (A) thickness, (B) geoid, (C) uplift, (D) spherical harmonic coefficients of smoothed uplift.
    Thick (resp.\ thin) gray curves show the full (resp.\ smoothed) FEM solutions obtained with CitcomSVE; black curves show thin shell solutions.
      Dotted lines show the results for the spherically symmetric model.
   In panel~D, the FEM coefficients of degree 8 and order 0 to 6 are set to zero because they do not appear in Table~2 of \citet{a2014}.
   See Section~\ref{TidalResponseGanymede} for details.
   }
   \label{FigGA}
\end{figure}

Next, I will compare FEM and thin shell results if Ganymede's shell thickness varies laterally (model GA of \textit{A14}; equatorial profile shown in Fig.~\ref{FigGA}A).
I work with the same thickness file as the authors (\textit{G.~A}, private communication) and similarly restrict tidal forcing to the component of zeroth harmonic order.
The fluid-crust Love numbers -- required to take into account the structure below the shell -- are computed with the numerical integration method of Section~\ref{LaterallyUniformThickShell}.
Though degree-0 effects and ocean compressibility are not discussed in \textit{A14},  I will assume $K_o=2\rm\,GPa$ so that $h_0^{\circ P}\approx-0.3$, which decreases the displacement by less than 0.1\% whereas the geoid is barely affected.
Fig.~\ref{FigGA} (panels B and C) shows the geoid and uplift along the equator as in Fig.~3 of \textit{A14}.
The normalization follows \textit{A14}, so that the uplift (resp.\ geoid) corresponds to $h_2$ (resp.\ $k_2$) in the limit of a spherically symmetric body.
As discussed in Appendix~A of \textit{A14} (and clearly visible in Fig.~\ref{FigGA}), short-wavelength noise affects the FEM solution because of the large lateral viscosity contrasts across the mesh elements.
Besides, numerical noise causes nonzero leakage from the dominant $(2,0)$ harmonic into other long wavelength harmonics.
Since leakage also affects the spherically symmetric case, \textit{A14} propose a smoothed solution by subtracting the spherically symmetric solution and keeping only the dominant spherical harmonic components (see Table~2 of \textit{A14}; the $(2,0)$ components of the uplift and geoid are 1.407 and 0.48 as in their Table~A1).
The smoothing significantly reduces the magnitude of the primary peaks (at longitudes $0^\circ$ and $180^\circ$).

Thin shell and FEM predictions differ in three important respects:
\begin{enumerate}
\item
There is a constant shift (along the equator) between the thin shell and FEM solutions.
For the uplift (resp.\ geoid), the shift is about $0.016$ (resp.\ $0.006$), i.e.\ the difference between the $h_2$ (resp.\ $k_2$) solutions of Table~\ref{TableLove2}.
Thus, the explanation for the shift must be the same whether the shell thickness is laterally variable or not.
\item
The maximum amplitude of variations (given by the primary peaks) is smaller in the thin shell approach: by about a factor of 2 for the uplift, by about one third for the geoid, if comparison is done with the full FEM solution.
The difference is however much smaller (especially for the geoid) if the thin shell solution is instead compared to the smoothed FEM solution.
\item
The thin shell solution is much smoother: there is no short-wavelength noise, and the amplitude ratio of the secondary peaks (at longitudes $90^\circ$ and $270^\circ$) to the primary peaks is much smaller.
In the FEM solution, the  amplitude ratio of secondary to primary peaks is larger than the corresponding ratio for shell thickness (Fig.~\ref{FigGA}A).
\end{enumerate}
These differences raise the following issues regarding the FEM method:
\begin{enumerate}
\item
Systematic bias ($1.6\%$ too low) in the FEM solution: does it have an impact on the corrections (which are of smaller amplitude) due to lateral thickness variations?
This issue does not arise in the thin shell approach because the problem is solved in the spectral domain: non-uniform corrections appear as perturbation terms and are protected to leading order from an error in the dominant degree~2 coefficient.
\item
Long and medium wavelength variations: why is the amplitude ratio of secondary to primary peaks larger than the corresponding ratio for shell thickness?
It could be expected that the elastic shell acts as a kind of filter, smoothing more the influence of short-wavelength thickness variations.
\item
Short-wavelength noise: does it have a negligible effect on the long-wavelength part of the solution?
Since the non-uniformity of the shell couples all wavelengths, there could be a cascade effect from short to long wavelengths.
\end{enumerate}
In spite of these differences, thin shell and FEM solutions share many similarities: Fig.~\ref{FigGA}D shows that the dominant spherical harmonic coefficients of the two solutions are well correlated, especially at the longest wavelengths.
At this stage, it is not possible to conclude whether the corrections due to non-uniform shell thickness are best predicted by FEM or by the thin shell approach.
Rigorous benchmarking should be done with simpler shell structures, in which the thickness variations are either zonal or sectoral.
If there is a problem with the thin shell method, it probably originates in the approximate tidal coupling, because the elastic part of the method compares well with other FEM methods (Section~\ref{PressurizedOcean} of this paper; \citet{kalousova2012}).
Finally, note that the disagreement between FEM and thin shell result is probably irrelevant to Enceladus since its tidal deformations are dominated by elastic effects (hard-shell regime).

%=========================================================================================
%=========================================================================================
\subsection{Results}
\label{Results}

In this section, I apply the thin shell approach to the non-uniform models (conductive and convective) described in Section~\ref{ShellStructure}.
The tidal deformations of Enceladus are small (of the order of $1\rm\,m$) and there is little hope to measure them in the near future, even with an orbiter around the satellite.
However, I find it useful to give a concrete example of a non-uniform thin shell solution (radial displacement and stress function).
It not only shows what the thin shell approach really does, but it can serve as a benchmark for other methods.
Physically, the most important prediction concerns the magnitude of tidal stresses at the surface, which may control the periodic opening and closing of the `tiger stripes' fractures at the south pole.

Instead of showing tidal deformations at different times during the orbit (of period $T$), I give them in the frequency domain. 
The conversion to the time domain is given by Eq.~(\ref{fourier}).
For example,
\begin{equation}
\tilde w(t) =  {\rm Re}(w) \cos\omega t - {\rm Im}(w) \sin\omega t  \, .
\end{equation}
Thus, the real part of $w$ corresponds the amplitude at $t=0$ and subsequent half-periods, while the imaginary part corresponds to the amplitude at $t=T/4$ plus half-periods.
I checked that the spectrum of the tidal response is consistent with the predictions of mode coupling (see Section~\ref{MembraneLimit}).
The degree-2 zonal harmonic in the shell thickness (Eq.~(\ref{ThicknessVarGRL})) induces a dominant non-degree-2  tidal response at degree~4 but with significant contributions from even degrees higher than~4.
By contrast, the degree-3 zonal harmonic induces a dominant tidal response at degrees 3 and 5, with subdominant contributions at degree~4 and at degrees higher than 5, both even and odd.
Spectral analysis is useful for benchmarking but physical interpretation is best done in the spatial domain because the lateral variations of Enceladus's shell thickness are too large to be well approximated by a low-order perturbative expansion.  

\begin{figure}
   \centering
   \includegraphics[width=15cm]{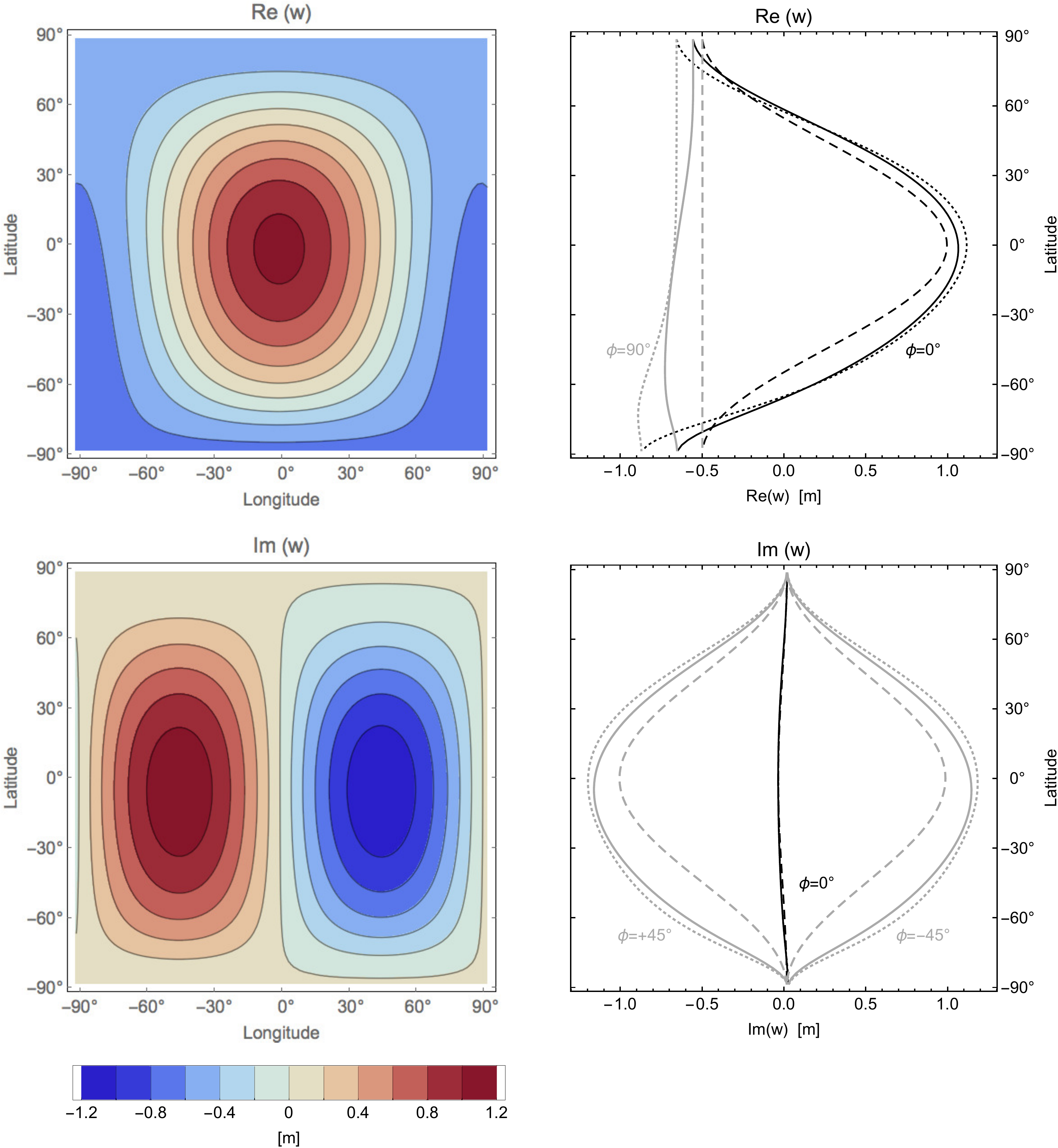}
   \caption[Radial displacement of a non-uniform conductive shell]
   {Radial displacement of a non-uniform conductive shell: spatial patterns (left panels) and meridional crosscuts (right panels).
   Upper (resp.\ lower) panels show the real (resp.\ imaginary) part of the radial displacement in the frequency domain.
   Solid (resp.\ dotted) crosscuts show the thin shell (resp.\ membrane) results, both for the laterally non-uniform shell, while dashed crosscuts show the results for the laterally uniform shell.
   The tidal axis is aligned with zero longitude.
   Spatial patterns repeat between longitudes $90^\circ$ and $270^\circ$.
   }
   \label{Figw}
\end{figure}

\begin{figure}
   \centering
   \includegraphics[width=15cm]{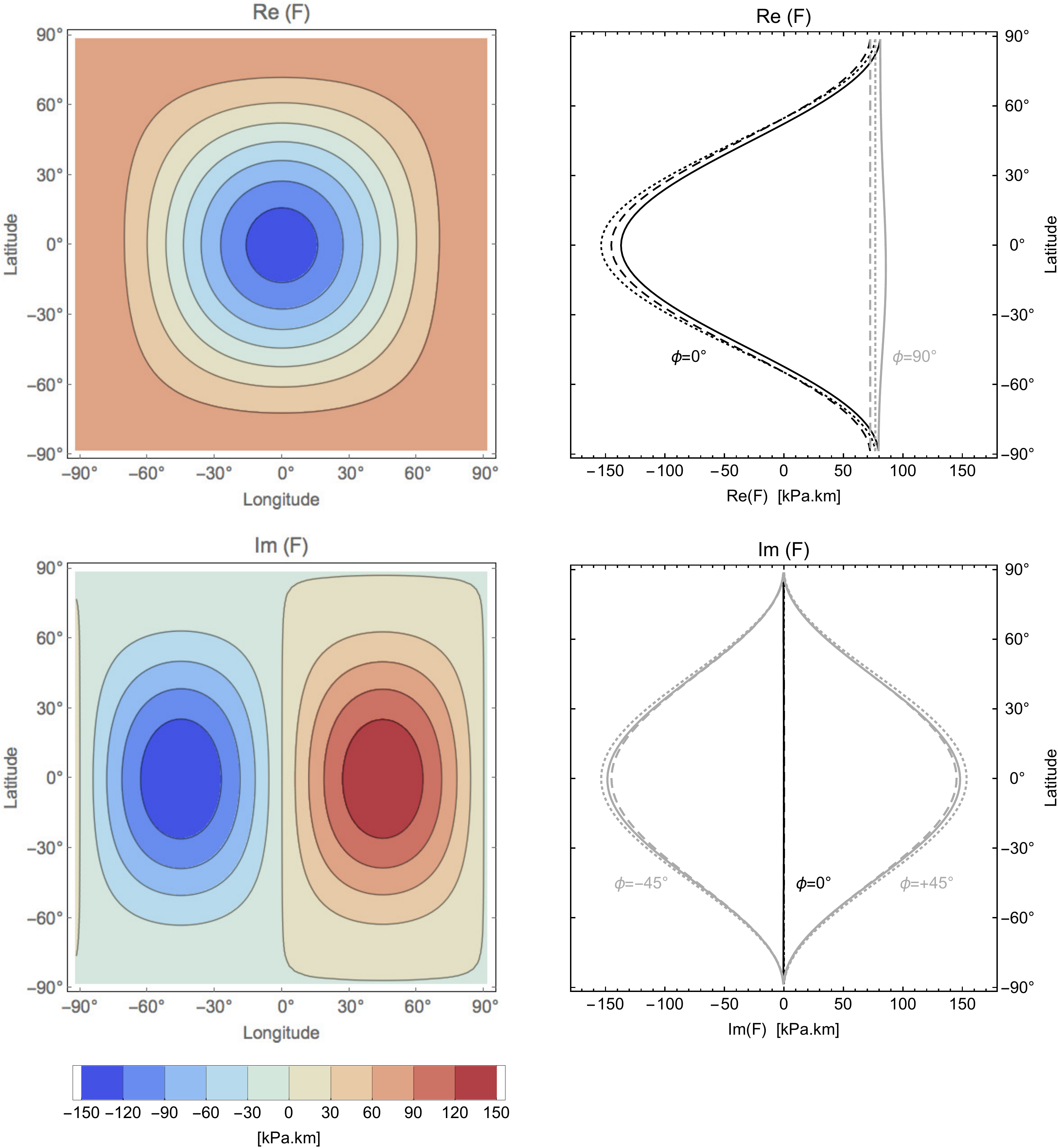}
   \caption[Stress function of a non-uniform conductive shell]
   {Stress function of a non-uniform conductive shell: spatial patterns (left panels) and meridional crosscuts (right panels).
   Upper (resp.\ lower) panels show the real (resp.\ imaginary) part of the stress function in the frequency domain.
   Other details as in Fig.~\ref{Figw}.
   }
   \label{FigF}
\end{figure}

Figs.~\ref{Figw} and \ref{FigF} (left panels) show the spatial patterns of the radial displacement and stress function for the non-uniform conductive model.
In the right panels, crosscuts of the left panels at specific longitudes ($0^\circ$ and either $45^\circ$ or $90^\circ$) are compared to the results of the laterally uniform conductive model (Eq.~(\ref{TidalThinShellSol}), dashed curves) and to the non-uniform membrane approximation (Eq.~(\ref{MemApprox}), dotted curves).
Viscoelasticity has a small effect in the conductive model, visible for example in the slight deviation from zero of the $\varphi=0^\circ$ cross-cut of ${\rm Im}(w)$.
The non-uniform shell structure has a significant (but not major) effect on the radial displacement: at the south pole, $w(t=0)$ changes from $-0.5\rm\,m$ to $-0.7\rm\,m$.
By contrast, lateral variations of the shell structure have a much smaller effect on the stress function.
This behaviour is expected from the dominance of membrane effects in tidal deformations (Section~\ref{MembraneLimit}).
If the shell convects, bending effects become significant at the south pole where ${\rm Re}(w)$ and ${\rm Im}(w)$ reach $-1.4\rm\,m$ and $0.5\rm\,m$, respectively (the membrane approximation fails in that case).
The stress function, however, is weakly affected by the softer rheology at the south pole.

\begin{figure}
   \centering
   \includegraphics[width=15cm]{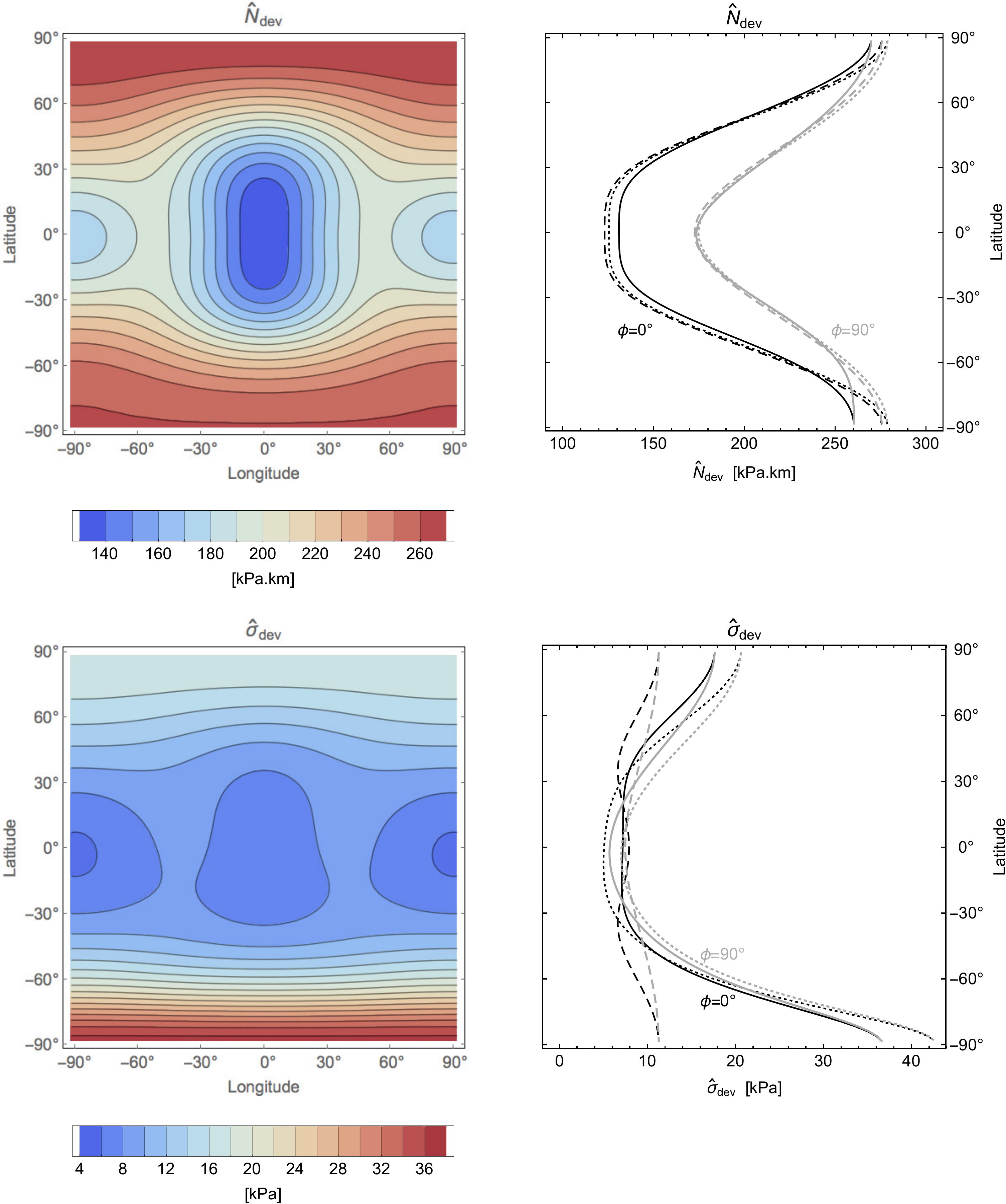}
   \caption[Stress invariants for a non-uniform conductive shell]
   {Stress invariants for a non-uniform conductive shell: deviatoric stress resultant (top) and surface deviatoric stress (bottom).
   Left panels show spatial patterns and right panels show meridional crosscuts.
   Other details as in Fig.~\ref{Figw}.
   }
   \label{FigDevStress}
\end{figure}

Fig.~\ref{FigDevStress} shows the spatial patterns and meridional crosscuts of the time-averaged deviatoric stress resultant $\widehat N_{\rm dev}$ and surface deviatoric stress $\widehat \sigma_{\rm dev}$ (Section~\ref{TidalStressSec}) for the non-uniform conductive model.
Similarly to the stress function, the stress resultant is weakly affected by lateral variations in shell structure, with the advantage of having a physical interpretation in terms of depth-integrated stresses.
The surface stress varies by a factor of 4 ($\varphi=0^\circ$) to 6 ($\varphi=90^\circ$) between the equator and the south pole, in parallel with the variation in shell thickness.
Again, this behaviour agrees with the membrane limit.

%\FloatBarrier

%=========================================================================================
%=========================================================================================
\subsection{Discussion}
\label{Discussion}

\begin{figure}
   \centering
    \includegraphics[width=7cm]{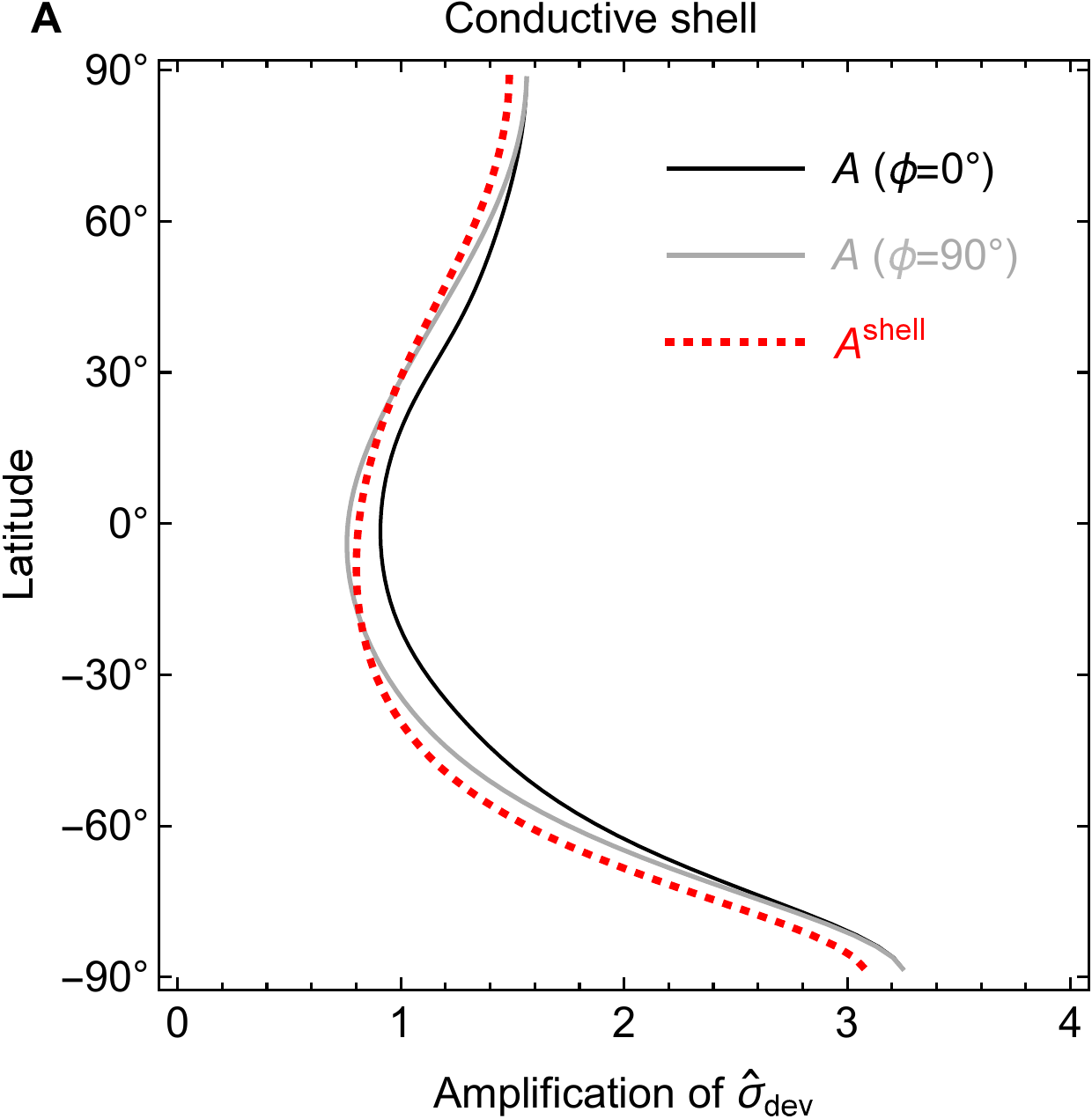}
      \hspace{2mm}
    \includegraphics[width=7cm]{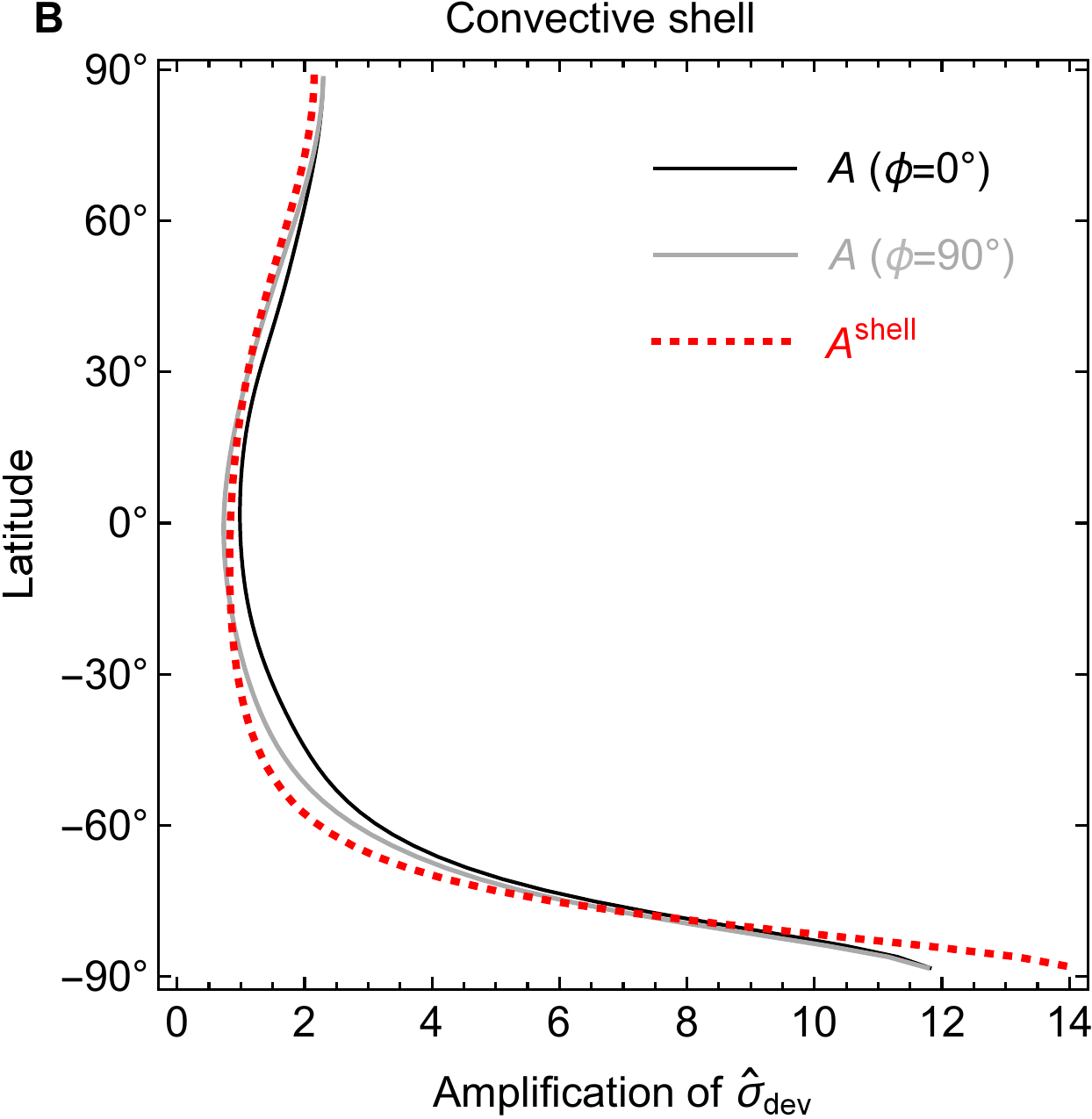}
   \caption[Stress amplification]
   {Amplification of the time-averaged surface deviatoric stress: (A) non-uniform conductive shell; (B) non-uniform convective shell.
   Solid curves show the stress ratio $A$ (Eq.~(\ref{AmpliStress})) while dotted curves show the extensibility ratio $A^{shell}$ (Eq.~(\ref{AmpliShell})).
   }
   \label{FigAmpli}
\end{figure}

The equator/pole stress difference is only partly due to the shell structure: tidal stresses vary by nearly a factor of 2 even if the shell is completely uniform (Fig.~\ref{FigsigdevUNI}).
Therefore, the influence of the shell structure is more obvious in the ratio of the surface stress for a non-uniform shell to the surface stress for the uniform case.
Fig.~\ref{FigAmpli}A shows the amplification of the surface stress defined by
\begin{equation}
A = \widehat \sigma_{\rm dev}^{\,\rm var}/\widehat \sigma_{\rm dev}^{\,\rm uni} \, ,
\label{AmpliStress}
\end{equation}
where the superscripts refer to the variable thickness model ({\small var}) and the laterally uniform model ({\small uni}), respectively.
The amplification of $\widehat \sigma_{\rm dev}$ closely follows the variation in magnitude of the extensibility $\alpha$, which is inversely proportional to the shell thickness.
The equatorial stress is a bit reduced whereas the south polar stress is amplified by a factor of more than 3.
The amplification of surface stress is well predicted by the extensibility ratio
\begin{equation}
A^{shell} = \left| \alpha^{\rm var}/\alpha^{\rm uni} \right| .
\label{AmpliShell}
\end{equation}
The quantity $d/A^{shell}$ corresponds to the effective elastic thickness of the lithosphere (\textit{Be15b}, Eq.~(49)).
If the shell convects, $\widehat \sigma_{\rm dev}$ is amplified by a factor of 12 at the south pole with respect to the laterally uniform conductive model (Fig.~\ref{FigAmpli}B).
The scaling rule is still applicable, though it overestimates by 20\% the stress at the south pole where bending effects become important.
Our conclusions about stress amplification agree with the recent work of \citet{behounkova2017}, who found that the stress in an elastic shell of variable thickness is approximately inversely proportional to the local shell thickness.
Using a FEM model, these authors also predict the effect of faults breaking the shell through all its depth, showing that it increases the stress at the tips of the faults (as shown in \citet{soucek2016}).
Their model, however, is purely elastic (viscoelasticity is introduced a posteriori to compute tidal heating), contrary to the thin shell approach developed in this paper.
Unfortunately, their 2D density plots do not lend themselves to benchmarking.

\begin{figure}
   \centering
    \includegraphics[width=7cm]{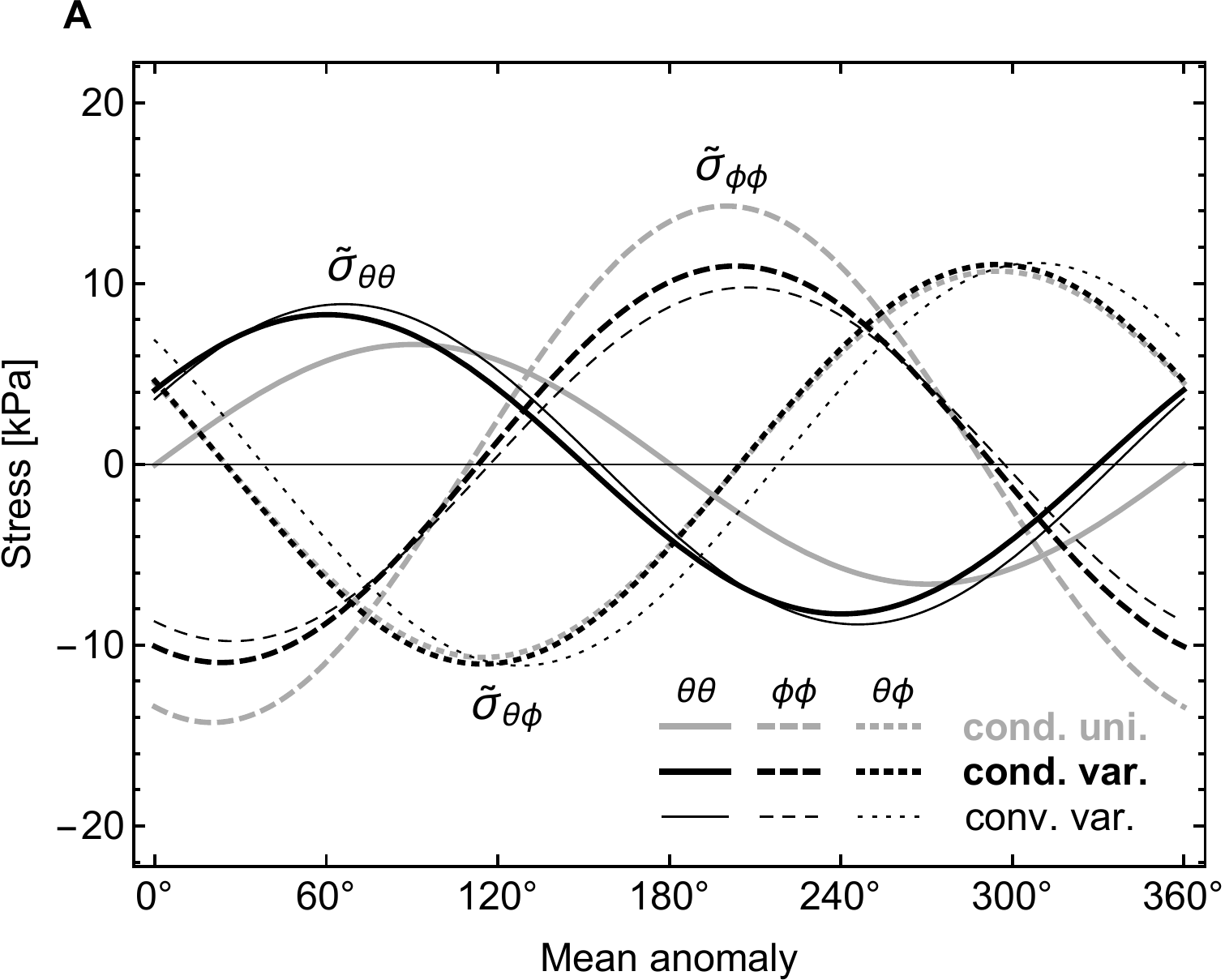}
    \hspace{2mm}
    \includegraphics[width=7cm]{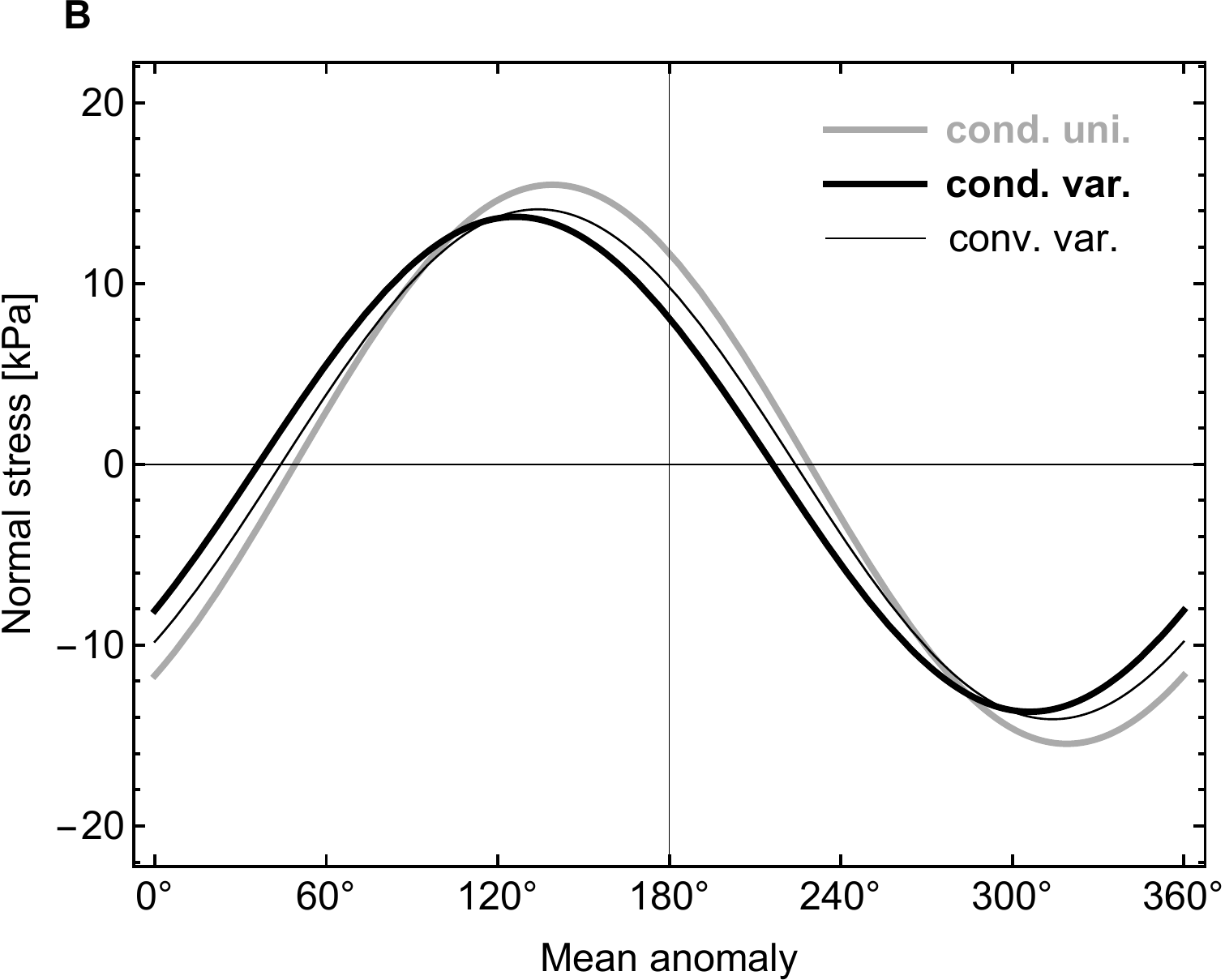}
   \caption[Surface stresses at $(70^\circ$S, $195^\circ$E)]
   {Surface stresses at $(70^\circ$S, $195^\circ$E): (A) stress components; (B) stress normal to the Alexandria fault (orientation of $43^\circ$ with respect to the radial direction).
   The labels `cond.' and `conv.' denote conductive and convective shells, while `uni.' and `var.' denote shells of uniform and variable thickness, respectively.
  Stresses are rescaled by the factor $1/A$ (Eq.~(\ref{AmpliStress})) with $A=2.4$ (resp.\ $A=5.1$) for the conductive (resp.\ convective) shell of variable thickness.
   }
   \label{FigStressLocal}
\end{figure}

Stress orientation is much less sensitive to variations in shell thickness and rheology than stress magnitude, as can be expected from the assumption of isotropic viscoelastic moduli.
As an illustration, Fig.~\ref{FigStressLocal}A shows the rescaled stress components, in the spatial domain, at $(70^\circ$S, $195^\circ$E) on the Alexandria tiger stripe (same location as in Fig.~A of \citet{nimmo2007} and Fig.~S2 of \citet{behounkova2015}).
Compared to the laterally uniform case, the magnitude of the rescaled stress components can be somewhat larger (${\theta\theta}$), equal (${\theta\varphi}$) or somewhat smaller (${\varphi\varphi}$), while their phase can be slightly shifted (${\theta\theta}$), but these effects are minor with respect to the overall amplification.
In particular, the stress normal to the fault is only slightly advanced in phase if the thickness is non-uniform (Fig.~\ref{FigStressLocal}B).

What are the implications for the formation of the tiger stripes and the geyser activity?
Most scenarios postulate that the tiger stripes form as tensile cracks, whether because of diurnal tidal stress \citep{nimmo2007}, ocean pressurization \citep{manga2007,johnston2017}, non-synchronous rotation \citep{patthoff2011}, or yet another cause.
If tidal stresses were responsible, the orientation of the tiger stripes should be close to the tidal axis because this direction maximises the normal tensile stress.
In fact, the fault segments have directions between $20^\circ\rm{}W$ and $45^\circ\rm{}W$ from the tidal axis.
The normal tensile is still near-maximum in these directions, but this does not explain why the tiger stripes are not oriented instead at $0^\circ$ or at $30^\circ\rm{}E$.
Crustal thinning does not help because the fault orientation maximizing the normal tensile stress is still along the tidal axis (at least in an elastic shell with axial symmetry).
On the other hand, stress amplification due to crustal thinning is absolutely necessary in order to increase the tidal stress beyond the yield strength of ice (see \citet{johnston2017}), which is otherwise larger than the tidal stress (about $15\rm\,kPa$ in a $23\rm\,km$-thick uniform shell).
Stress amplification also explains why the largest and most recent cracks appear at the south pole but not elsewhere, except maybe at the north pole (see Fig.~2 of \citet{johnston2017}).
More generally, crustal thinning is the most obvious explanation for south polar faulting in all models assuming global deformations, such as those resulting from a pressurized ocean or from non-synchronous rotation.
This being said, the case for tiger stripes forming as tensile cracks is far from being proven.
On the contrary, \citet{yin2015} argue that these faults originate in left-slip bookshelf faulting caused by gravitational spreading due to the higher elevation of the leading-edge margin of the South Polar Terrain. 

The case for tidal stresses influencing the geyser activity is more solid.
Three independent studies, using either Cassini's Visible and Infrared Mapping Spectrometer \citep{hedman2013} or its Imaging Science Subsystem \citep{nimmo2014,ingersoll2017}, found that the brightness of the south polar plume is minimum and maximum around the pericentre and apocentre, respectively.
The diurnal periodicity and the occurrence of the maximum peak brightness at apocentre both strongly suggest that tidal stresses modulate the activity of geysers.
The simplest approach consists in correlating the activity with the opening/closing of faults: in a viscoelastic model, faults open if the stress normal to the fault is tensile \citep{hurford2007}.
As the tiger stripes are not aligned with the tidal axis, the normal tensile stress peaks significantly before the apocentre (Fig.~\ref{FigStressLocal}B); this advance is even larger if the thickness is non-uniform (\citet{behounkova2017} find the same effect in an elastic model without faults).
\citet{nimmo2014} suggested that the viscoelasticity of the shell could be responsible for the delay, but the required phase shift is large: about $56^\circ$ for all the faults together.
In a model with a global ocean, it can be attributed to viscoelasticity only if the whole shell is very thick ($60-70\rm\,km$) and extremely soft with the exception of a thin lithosphere \citep{behounkova2015}.
By contrast, convection restricted to the polar regions only slightly delays the response of a laterally non-uniform shell (Fig.~\ref{FigStressLocal}B).
In conclusion, viscoelasticity cannot save the simplest tidal stress model in which geyser activity varies in tandem with the stress normal to the tiger stripes.
The time delay could instead be due to turbulent flow in the fissures connecting the ocean and the surface \citep{kite2016}, or to more complicated faults mechanics (e.g.\ the influence of cross-cutting fractures, see \citet{helfenstein2015}).

%\FloatBarrier

%%%%%%%%%%%%%%%%%%%%%%%%%%%%%%%%%%%%%%%%%%%%%%%%%%%%%%%%%%%
%%%%%%%%%%%%%%%%%%%%%%%%%%%%%%%%%%%%%%%%%%%%%%%%%%%%%%%%%%%
%%%%%%%%%%%%%%%%%%%%%%%%%%%%%%%%%%%%%%%%%%%%%%%%%%%%%%%%%%%
\section{Summary}

According to recent gravity and libration data, Enceladus harbours a global ocean under an icy shell having an average thickness less than 10\% of the radius.
Large variations of shell thickness (and probably rheology) occur, mostly in latitude, and result in a few km thick (and probably soft) crust at the south pole.
Lateral variations in rheology are probably present although they cannot be directly deduced from the data, and the existence of a subsurface ocean implies even larger variations of rheology with depth.
Tidal stresses are enhanced by crustal thinning at the south pole, so that realistic models of tidal tectonics and dissipation should take into account spatial variations of shell thickness and rheology.

For this purpose, I developed the theory of \textit{non-uniform viscoelastic thin shells}, assuming that the shell is thin, viscoelasticity is linear, and Poisson's ratio $\nu$ is uniform.
Although this theory is essentially two-dimensional, it includes 3D features by taking into account the dependence of rheology on depth, as well as lateral variations in rheology and shell thickness.
The variations of the shell structure are not treated as perturbations and can thus be large.
The theory is governed by two differential equations of the 4th order depending on two scalar variables and having non-constant coefficients (Eq.~(\ref{GoverningEq})).
The variables are the \textit{radial displacement} $w$ (constant with depth inside the shell) and the \textit{stress function} $F$, from which all other quantities (stress, strain, lateral displacement) are derived.
Besides Poisson's ratio, three 2D viscoelastic parameters are defined in terms of the moments of the complex shear modulus: the \textit{extensibility} $\alpha$ and the \textit{bending rigidity} $D$ generalize the parameters of thin elastic shell theory, and the third parameter $\chi$ is close to one.
While a reference surface needs to be defined, as always in thin shell theory, observable quantities (such as displacements and stresses) do not depend on that choice (Appendices~\ref{Covariance}, \ref{StrainFw}, and \ref{StressFw}).
For tidal coupling, the best choice for the reference surface is the sphere whose radius coincides with the mean surface radius of the body.

Enceladus's shell is coupled to tides under the assumption that the shell and ocean have a negligible density contrast (Section~\ref{MasslessShell}), leading to an error of $1-2\%$ on the Love numbers $(h_2,l_2)$ if the shell is laterally uniform and $23\rm\,km$ thick.
The resulting \textit{tidal thin shell equations} (Eq.~(\ref{TidalThinShellEq})) include self-gravity and allow for realistic interior models:
the core can be viscoelastic and stratified in density (as well as the ocean), as long as the structure below the shell is spherically symmetric.
If the shell is laterally uniform, the theory reduces to the analytical thin shell formulas for the tidal Love numbers $h_n$, $k_n$, and $l_n$ (Section~\ref{LaterallyUniformThinShell}).
Thick shell corrections are partly included in the form of bending corrections, which lead to a much better estimate of the tangential displacement than in the membrane limit (Sections~\ref{UniformThickShell} and \ref{LaterallyUniformThickShell}).
The error on the radial and tangential displacements is less than 5\% and 2\%, respectively, if the shell is laterally uniform with a thickness less than 10\% of the radius (Fig.~\ref{Figh2l2Conductive}); the error on the time-averaged deviatoric stress varies laterally between less than 1\% at the poles and 15\% along the tidal axis (Fig.~\ref{FigsigdevUNI}).

If the shell is laterally non-uniform, the tidal thin shell equations could in principle be solved on a grid, but the spherical geometry and the spherical differential operators are easier to handle in the spectral domain.
In a spherical harmonic basis, the tidal thin shell equations become a system of coupled linear equations which can be solved with standard matrix methods (Section~\ref{NumericalMethod}).
This method is very efficient if the shell structure varies only in latitude.
For arbitrary variations of the shell structure, the matrices increase in size as the square of the maximum harmonic degree required to characterize the variations.
In the future, one could investigate other methods restricted to the spatial domain.
I compared the results for a laterally non-uniform shell with two previous studies using finite element methods (FEM).
The first comparison does not involve tides, but overpressure due to the crystallization of Enceladus's internal ocean \citep{johnston2017}.
The excellent agreement between the thin shell approach and the FEM results (Fig.~\ref{FigJM}) confirms the validity of the thin elastic shell equations with variable thickness, which were already validated by \citet{kalousova2012}.
The second comparison concerns the tides of Ganymede, a much larger satellite for which the effect of variable shell thickness is below 1\% of the dominant degree-2 response \citep{a2014}.
In spite of many similarities, the thin shell and FEM results do not fully agree (Fig.~\ref{FigGA}).
No definite conclusion can be drawn without a more detailed benchmarking, but the thin shell solution appears to be less biased and is certainly less noisier.
Benchmarking against the FEM model of \citet{behounkova2017} will be done in the second paper of the series.

If the variations in shell structure are not too steep, the tidal deformations of the shell are of long wavelength so that the membrane limit is a good first-order approximation.
In that limit, the shell can be said to be soft or hard (Sections~\ref{SoftHardShells} and \ref{MembraneLimit}).
In contrast to Europa and Ganymede, Enceladus has a hard shell which implies that the surface deformation is controlled by the laterally varying structure of the shell, whereas the stress function (and thus the depth-integrated stress) is nearly independent of shell properties.
In good approximation, surface stresses are inversely proportional to the absolute value of the depth-integrated shear modulus $\mu_0d$ (Fig.~\ref{FigAmpli}).
If Enceladus's shell is conductive with isostatic thickness variations, crustal thinning at the poles strongly enhances the surface stress, whereas the surface deformation is only moderately enhanced.
For the isostatic shell shown in Fig.~\ref{FigThickness}, the amplification is about 60\% at the north pole and about 3.3 at the south pole. 
If the shell is softened by convection below the poles, the stress can be amplified in total by a factor of 10 at the south pole.
However, a locally viscous shell does not explain the time lag between the maximum geyser activity and the maximum of normal tension along the faults (Section~\ref{Discussion}).
In conclusion, crustal thinning and rheology softening probably play a major role in the localization of the tectonic activity at the south pole of Enceladus.
In a second paper, I will examine how the non-uniformity of the crust affects tidal dissipation.

\small
\section*{Acknowledgments}
This work is financially supported by the Belgian Federal Science Policy Office through the Brain Pioneer contract BR/314/PI/LOTIDE.
I thank Laurent Mont\'esi for answering many questions and sending me his results in \textit{Matlab} format.
I am grateful to Geruo A and Shijie Zhong for sharing with me the shell thickness file used in their paper and answering questions about their method.
\normalsize

%%%%%%%%%%%%%%%%%%%%%%%%%%%%%%%%%%%%%%%%%%%%%%%%%%%%%%%%%%%
%%%%%%%%%%%%%%%%%%%%%%%%%%%%%%%%%%%%%%%%%%%%%%%%%%%%%%%%%%%
%%%%%%%%%%%%%%%%%%%%%%%%%%%%%%%%%%%%%%%%%%%%%%%%%%%%%%%%%%%
% APPENDICES
%%%%%%%%%%%%%%%%%%%%%%%%%%%%%%%%%%%%%%%%%%%%%%%%%%%%%%%%%%%
%%%%%%%%%%%%%%%%%%%%%%%%%%%%%%%%%%%%%%%%%%%%%%%%%%%%%%%%%%%
%%%%%%%%%%%%%%%%%%%%%%%%%%%%%%%%%%%%%%%%%%%%%%%%%%%%%%%%%%%
\begin{appendices}

%%%%%%%%%%%%%%%%%%%%%%%%%%%%%%%%%%%%%%%%%%%%%%%%%%%%%%%%%%%
%%%%%%%%%%%%%%%%%%%%%%%%%%%%%%%%%%%%%%%%%%%%%%%%%%%%%%%%%%%
%%%%%%%%%%%%%%%%%%%%%%%%%%%%%%%%%%%%%%%%%%%%%%%%%%%%%%%%%%%
\section{Spherical differential operators of order~2}
\label{OperatorsOrder2}
\renewcommand{\theequation}{A.\arabic{equation}} % redefine the command that creates the equation no.
\setcounter{equation}{0}  % reset counter 

Tensorial differential operators of order~2 are defined on the surface of the sphere by (\textit{Be08}, Eq.~(15))
\begin{eqnarray}
{\cal O}_1 &=& \partial_\theta^2 + 1 \, ,
\nonumber \\
{\cal O}_2 &=& \csc^2\theta \, \partial_\varphi^2 + \cot \theta \, \partial_\theta + 1 \, ,
\nonumber \\
{\cal O}_3 &=& \csc\theta \left( \partial_\theta \partial_\varphi - \cot \theta \, \partial_\varphi \right) \, .
\label{OpO}
\end{eqnarray}
These operators give zero when applied on spherical harmonics of degree~1.
Their action on other spherical harmonics can be evaluated by replacing derivatives of associated Legendre functions with recurrence formulas (see Eqs.~(B.5)-(B.8) of \textit{Be10} for zonal harmonics).

The scalar differential operator of order~2 is given by
\begin{eqnarray}
\Delta' &=& {\cal O}_1 + {\cal O}_2
\nonumber \\
&=& \Delta + 2 \, ,
\label{DeltaPrime}
\end{eqnarray}
where
$\Delta$ is the spherical Laplacian.
Spherical surface harmonics of degree $n$ are eigenfunctions of $\Delta$ and $\Delta'$ with eigenvalues
\begin{eqnarray}
\delta_n &=& - n \left(n+1\right) ,
\nonumber \\
\delta'_n &=& - n \left(n+1\right) + 2 \, .
\label{eigendelta}
\end{eqnarray}

%%%%%%%%%%%%%%%%%%%%%%%%%%%%%%%%%%%%%%%%%%%%%%%%%%%%%%%%%%%
%%%%%%%%%%%%%%%%%%%%%%%%%%%%%%%%%%%%%%%%%%%%%%%%%%%%%%%%%%%
%%%%%%%%%%%%%%%%%%%%%%%%%%%%%%%%%%%%%%%%%%%%%%%%%%%%%%%%%%%
\section{Spherical differential operators of order~4}
\label{OperatorsOrder4}
\renewcommand{\theequation}{B.\arabic{equation}} % redefine the command that creates the equation no.
\setcounter{equation}{0}  % reset counter 

%=========================================================================================
%=========================================================================================
\subsection*{Definitions}

Scalar differential operators of order~4 are defined on the surface of the sphere by (\textit{Be08}, Eq.~(33) and \textit{Be10}, Eq.~(A.7))
\begin{eqnarray}
{\cal C}(a\,;b) &=& \Delta' \left( a \, \Delta' \,  b \right) \, ,
\label{defC} \\
{\cal A}(a\,;b) &=& ({\cal O}_1 \, a)({\cal O}_2 \, b) +  ({\cal O}_2 \, a)({\cal O}_1 \, b) - 2 \, ({\cal O}_3 \, a)({\cal O}_3 \, b) \, .
\label{defA1}
\end{eqnarray}
An operator closely related to ${\cal A}$ is given by
\begin{eqnarray}
{\cal D}_4'(a\,;b) &=& ({\cal O}_1 \, a)({\cal O}_1 \, b) +  ({\cal O}_2 \, a)({\cal O}_2 \, b) + 2 \, ({\cal O}_3 \, a)({\cal O}_3 \, b)
\nonumber \\
&=& \left( \Delta' a \right) \left( \Delta' b \right) - {\cal A}(a\,;b) \, .
\label{defD4}
\end{eqnarray}
The operator ${\cal J}$ is defined by (\textit{Be08}, Eq.~(62))
\begin{equation}
{\cal J} (X;Y;Z)  = 
\csc^2\theta \, \Big(
 \partial_\theta \left( \sin^2\theta \, \partial_\theta X \right)
+ \partial_\varphi^2 \, Y
+ 2 \, \partial_\theta \left( \sin\theta \, \partial_\varphi Z \right)
\Big)
- \cot\theta \, \partial_\theta Y
+ 2 Y \, .
\label{defJ}
\end{equation}
If $a$ and $b$ are scalar functions, then (\textit{Be08}, Eq.~(63))
\begin{eqnarray}
{\cal J}(a\,;a\,;0) &=& \Delta' a \, ,
\nonumber \\
{\cal J}( a \, {\cal O}_2 b \,; a \, {\cal O}_1 b \,; - a \, {\cal O}_3 b ) &=& {\cal A}(a \,; b) \, .
\label{Jeval}
\end{eqnarray}

%=========================================================================================
%=========================================================================================
\subsection*{Properties}

${\cal A}$ is symmetric:
\begin{equation}
{\cal A}(a\,;b) = {\cal A}(b\,;a) \, .
\label{sym}
\end{equation}
${\cal A}$ is of order two if $a$ is constant:
\begin{equation}
{\cal A}(a\,;b) = a \, \Delta'b \hspace{5mm} \mbox{($a$ is constant)} \, .
\label{opAred}
\end{equation}
${\cal C}$ vanishes if $b$ is of degree~1; ${\cal A}$ vanishes if either $a$ or $b$ is of degree~1:
\begin{equation}
{\cal C}(a\,;Y_{1m}) = {\cal A}(a\,;Y_{1m}) = {\cal A}(Y_{1m}\,;b) =0 \, .
\label{nodeg1}
\end{equation}
The images of ${\cal C}$ and ${\cal A}$ do not include spherical harmonics of degree~1 (see proof below Eq.~(\ref{lambdacoeff})):
\begin{equation}
\int_S {\cal C}(a\,;b)\, Y_{1m}^* \, d\Omega = \int_S {\cal A}(a\,;b)\, Y_{1m}^* \, d\Omega = 0 \, .
\label{nodeg1image}
\end{equation}

%=========================================================================================
%=========================================================================================
\subsection*{Evaluation}

The operator ${\cal A}$ can be rewritten as an expression where the only intervening operator is $\Delta'$ (\textit{Be10}, Eq.~(G.5)): 
\begin{eqnarray}
{\cal A}(a\,;b) &=& \frac{1}{4}  \left[
\, - \Delta' \Delta' (ab) - (\Delta' \Delta' \, a) \, b - a \, (\Delta' \Delta' \, b)
\right. \nonumber \\
&& \hspace{5mm} + \, 2 \, (\Delta' \, a)(\Delta' \, b) + 2 \, \Delta' ( (\Delta' \, a) \, b  +  a \, (\Delta' \, b) )
\nonumber \\
&&\hspace{5mm}   \left. - \, 2 \, \Delta' (ab) - 2 \, (\Delta' \, a) \, b - 2 \, a \, ( \Delta' \, b) + 8 \, ab \, \right] \, .
\label{defA2}
\end{eqnarray}
The action of ${\cal A}$ (as of ${\cal C}$) can thus be evaluated by applying $\Delta'$ in the spectral domain and performing multiplications in the spatial domain. 
If $a$ and $b$ are spherical surface harmonics of degree $m$ and $n$ (noted $a_m$ and $b_n$), I can write (\textit{Be10}, Eqs.~(G.6)-(G.9)):
\begin{eqnarray}
{\cal C}(a_m\,;b_n) &=&
\delta'_n \, \Delta' \, (a_m b_n)  \, ,
\nonumber \\
{\cal A}(a_m\,;b_n) &=& 
- \frac{1}{4} \left( \Delta' \Delta' + \kappa_{mn} \, \Delta'  + \lambda_{mn} \right) (a_m b_n) \, ,
\label{Aharmonics}
\end{eqnarray}
where $\delta'_n$ is given by Eq.~(\ref{eigendelta}) and
\begin{eqnarray}
\kappa_{mn} &=& 2 \left(1 - \delta'_m - \delta'_n \right) \, ,
\nonumber \\
\lambda_{mn} &=& \left( \delta'_m - \delta'_n \right)^2 + 2 \left( \delta'_m + \delta'_n \right) - 8
\nonumber \\
&=& \left(m-n-1 \right) \left(m-n+1 \right) \left(m+n \right) \left( m+n+2 \right) .
\label{lambdacoeff}
\end{eqnarray}
With these evaluation formulas, it is simple to prove that degree-1 harmonics do not belong to the image of ${\cal C}$ and ${\cal A}$ (Eq.~(\ref{nodeg1image})).
This statement is obviously true for all terms including $\Delta'$ in Eq.~(\ref{Aharmonics}) since $\Delta'$ is a diagonal operator yielding zero when applied on spherical harmonics of degree~1.
The only possible problem could arise from the $\lambda_{mn}$ term if the expansion of $a_m b_n$ includes a degree-1 component.
However, this can only occur if $m=n\pm1$ in which case Eq.~(\ref{lambdacoeff}) shows that $\lambda_{mn}=0$ (this proof is much shorter than the one given in Appendix~7.8 of \textit{Be08}).

%%%%%%%%%%%%%%%%%%%%%%%%%%%%%%%%%%%%%%%%%%%%%%%%%%%%%%%%%%%
%%%%%%%%%%%%%%%%%%%%%%%%%%%%%%%%%%%%%%%%%%%%%%%%%%%%%%%%%%%
%%%%%%%%%%%%%%%%%%%%%%%%%%%%%%%%%%%%%%%%%%%%%%%%%%%%%%%%%%%
\section{Strain in thin shell theory}
\label{StrainThinShell}
\renewcommand{\theequation}{C.\arabic{equation}} % redefine the command that creates the equation no.
\setcounter{equation}{0}  % reset counter 

 %=========================================================================================
 %========================================================================================
\subsection*{Assumptions}

The standard assumptions of the thin shell theory are \citep{kraus1967}:
\begin{enumerate}
\item The shell is thin, say less than one tenth of the radius of the sphere.
\item The deflections of the shell are small, so that linear perturbations make sense.
\item The transverse normal stress is negligible ($\sigma_{\zeta\zeta}=0$) so that, from Hooke's law,
\begin{equation}
\epsilon_{\zeta\zeta} = - \frac{\nu}{1-\nu} \left( \epsilon_{\theta\theta} + \epsilon_{\varphi\varphi} \right) .
\label{RadialStrain}
\end{equation}
\item Normals to the reference surface of the shell remain normal to it and undergo no change of length during deformation:
\begin{eqnarray}
\epsilon_{\zeta\zeta} &=& 0 \, ,
\label{epszz}\\
\epsilon_{\theta\zeta}=\epsilon_{\varphi\zeta} &=& 0 \, .
\label{ShearStrain}
\end{eqnarray}
\end{enumerate}
The condition $\epsilon_{\zeta\zeta}=0$ imposes that the tranverse displacement $w$ is constant through the shell thickness.
As is well-known, Eqs.~(\ref{RadialStrain}) and (\ref{epszz}) are not consistent, though it does not pose a real problem for thin shell theory \citep{kraus1967}.
The tidal dissipation rate, however, depends on all strains so that I need to choose which condition $\epsilon_{\zeta\zeta}$ satisfies.
For that purpose, it is important to require that Eq.~(\ref{RadialStrain}) is strictly true because it is required for the \textit{plane stress state} (see Eq.~(\ref{StressStrain})).
The constraints $\sigma_{\theta\zeta}=\sigma_{\varphi\zeta}=0$, also required for the plane stress state, result from Eq.~(\ref{ShearStrain}).

%=========================================================================================
%=========================================================================================
\subsection*{Linearization of strain}

Applying the second and fourth assumptions to the fundamental strain-displacement equations, I can express the strains in terms of quantities defined on the reference surface of the shell (\textit{Be08}, Eq.~(12)):
\begin{equation}
\left(
\begin{array}{c}
\epsilon_{\theta\theta} \\
\epsilon_{\varphi\varphi} \\
\epsilon_{\theta\varphi}
\end{array}
\right)
=
\left(
\begin{array}{c}
\epsilon_\theta^0 \\
\epsilon_\varphi^0  \\
\gamma_{\theta\varphi}^0/2 
\end{array}
\right)
+ \frac{\zeta}{1+\zeta/R_0} \,
\left(
\begin{array}{c}
\kappa_\theta^0 \\
\kappa_\varphi^0  \\
\tau^0/2 
\end{array}
\right) ,
\label{StrainDispl}
\end{equation}
where the membrane (or extensional) strains are the tangential strains on the reference surface (\textit{Be08}, Eqs.~(12)-(13)),
\begin{eqnarray}
\epsilon_\theta^0 &=& R_0^{-1} \left( \partial_\theta v_\theta + w \right) ,
\nonumber \\
\epsilon_\varphi^0 &=& R_0^{-1} \left( \csc\theta \, \partial_\varphi v_\varphi + \cot\theta \, v_\theta + w \right) ,
\nonumber \\
\gamma_{\theta\varphi}^0 &=& R_0^{-1} \left( \partial_\theta v_\varphi - \cot \theta \, v_\varphi + \csc\theta \, \partial_\varphi v_\theta \right) ,
\label{MemStrain}
\end{eqnarray}
while the bending (or flexural) strains on the reference surface are given by (\textit{Be08}, Eq.~(14))
\begin{equation}
\left( \kappa_\theta^0 \, , \, \kappa_\varphi^0 \, , \, \tau^0 \right) = - R_0^{-2} \left( {\cal O}_1 , \, {\cal O}_2 \, , 2{\cal O}_3 \right) w \, .
\label{BenStrain}
\end{equation}
In these equations, $w$ is the radial displacement (positive outward) which is constant through the shell thickness, $(v_\theta,v_\varphi)$ are the components of the tangential displacement on the reference surface, and ${\cal O}_{1,2,3}$ are the spherical differential operators defined by Eq.~(\ref{OpO}).

%%%%%%%%%%%%%%%%%%%%%%%%%%%%%%%%%%%%%%%%%%%%%%%%%%%%%%%%%%%
%%%%%%%%%%%%%%%%%%%%%%%%%%%%%%%%%%%%%%%%%%%%%%%%%%%%%%%%%%%
%%%%%%%%%%%%%%%%%%%%%%%%%%%%%%%%%%%%%%%%%%%%%%%%%%%%%%%%%%%
\section{Stress in thin shell theory}
\label{StressThinShell}
\renewcommand{\theequation}{D.\arabic{equation}} % redefine the command that creates the equation no.
\setcounter{equation}{0}  % reset counter

%=========================================================================================
%=========================================================================================
\subsection*{Plane stress}

The third and fourth assumptions of thin shell theory ($\sigma_{\zeta\zeta}=\sigma_{\theta\zeta}=\sigma_{\varphi\zeta}=0$) imply a state of plane stress (\textit{Be08}, Eq.~(16)):
\begin{equation}
\left(
\begin{array}{c}
\sigma_{\theta\theta} \\
\sigma_{\varphi\varphi} \\
\sigma_{\theta\varphi}
\end{array}
\right)
=
\frac{E}{1-\nu^2} \left(
\begin{array}{ccc}
1 & \nu & 0 \\
\nu & 1 & 0 \\
0 & 0 & 1-\nu
\end{array}
\right)
\left(
\begin{array}{c}
\epsilon_{\theta\theta} \\
\epsilon_{\varphi\varphi} \\
\epsilon_{\theta\varphi}
\end{array}
\right) .
\label{StressStrain}
\end{equation}

%=========================================================================================
%=========================================================================================
\subsection*{Stress and moment resultants}

In thin shell theory, the equations of equilibrium involve stress and moment resultants which result from integrating the microscopic stress over the shell thickness (\textit{Be08}, Eq.~(17)):
\begin{eqnarray}
N_{ij} &=& \int_d \sigma_{ij} \, (1+\zeta/R_0) \, d\zeta \hspace{1cm} (i=\theta,\varphi) \, , 
\nonumber \\
Q_i &=& \int_d \sigma_{i\zeta} \, (1+\zeta/R_0) \, d\zeta \hspace{1cm} (i=\theta,\varphi) \, ,  
\nonumber  \\
M_{ij} &=& \int_d \sigma_{ij} \, (1+\zeta/R_0) \, \zeta \, d\zeta \hspace{1cm} (i=\theta,\varphi) \, , 
\label{resultants}
\end{eqnarray}
where $N_{ij}$ are the tangential stress resultants, $Q_i$ are the shear stress resultants, and $M_{ij}$ are the moment resultants, all defined per unit of length on the reference surface.

%=========================================================================================
%=========================================================================================
\subsection*{Equations of equilibrium}

The equations of equilibrium for the stress and moment resultants are
\begin{eqnarray}
\partial_\theta \left( \sin\theta \, N_{\theta\theta} \right) + \partial_\varphi N_{\theta\varphi} - \cos \theta \, N_{\varphi\varphi} + \sin\theta \, Q_\theta + R_0 \sin\theta \, q_\theta &=&  0 \, ,
\nonumber \\
\partial_\theta \left( \sin\theta \, N_{\theta\varphi} \right) + \partial_\varphi N_{\varphi\varphi} + \cos\theta \, N_{\theta\varphi} + \sin\theta \, Q_\varphi  + R_0 \sin \theta \, q_\varphi &=& 0 \, ,
\nonumber \\
\partial_\theta \left( \sin\theta \, Q_\theta \right) + \partial_\varphi Q_\varphi - \sin\theta \, \left( N_{\theta\theta} + N_{\varphi\varphi} \right) + R_0 \, \sin \theta \, q &=& 0 \, ,
\nonumber \\
\partial_\theta \left( \sin\theta \, M_{\theta\theta} \right) + \partial_\varphi M_{\theta\varphi} - \cos\theta \, M_{\varphi\varphi} + R_0 \sin\theta \left( m_\theta - Q_\theta \right) &=& 0 \, ,
\nonumber \\
\partial_\theta \left( \sin\theta \, M_{\theta\varphi} \right) + \partial_\varphi M_{\varphi\varphi} + \cos\theta \, M_{\theta\varphi} + R_0 \sin\theta \left( m_\varphi - Q_\varphi \right) &=& 0 \, ,
\label{EqEquilibrium}
\end{eqnarray}
where $q$ is the transverse load (positive outward), $(q_\theta,q_\varphi)$ are the components of the tangential load, and $(m_\theta,m_\varphi)$ are the moments of the surface forces.
These equations are identical to Eqs.~(23)-(28) of \textit{Be08}, except for the additional moments of the surface forces (see \citet{kraus1967}, Eq.~(3.42)).
In thin shell theory, the moments of the tangential loads are neglected in the first-order approximation, but must be kept in the second-order approximation (they should have been included in \textit{Be08}).
The loads and their moments are measured per unit area of the reference surface (\citet{kraus1967}, Eq.~(3.37a)):
\begin{eqnarray}
R_0^2 \, q &=& \left( R_0 + \zeta^+ \right)^2 q^+ + \left( R_0 + \zeta^- \right)^2 q^- \, ,
\nonumber \\
R_0^2 \, q_i &=& \left( R_0 + \zeta^+ \right)^2 q_i^+ + \left( R_0 + \zeta^- \right)^2 q_i^- \hspace{10mm} (i=\theta,\varphi) \, ,
\nonumber \\
R_0^2 \, m_i &=& \zeta^+ \left( R_0 + \zeta^+ \right)^2 q_i^+ + \zeta^- \left( R_0 + \zeta^- \right)^2 q_i^- \hspace{5mm} (i=\theta,\varphi) \, ,
\label{defLoads}
\end{eqnarray}
where $(q^+,q_i^+)$ and $(q^-,q_i^-)$ are the loads on the top boundary $(R_0+\zeta^+)$ and bottom boundary $(R_0+\zeta^-)$ of the shell, respectively.

%=========================================================================================
%==========================================================================================
\subsection*{Stress-strain relation}

In order to derive the thin shell equations, we must express the stress and moment resultants in terms of the membrane and bending strains.
The resulting relations are much simpler if Poisson's ratio is constant throughout the shell, radially and laterally..
In that case, the diagonal stress resultants and moment resultants can be written as
\begin{eqnarray}
\left(
\begin{array}{cc}
N_{\theta\theta}/d & N_{\varphi\varphi}/d \\
M_{\theta\theta}/d^2 & M_{\varphi\varphi}/d^2 \\
\end{array}
\right)
=  \frac{1}{1-\nu^2} \, \mathbf{Z}
\left(
\begin{array}{cc}
\epsilon_\theta^0 + \nu \, \epsilon_\varphi^0 & \epsilon_\varphi^0 + \nu \, \epsilon_\theta^0 \\
d \left( \kappa_\theta^0 + \nu \, \kappa_\varphi^0 \right) & d \left( \kappa_\varphi^0 + \nu \, \kappa_\theta^0 \right) \\
\end{array}
\right) ,
\label{resultantsdiag}
\end{eqnarray}
while the nondiagonal components read
\begin{equation}
\left(
\begin{array}{c}
N_{\theta\varphi}/d \\
M_{\theta\varphi}/d^2
\end{array}
\right)
=  \frac{1}{2(1+\nu)} \, \mathbf{Z} 
\left(
\begin{array}{c}
\gamma_{\theta\varphi}^0 \\
d \, \tau^0
\end{array}
\right) ,
\label{resultantsnondiag}
\end{equation}
where
\begin{equation}
\mathbf{Z} = 
\left(
\begin{array}{cc}
E_0+\varepsilon E_1 & E_1 \\
E_1+\varepsilon E_2 & E_2 \\
\end{array}
\right) ,
\label{defZ}
\end{equation}
in which $\varepsilon=d/R_0$ and $(E_0,E_1,E_2)$ are the zeroth, first, and second moments of Young's modulus defined by an equation similar to Eq.~(\ref{momentsmu}).
Since $\nu$ is uniform, they are related to the moments of the shear modulus by
\begin{equation}
E_p = 2 \left( 1+\nu \right) \mu_p \, .
\label{momentYoung}
\end{equation}

%%%%%%%%%%%%%%%%%%%%%%%%%%%%%%%%%%%%%%%%%%%%%%%%%%%%%%%%%%%
%%%%%%%%%%%%%%%%%%%%%%%%%%%%%%%%%%%%%%%%%%%%%%%%%%%%%%%%%%%
%%%%%%%%%%%%%%%%%%%%%%%%%%%%%%%%%%%%%%%%%%%%%%%%%%%%%%%%%%%
\section{The first thin shell governing equation}
\label{FirstGoverningEquation}
\renewcommand{\theequation}{E.\arabic{equation}} % redefine the command that creates the equation no.
\setcounter{equation}{0}  % reset counter 

In the stress function approach, the equations of equilibrium for a thin shell are solved in terms of the radial displacement $w$ and an unknown stress function $F$.
Following \textit{Be08}, the first step consists in combining the stress and moment resultants (Eqs.~(\ref{resultantsdiag}) and (\ref{resultantsnondiag})) so as to eliminate the membrane strains, while the bending strains are expressed in terms of $w$ (Eq.~(\ref{BenStrain})):
 \begin{eqnarray}
M_{\theta\theta} &=& - R_0^{-2} \, D_{\rm v} \left( {\cal O}_1 + \nu {\cal O}_2 \right) w + R_0 \, \xi^{-1} N_{\theta\theta} \, ,
\nonumber \\
M_{\varphi\varphi} &=& - R_0^{-2} \, D_{\rm v} \left( {\cal O}_2 + \nu {\cal O}_1 \right) w + R_0 \, \xi^{-1} N_{\varphi\varphi} \, ,
\nonumber \\
M_{\theta\varphi} &=&  - R_0^{-2} \, D_{\rm v} \left( 1-\nu \right) {\cal O}_3 \, w + R_0 \, \xi^{-1} N_{\theta\varphi} \, ,
\label{moment}
\end{eqnarray}
where $D_{\rm v}$ is the viscoelastic analog of the elastic bending rigidity (\textit{Be08}, Eq.~(21)):
\begin{equation}
D_{\rm v} = \frac{d^3}{1-\nu^2} \, \frac{E_0 \, E_2 - (E_1)^2}{E_0 + \varepsilon E_1} \, .
\label{defD}
\end{equation}
The nondimensional parameter $\xi$ (not to be confused with $\xi_n$ in Eq.~(\ref{xin})), which is a large quantity, is defined by 
 \begin{equation}
\frac{1}{\xi} =  \frac{E_1+\varepsilon E_2}{E_0+\varepsilon E_1} \, \varepsilon \, .
\label{defxi}
\end{equation}
An associated parameter, close to one, is defined by
\begin{equation}
\chi = \frac{1}{1+1/\xi} \, .
\label{defchi}
\end{equation}
Eqs.~(\ref{defD}) and (\ref{defchi}) reduce to the equations for the elastic case, Eqs.~(\ref{defalphaDH})-(\ref{defetaxiH}), if 3 conditions are met:
the shell is radially homogeneous, thickness variations are symmetric with respect to the middle surface of the shell, and the middle surface is chosen as the reference surface.

Interestingly, the relations between stress and moment resultants (Eq.~(\ref{moment})) have the same form as those of \textit{Be08} (see Eq.~(38) of that paper).
The differences are hidden in the more general definitions of $D_{\rm v}$ and $\xi$.
The equations of equilibrium can thus be solved exactly as in Section~3.3.1 of \textit{Be08}.
In this paper, I model tidal forcing as a transverse load on the shell, so that tangential loads are irrelevant.
However, tangential loading will be required when benchmarking the thin shell approach against a surface loading model (see Section~\ref{FEM}).
For that purpose, it will be sufficient to assume that the tangential load and its moment are the surface gradients of the scalar potentials $\Omega$ and $\Omega_M$, respectively:
\begin{eqnarray}
\left( q_\theta \, , \, q_\varphi \right) &=& - R_0^{-1} \left( \partial_\theta \, , \, \csc\theta \, \partial_\varphi \right) \Omega \, ,
\nonumber \\
\left( m_\theta \, , \, m_\varphi \right) &=& - \left( \partial_\theta \, , \, \csc\theta \, \, \partial_\varphi \right) \Omega_M \, .
\label{defOmega}
\end{eqnarray}
In other words, the tangential load and the associated moments have no toroidal component (see \textit{Be08}, Eq.~(45)).
This is for example the case if both the shell and the forcing are axially symmetric.
Note that $\Omega_M$ is smaller than $\Omega$ by one order of magnitude in $\varepsilon=d/R_0$.

To begin with, the first two and the last two equations of equilibrium (in Eq.~(\ref{EqEquilibrium})) are combined in order to eliminate the shear stress resultants $(Q_\theta,Q_\varphi)$, before being solved in terms of $w$, a stress function $F$ to be determined later, and the load potentials $(\Omega,\Omega_M)$:
\begin{eqnarray}
N_{\theta\theta} &=& \chi \, \Big( {\cal O}_2 \, F + R_0^{-3} \, D_{\rm v} \left( {\cal O}_1 +\nu {\cal O}_2 \right) w + \Omega + \Omega_M \Big) \, ,
\nonumber \\
N_{\varphi\varphi} &=&  \chi \, \Big( {\cal O}_1 \, F + R_0^{-3} \, D_{\rm v} \left( {\cal O}_2 +\nu {\cal O}_1 \right) w  + \Omega + \Omega_M \Big) \, ,
\nonumber \\
N_{\theta\varphi} &=& \chi \, \Big( - {\cal O}_3 \, F  + (1-\nu) \, R_0^{-3} \, D_{\rm v} \, {\cal O}_3 \, w \Big) \, .
\label{eqN}
\end{eqnarray}
These equations differ from Eq.~(52) of \textit{Be08} by the presence of the moment potential $\Omega_M$ and the absence of the toroidal stress function $H$ (the latter is associated with the toroidal component of the tangential load).

Next, the first two equations of equilibrium are used in order to express the shear stress resultants $(Q_\theta,Q_\varphi)$ in terms of $F$ and $w$.
The resulting expressions have the same form as Eqs.~(39)-(40) and Eqs.~(53)-(54) of \textit{Be08}, except for additional terms depending on the moment potential: $-\partial_\theta(\chi\Omega_M)$ in $Q_\theta$ and $-\partial_\varphi(\chi\Omega_M)$ in $Q_\varphi$.
After substituting these expressions in the third equation of equilibrium, I get the first equation relating $F$ and $w$:
\begin{equation}
\Delta' \left( \chi D_{\rm v} \, \Delta' w \right) - (1-\nu) \, {\cal A}(\chi D_{\rm v} \,; w) + R_0^3 \, {\cal A}(\chi \,; F) 
= R_0^4 \, q + R_0^3 \left( \Delta \Omega - \Delta' \left( \chi \left(\Omega+\Omega_M \right) \right) \right)  .
\label{master1}
\end{equation}
Apart from the additional potential $\Omega_M$, this equation has the same form as the first governing equation for a homogeneous elastic shell (Eq.~(58) of \textit{Be08}) in which the toroidal stress function is set to zero and the elastic rigidity is replaced by  $D_{\rm v}$.
If $\Omega=\Omega_M=0$, it reduces to the first line of Eq.~(\ref{ElasticGoverningEq}) in which $D_{\rm e}$ is replaced by $D_{\rm v}$.
$D_{\rm v}$ being always paired with $\chi$, I will define the rigidity  in the main text as $D=\chi{}D_{\rm v}$.

Projecting Eq.~(\ref{master1}) on spherical harmonics of degree~1 yields, with the help of Eq.~(\ref{nodeg1image}), a constraint on the degree-1 load,
\begin{equation}
q_1 - \left(2/R_0 \right) \Omega_1 = 0 \, ,
\label{qOmega1}
\end{equation}
meaning that the external forces acting on the shell sum to zero (\textit{Be08}, Eq.~(79)).

%%%%%%%%%%%%%%%%%%%%%%%%%%%%%%%%%%%%%%%%%%%%%%%%%%%%%%%%%%%
%%%%%%%%%%%%%%%%%%%%%%%%%%%%%%%%%%%%%%%%%%%%%%%%%%%%%%%%%%%
%%%%%%%%%%%%%%%%%%%%%%%%%%%%%%%%%%%%%%%%%%%%%%%%%%%%%%%%%%%
\section{The second thin shell governing equation}
\label{SecondGoverningEquation}
\renewcommand{\theequation}{F.\arabic{equation}} % redefine the command that creates the equation no.
\setcounter{equation}{0}  % reset counter 
 
 In the stress function approach, a second equation relating $F$ and $w$ results from the following \textit{compatibility relation} (\textit{Be08}, Eq.~(59)):
 \begin{equation}
\partial_\theta \left( \sin\theta \, \partial_\varphi \gamma_{\theta\varphi}^0 \right)
= \partial_\theta \left( \sin^2 \theta \, \partial_\theta \, \epsilon_\varphi^0 \right)
+ \partial_\varphi^2 \, \epsilon_\theta^0
- \sin\theta \cos\theta \; \partial_\theta \, \epsilon_\theta^0
+ 2 \sin^2 \theta \, \epsilon_\theta^0 
-  R_0^{-1} \sin^2 \theta \, \Delta' w \, .
\label{compatibility1}
\end{equation}
Membrane strains are related to stress resultants by Eqs.~(\ref{resultantsdiag}) and (\ref{resultantsnondiag}):
\begin{eqnarray}
\epsilon_\theta^0 &=& \alpha_{\rm v} \left( N_{\theta\theta} - \nu N_{\varphi\varphi} \right) + R_0^{-1} \beta \, {\cal O}_1 \, w \, ,
\nonumber \\
\epsilon_\varphi^0 &=& \alpha_{\rm v} \left( N_{\varphi\varphi} - \nu N_{\theta\theta} \right) + R_0^{-1} \beta \, {\cal O}_2 \, w \, ,
\nonumber \\
\gamma_{\theta\varphi}^0 &=& 2 \alpha_{\rm v} \left( 1+\nu \right) N_{\theta\varphi} + 2 R_0^{-1} \beta \, {\cal O}_3 \, w \, ,
\label{strainsinverted}
\end{eqnarray}
where $\alpha_{\rm v}$ is the viscoelastic analog of the elastic extensibility (\textit{Be08}, Eq.~(61)), 
\begin{equation}
\alpha_{\rm v} = \frac{1}{\left(E_0+\varepsilon E_1 \right) d} \, ,
\label{defalpha}
\end{equation}
while $\beta$ is a nondimensional parameter proportional to the first moment of $E$:
\begin{equation}
\beta =  \frac{E_1 \varepsilon}{E_0 + \varepsilon E_1} \, .
\end{equation}
The terms of Eq.~(\ref{strainsinverted}) proportional to $\beta$ are new with respect to \textit{Be08}.

The compatibility relation can now be rewritten as
\begin{equation}
\Delta' \left( \alpha \left( N_{\theta\theta} + N_{\varphi\varphi} \right) \right)
- (1+\nu) \, {\cal J}_\alpha
+ {\cal J}_\beta
- R_0^{-1} \, \Delta' w = 0 \, ,
\label{compatibility2}
\end{equation}
 where $({\cal J}_\alpha,{\cal J}_\beta)$ result from the application of the operator ${\cal J}$ (Eq.~(\ref{defJ})):
 \begin{eqnarray}
{\cal J}_\alpha &=& {\cal J} \left(  \alpha_{\rm v} \, N_{\theta\theta} \,;  \alpha_{\rm v} \, N_{\varphi\varphi} \,;  \alpha_{\rm v} \, N_{\theta\varphi} \right) ,
\nonumber \\
{\cal J}_\beta &=& R_0^{-1} \, {\cal J} \left( \beta \, {\cal O}_2 \, w \,;  \beta \, {\cal O}_1 \, w \,;  -\beta \, {\cal O}_3 \, w \right) .
\end{eqnarray}
Eq.~(\ref{compatibility2}) coincides with Eq.~(60) of \textit{Be08} except for the new term ${\cal J}_\beta$.
As in \textit{Be08}, I substitute in the first two terms of Eq.~(\ref{compatibility2}) the stress resultants expressed in terms of $F$ and $w$ (Eq.~(\ref{eqN})).
The term ${\cal J}_\beta$ can be evaluated with Eq.~(\ref{Jeval}).
In this way, the compatibility relation yields an equation relating $F$ and $w$:
\begin{eqnarray}
&&
\Delta' \left( \chi\alpha_{\rm v} \, \Delta' F \right) - (1+\nu) \, {\cal A}(\chi\alpha_{\rm v} \,; F) + (1-\nu) \, \Delta' (\chi\alpha_{\rm v} \, \Omega)
\nonumber \\
&& \hspace{-5mm} + \,\, R_0^{-3} \left( 1-\nu^2\right) {\cal A}(\chi \alpha_{\rm v} D_{\rm v} \,; w)
+ R_0^{-1} \, {\cal A}(\beta \,; w)
- R_0^{-1} \, \Delta' w = 0 \, .
 \label{diffeq}
\end{eqnarray}
The terms depending on $w$ can be combined with the following identity:
\begin{equation}
R_0^{-2} \left( 1-\nu^2 \right) \chi \alpha_{\rm v} D_{\rm v} + \beta = 1-\chi \, .
\label{identitybetachi}
\end{equation}
The final expression for the second equation relating $F$ and $w$ reads
 \begin{equation}
\Delta' \left( \chi\alpha_{\rm v} \, \Delta' F \right) - (1+\nu) \, {\cal A}(\chi\alpha_{\rm v} \,; F) - R_0^{-1} \, {\cal A}(\chi \,; w)
= - (1-\nu) \, \Delta' \left( \chi\alpha_{\rm v} \left(\Omega+\Omega_M \right) \right)  \, .
\label{master2}
\end{equation}
Apart from the additional potential $\Omega_M$, this equation has the same form as the second governing equation for a homogeneous elastic shell (Eq.~(66) of \textit{Be08}) in which the toroidal stress function is set to zero and the elastic extensibility is replaced by  $\alpha_{\rm v}$.
If $\Omega=\Omega_M=0$, it reduces to the second line of Eq.~(\ref{ElasticGoverningEq}) in which $\alpha_{\rm e}$ is replaced by $\alpha_{\rm v}$.
In the main text, I will define the extensibility as $\alpha=\chi{}\alpha_{\rm v}$ since $\alpha_{\rm v}$ is always paired with $\chi$.

%%%%%%%%%%%%%%%%%%%%%%%%%%%%%%%%%%%%%%%%%%%%%%%%%%%%%%%%%%%
%%%%%%%%%%%%%%%%%%%%%%%%%%%%%%%%%%%%%%%%%%%%%%%%%%%%%%%%%%%
%%%%%%%%%%%%%%%%%%%%%%%%%%%%%%%%%%%%%%%%%%%%%%%%%%%%%%%%%%%
\section{Covariance of governing equations}
\label{Covariance}
\renewcommand{\theequation}{G.\arabic{equation}} % redefine the command that creates the equation no.
\setcounter{equation}{0}  % reset counter 

I show here that the governing equations are covariant under a change of the reference surface.
Covariance means that the equations do not change their form, even though parameters and variables are transformed.
This symmetry ensures that the governing equations are valid with different choices of the reference surface.
Suppose that the radius of the reference surface is shifted by a constant:
\begin{eqnarray}
R_0 &\rightarrow& R'_0 = R_0 + \delta{}R \, ,
\nonumber \\
\zeta &\rightarrow& \,\,\, \zeta' = \zeta - \delta{}R \, .
\label{shiftcoord}
\end{eqnarray}
The moments of the shear modulus $\mu_p$ (Eq.~(\ref{momentsmu})) transform as
\begin{eqnarray}
\mu_0 &\rightarrow& \mu_0' = \mu_0 \, ,
\nonumber \\
\mu_1 &\rightarrow& \mu_1' = \mu_1 - (\delta{}R/d) \, \mu_0 \, ,
\nonumber \\
\mu_2 &\rightarrow& \mu_2' = \mu_2 - 2 (\delta{}R/d) \, \mu_1 + (\delta{}R/d)^2 \, \mu_0 \, ,
\label{transfoparam}
\end{eqnarray}
and similar equations hold for the moments of Young's modulus $E_p$.
The following parameters (Eqs.~(\ref{muinv}) and (\ref{defalphaD})) are invariant,
\begin{eqnarray}
\left( \mu_{\rm inv} \, , \alpha_{\rm inv} \, , D_{\rm inv} \right)' &=& \left( \mu_{\rm inv} \, , \alpha_{\rm inv} \, , D_{\rm inv} \right) .
\label{parinv}
\end{eqnarray}
The transformations of $\chi$ and $\psi$ are obtained by expanding Eq.~(\ref{defchiparam}) to second order in $d/R_0$ and substituting Eqs.~(\ref{shiftcoord})-(\ref{transfoparam}),
yielding
\begin{equation}
\left( \chi \, , \psi \right)' \approx  \left( 1 + \delta R/R_0 \right)  \left( \chi \, , \psi \right) .
\label{covchi}
\end{equation}
The following quantities are invariant:
\begin{eqnarray}
\left( w \, , R_0 \chi F \right)' &=& \left( w \, , R_0 \chi F \right) ,
\nonumber \\
\left( R_0^2 \, q \, , R_0 \Omega \right)' &=& \left( R_0^2 \, q \, , R_0 \Omega \right) ,
\nonumber \\
\left( R_0^2 \left( \Omega+\Omega_M \right) \right)' &=& R_0^2 \left( \Omega+\Omega_M \right) .
\label{varinv}
\end{eqnarray}
First, the radial displacement $w$ is invariant because it is constant with depth, while $R_0 \chi F$ is invariant because Eq.~(\ref{eqN}) relates $\chi F$ to the stress resultant which is measured per unit of arc length on the reference surface (Eq.~(\ref{resultants})).
Second, $(q,q_\theta,q_\varphi)$ are measured per unit of area on the reference surface (Eq.~(\ref{defLoads})) and $(q_\theta,q_\varphi)$ is the surface gradient of $-\Omega/R_0$ (Eq.~(\ref{defOmega})).
Third, the moment potential transforms as $(R_0^2\,\Omega_M)'=R_0^2\,\Omega_M-R_0\delta{}R\,\Omega$ (see Eq.~(\ref{defLoads})) so that $R_0^2( \Omega+\Omega_M)$ is invariant, taking into account that $\Omega_M$ is already a second-order quantity.

From Eqs.~(\ref{parinv}) to (\ref{varinv}), I conclude that the governing equations in the second-order approximation (Eqs.~(\ref{master1}) and (\ref{master2})) are covariant under a change of the reference surface:
the terms of the first equation scale as $R_0^2$ whereas the second equation is invariant.

%%%%%%%%%%%%%%%%%%%%%%%%%%%%%%%%%%%%%%%%%%%%%%%%%%%%%%%%%%%
%%%%%%%%%%%%%%%%%%%%%%%%%%%%%%%%%%%%%%%%%%%%%%%%%%%%%%%%%%%
%%%%%%%%%%%%%%%%%%%%%%%%%%%%%%%%%%%%%%%%%%%%%%%%%%%%%%%%%%%
\section{Thin Shell Expansion (TSE) rule}
\label{TSErule}
\renewcommand{\theequation}{H.\arabic{equation}} % redefine the command that creates the equation no.
\setcounter{equation}{0}  % reset counter 

There is a lot of confusion in the literature about retaining or not small terms in thin shell equations.
For example,  the factor $\zeta/(1+\zeta/R_0)$ can be linearized or not (see \textit{Be08} and the second-order theories discussed in \citet{kraus1967}).
Moreover, the strain-displacement equations (Eq.~(\ref{StrainDispl})) can be modified by terms smaller by one order of magnitude in $d/R_0$ compared to the terms already present, depending on the way thin shell conditions are applied.
This ambiguity arises because of the inconsistency between thin shell assumptions mentioned in Appendix~\ref{StrainThinShell}. 

To clear up the confusion, it helps a lot to estimate the magnitude of the terms in the right-hand side of Eq.~(\ref{StrainDispl}).
Using Eqs.~(\ref{MemStrain})-(\ref{BenStrain}), I estimate the magnitude of membrane and bending strains in terms of the radial displacement:
\begin{eqnarray}
{\cal O}(\epsilon_i^0) &\sim& {\cal O}(w/R_0) \, ,
\nonumber \\
{\cal O}(d \, \kappa_i^0 ) &\sim& {\cal O}(\varepsilon \, \Delta (w/R_0) ) \, ,
\end{eqnarray}
with similar relations for $\gamma_{\theta\varphi}^0$ and $d\tau^0$.
Recall that $\Delta$ is the spherical Laplacian (Eq.~(\ref{DeltaPrime})) with eigenvalue $-n(n+1)$ and $\varepsilon=d/R$ is the relative shell thickness.
To consider the worst case in which a maximum of terms contribute, I assume that the shell is in the mixed regime, in which membrane and bending strains are of comparable magnitude (Eq.~(\ref{MembraneBendingTransition})):
\begin{equation}
{\cal O}(\epsilon_i^0) \sim {\cal O}(d \, \kappa_i^0) \, ,
\label{mixed}
\end{equation}
which implies that
\begin{equation}
{\cal O}(\varepsilon \, \Delta (w/R_0)) \sim {\cal O}(w/R_0) \, .
\end{equation}
Regarding the magnitude of the stress function, Eq.~(\ref{eqN}) yields
\begin{equation}
{\cal O}(\alpha \, \Delta F) \sim {\cal O}(w/R_0) \, .
\label{magnitudeF}
\end{equation}
Thus, I can state the following Thin Shell Expansion rule, or TSE rule:

\textit{In thin shell theory, the terms of any equation can be classified by order of magnitude in the parameter $\varepsilon=d/R_0$ with the following rules:
\begin{enumerate}
\item a double derivative compensates one power in $\varepsilon$;
\item
the magnitudes of $F$ and $w$ are related by ${\cal O}(\alpha F) \sim {\cal O}(\varepsilon w/R_0)$.
\end{enumerate}
In the spectral domain, the same rule holds with double derivatives replaced by the squared harmonic degree $n^2$.}

%%%%%%%%%%%%%%%%%%%%%%%%%%%%%%%%%%%%%%%%%%%%%%%%%%%%%%%%%%%
%%%%%%%%%%%%%%%%%%%%%%%%%%%%%%%%%%%%%%%%%%%%%%%%%%%%%%%%%%%
%%%%%%%%%%%%%%%%%%%%%%%%%%%%%%%%%%%%%%%%%%%%%%%%%%%%%%%%%%%
\section{First-order approximation}
\label{FirstOrderApprox}
\renewcommand{\theequation}{I.\arabic{equation}} % redefine the command that creates the equation no.
\setcounter{equation}{0}  % reset counter 

While the theory of uniform thin elastic shells is usually restricted to the first-order approximation, I show here that doing the same for non-uniform thin shells presents no computational advantage and lacks consistency.
With the TSE rule of Appendix~\ref{TSErule}, we can check that the first-order approximation of Eq.~(\ref{GoverningEq}) consists in setting $\chi\approx\psi\approx1$ everywhere except in the last term of each equation:
\begin{eqnarray}
{\cal C}( D_{\rm inv} \,; w ) - (1-\nu) \, {\cal A}( D_{\rm inv} \,; w) + R_0^3 \, {\cal A}(\chi \,; F) &=& R_0^4 \, q \, ,
\nonumber \\
{\cal C}( \alpha_{\rm inv}  \,;  F ) - (1+\nu) \, {\cal A}(\alpha_{\rm inv} \,; F) - R_0^{-1} \, {\cal A}(\chi \,; w) &=& 0 \, .
\label{GoverningEqApprox}
\end{eqnarray}
For example, the first two terms of the first equation are of magnitude
\begin{eqnarray}
{\cal C}( D_{\rm inv} \,; w ) &\sim& {\cal A}( D_{\rm inv} \,; w ) \,\, \sim \,\, {\cal O}\left( \varepsilon{}\mu_e{}R_0^3w \right) \, ,
\label{orderA}
\end{eqnarray}
while the third term of the same equation is of magnitude
\begin{eqnarray}
R_0^3 \, {\cal A}(\chi \,; F) &\sim& {\cal O}\left( \varepsilon\chi^u\,\mu_eR_0^3w \right)  \hspace{5mm} \mbox{if $\chi$ is uniform} \, ,
\nonumber \\
&\sim&  {\cal O}\left( \,\, \chi^\varepsilon \,\, \mu_e{}R_0^3w \right) \hspace{5mm} \mbox{if $\chi$ is laterally variable} \, ,
 \label{orderB}
\end{eqnarray}
where $\chi^u\sim{\cal O}(1)$ and $\chi^\varepsilon\sim{\cal O}(\varepsilon)$ are the uniform and laterally variable parts of $\chi$, respectively.
The cases of uniform and variable $\chi$ must be distinguished because the order of the operator ${\cal A}$ decreases from 4 to 2 if the first argument of ${\cal A}$ is constant.
Eqs.~(\ref{orderA})-(\ref{orderB}) show that the laterally variable part of $\chi$ must be taken into account even if it is of ${\cal O}(\varepsilon)$.
By contrast, for a homogeneous elastic shell, $\chi$ differed from unity by a term of ${\cal O}(\varepsilon^2)$ so that the approximation $\chi\approx1$ was possible everywhere (Eq.~(\ref{ElasticGoverningEqApprox})).

As $\chi$ still appears in Eq.~(\ref{GoverningEqApprox}), there is no computational advantage in doing the first-order approximation.
Moreover, keeping $\chi$ in some terms but not in others lacks consistency because the spatially uniform part of $\chi$ introduces second-order corrections whereas other ones are neglected.

%%%%%%%%%%%%%%%%%%%%%%%%%%%%%%%%%%%%%%%%%%%%%%%%%%%%%%%%%%%
%%%%%%%%%%%%%%%%%%%%%%%%%%%%%%%%%%%%%%%%%%%%%%%%%%%%%%%%%%%
%%%%%%%%%%%%%%%%%%%%%%%%%%%%%%%%%%%%%%%%%%%%%%%%%%%%%%%%%%%
\section{Tangential strain in terms of $(F,w)$}
\label{StrainFw}
\renewcommand{\theequation}{J.\arabic{equation}} % redefine the command that creates the equation no.
\setcounter{equation}{0}  % reset counter

The tangential strains at a given depth within the shell are related to the strains on the reference surface by Eq.~(\ref{StrainDispl}).
Substituting Eq.~(\ref{eqN}) into Eq.~(\ref{strainsinverted}), applying the identity (\ref{identitybetachi}), and substituting the resulting expression into Eq.~(\ref{StrainDispl}), I get
the tangential strains in terms of $(F,w)$ and $(\Omega,\Omega_M)$:
\begin{eqnarray}
\epsilon_{\theta\theta} &=& \alpha \left( \left( {\cal O}_2 - \nu {\cal O}_1 \right) F + \Omega + \Omega_M \right) - \left(z/R_0\right) {\cal O}_1 \, w \, ,
\nonumber \\
\epsilon_{\varphi\varphi} &=& \alpha \left( \left( {\cal O}_1 - \nu {\cal O}_2 \right) F + \Omega + \Omega_M \right) - \left(z/R_0\right) {\cal O}_2 \, w \, ,
\nonumber \\
\epsilon_{\theta\varphi} &=& - \alpha \left( 1+\nu \right) {\cal O}_3 \, F - \left(z/R_0\right) {\cal O}_3 \, w \, ,
\label{strainsFw}
\end{eqnarray}
where $\alpha=\chi\alpha_{\rm v}$ (Eq.~(\ref{defalphaD})) and $z$ is a complex parameter of ${\cal O}(\varepsilon)$:
\begin{eqnarray}
z &=& \frac{\zeta}{R_0+\zeta} - \left( 1-\chi \right) .
\label{zCoord}
\end{eqnarray}
Under a radial shift of the reference surface, $z$ transforms to ${\cal O}(\varepsilon^2)$ as
\begin{equation}
z' \approx  \left( 1 + \delta R/R_0 \right) z \, ,
\label{covz}
\end{equation}
as can be checked with Eqs.~(\ref{shiftcoord}) and (\ref{covchi}).
The formulas for the tangential strains given above (Eq.~(\ref{strainsFw})) are thus invariant.

The tangential displacements can be expressed in terms of 2D consoidal and toroidal displacement potentials $S(\theta,\varphi)$ and $T(\theta,\varphi)$:
$v_\theta=\partial_\theta{}S+\csc\theta\,\partial_\varphi{}T$ and $v_\varphi=\csc\theta\,\partial_\varphi{}S-\partial_\theta{}T$  (\textit{Be08}, Eqs.~(68)-(69)).
The tangential strains on the reference surface (Eq.~(\ref{MemStrain})) become
\begin{eqnarray}
\epsilon_\theta^0 &=& R_0^{-1} \left( \left( {\cal O}_1 - 1 \right) S + {\cal O}_3 \, T + w \right) ,
\nonumber \\
\epsilon_\varphi^0 &=& R_0^{-1} \left( \left( {\cal O}_2 - 1 \right) S - {\cal O}_3 \, T + w \right) ,
\nonumber \\
\gamma_{\theta\varphi}^0 &=& R_0^{-1} \left(  2 \, {\cal O}_3 S +  \left( {\cal O}_2 - {\cal O}_1 \right) T \right) .
\label{strainsPot}
\end{eqnarray}
Equating the trace of the strains ($\epsilon_\theta^0+\epsilon_\varphi^0$) in Eq.~(\ref{strainsPot}) and in Eq.~(\ref{strainsFw}) in which $\zeta=0$ for the reference surface, I can relate the consoidal potential $S$ to $(F,w)$ and $(\Omega,\Omega_M)$:
\begin{equation}
\Delta S = R_0 \, \alpha \left( 1- \nu \right) \left( \Delta' F + 2 \left( \Omega + \Omega_M \right) \right) + \left(1-\chi \right) \, \Delta' \, w - 2 w \, ,
\label{Spot}
\end{equation}
which is the viscoelastic generalization of Eq.~(71) in \textit{Be08} (with the additional potential $\Omega_M$).
If needed, one can derive a scalar differential equation for the toroidal potential $T$ (see \textit{Be08}, Eq.~(72)).
Strains are invariant under rigid displacements.
In particular, rigid translations satisfy the constraint (see Section~7.6 of \textit{Be08})
\begin{equation}
S_1 = w_1 \, .
\label{S1w1}
\end{equation}

%%%%%%%%%%%%%%%%%%%%%%%%%%%%%%%%%%%%%%%%%%%%%%%%%%%%%%%%%%%
%%%%%%%%%%%%%%%%%%%%%%%%%%%%%%%%%%%%%%%%%%%%%%%%%%%%%%%%%%%
%%%%%%%%%%%%%%%%%%%%%%%%%%%%%%%%%%%%%%%%%%%%%%%%%%%%%%%%%%%
\section{Tangential stress in terms of $(F,w)$}
\label{StressFw}
\renewcommand{\theequation}{K.\arabic{equation}} % redefine the command that creates the equation no.
\setcounter{equation}{0}  % reset counter 

The substitution of Eq.~(\ref{strainsFw}) into Eq.~(\ref{StressStrain}) yields the tangential stresses in terms of $(F,w)$ and $(\Omega,\Omega_M)$:
\begin{eqnarray}
\sigma_{\theta\theta} &=& E \left( \alpha \left( {\cal O}_2 \, F + \Omega + \Omega_M \right) - \frac{z}{R_0(1-\nu^2)}  \left( {\cal O}_1 + \nu {\cal O}_2 \right) w \right) ,
\nonumber \\
\sigma_{\varphi\varphi} &=& E \left( \alpha \left( {\cal O}_1 \, F + \Omega + \Omega_M \right) - \frac{z}{R_0(1-\nu^2)}  \left( {\cal O}_2 + \nu {\cal O}_1 \right) w \right) ,
\nonumber \\
\sigma_{\theta\varphi} &=& - E \left( \alpha \, {\cal O}_3 \, F + \frac{z}{R_0(1+\nu)} \, {\cal O}_3 \, w \right) .
\label{stressesFw}
\end{eqnarray}
Stresses depend on depth through Young's modulus $E$ and the parameter $z$, while they depend on the (local) shell thickness $d$ through $\alpha$ and $z$.
Similarly to the strains (Eq.~(\ref{strainsFw})), the formulas for the tangential stresses are invariant under a radial shift of the reference surface.
If the shell is laterally uniform and without tangential loads, Eq.~(\ref{stressesFw}) yields the same surface stresses as Eq.~(\ref{tidalstress}) if one substitutes in the former $w_n=h_nU_n/g$ and the $l_n-h_n$ relation for a thin shell (Eq.~(\ref{lnhn})).

While the stress components depend on the choice of coordinates, some stress-dependent quantities are scalar, i.e. invariant under coordinate transformations.
A few examples are the first and second invariants of the deviatoric stress, and the principal stresses at the surface.
Scalar quantities that are at most quadratic in the stress can be written in terms of the trace invariants ${\rm Tr}\,\bm{\sigma}$, ${\rm Tr}\,(\bm{\sigma} \cdot \bm{\sigma})$, and ${\rm Tr}\,(\bm{\sigma} \cdot \bm{\sigma}^*)$ (see Section~\ref{TidalStressSec}).
In thin shell theory,  radial components of the stress are negligible so that
\begin{eqnarray}
{\rm Tr}\,\bm{\sigma} &=& \sigma_{\theta\theta} + \sigma_{\varphi\varphi} \, ,
\nonumber \\
{\rm Tr} \left( \bm{\sigma} \cdot \bm{\sigma}^* \right) &=& \sigma_{\theta\theta} \, \sigma_{\theta\theta}^* + \sigma_{\varphi\varphi} \, \sigma_{\varphi\varphi}^* + 2 \, \sigma_{\theta\varphi} \, \sigma_{\theta\varphi}^* \, .
\label{sigmatens}
\end{eqnarray}
The formula for ${\rm Tr}\,(\bm{\sigma} \cdot \bm{\sigma})$ can be obtained by removing stars from the last equation.

The trace invariants can be expressed in terms of $(F,w)$ by substituting Eq.~(\ref{stressesFw}) into the previous equation and combining the ${\cal O}_i$ operators with Eqs.~(\ref{DeltaPrime}), (\ref{defA1}), and (\ref{defD4}).
Assuming there is no tangential load, I obtain
\begin{eqnarray}
{\rm Tr}\,\bm{\sigma} &=& E \alpha \, \Delta' F - \frac{E z}{R_0(1-\nu)} \, \Delta' w \, ,
\nonumber \\
{\rm Tr} \left( \bm{\sigma} \cdot \bm{\sigma}^* \right) &=& \sigma_2^{FF^*} + \sigma_2^{Fw^*} + \sigma_2^{F^*w} + \sigma_2^{ww^*} \, ,
\label{inv2Fw}
\end{eqnarray}
where
\begin{eqnarray}
\sigma_2^{FF^*} &=& \left| E \alpha \right|^2 \left( \left| \Delta' F \right|^2 - {\cal A}(F \,; F^*) \right) ,
\nonumber \\
\sigma_2^{Fw^*} &=& - \frac{ |E|^2 \alpha z^* }{R_0(1-\nu^2)}  \left( \nu \left( \Delta' F \right) \left( \Delta' w^* \right) + \left(1-\nu \right) {\cal A}(F \,; w^*) \right) ,
\nonumber \\
\sigma_2^{ww^*} &=& \frac{\left| E z \right|^2 }{R_0^2(1-\nu^2)^2} \left( \left(1+\nu^2\right) \left| \Delta' w \right|^2 - \left(1-\nu\right)^2{\cal A}(w \,; w^*) \right) ,
\label{inv2Comp}
\end{eqnarray}
and $\sigma_2^{F^*w}=(\sigma_2^{Fw^*})^*$.
The corresponding formulas for ${\rm Tr}\,(\bm{\sigma} \cdot \bm{\sigma})$ can be obtained by removing stars and absolute values from Eqs.~(\ref{inv2Fw})-(\ref{inv2Comp}).

%%%%%%%%%%%%%%%%%%%%%%%%%%%%%%%%%%%%%%%%%%%%%%%%%%%%%%%%%%%
%%%%%%%%%%%%%%%%%%%%%%%%%%%%%%%%%%%%%%%%%%%%%%%%%%%%%%%%%%%
%%%%%%%%%%%%%%%%%%%%%%%%%%%%%%%%%%%%%%%%%%%%%%%%%%%%%%%%%%%
\section{Analytical formulas for Love numbers}
\label{AnalyticalFormulas}
\renewcommand{\theequation}{L.\arabic{equation}} % redefine the command that creates the equation no.
\setcounter{equation}{0}  % reset counter 

%=========================================================================================
%=========================================================================================
\subsection*{Notation}

Consider a 3-layer body made of an incompressible  viscoelastic core, an ocean, and a floating shell with the same density as the ocean:
$\rho_b$ is the bulk density, $\rho$ is the density of the ocean and icy shell, $\xi_n$ is the degree-$n$ ocean-to-bulk density ratio (Eq.~(\ref{xin})) so that $\xi_1=\rho/\rho_b$.
The relative radii of the ocean and core are denoted $x=1-d/R$ and $y=R_c/R$, respectively.
The reduced shear moduli of the shell and core are denoted $\hat\mu=\mu/(\rho{}gR)$ and $\hat\mu_m=\mu_m/(\rho_bgR)$, respectively.
The reduced bulk modulus of the ocean is denoted $\hat K_o=K_o/(\rho_bgR)$.

%=========================================================================================
%=========================================================================================
\subsection*{Fluid-crust tidal Love numbers}

The fluid-crust model associated with the 3-layer model is the 2-layer body made of an incompressible  viscoelastic core below a surface ocean.
Ocean compressibility is not relevant to static tides \citep{saito1974}.
The fluid-crust tidal Love numbers ($n\geq2$) read
\begin{equation}
h_n^\circ = k_n^\circ + 1 = 
\frac{ A_n + \left(2n+1\right) y^4 \, \hat\mu_m }{ B_n  + \left(2n+1\right) \left(1-\xi_n \right) y^4 \, \hat\mu_m } \, ,
\label{hn0visco}
\end{equation}
where
$(A_n,B_n)$ are polynomials in $(y,\xi_1)$ given by Eq.~(C.5) of \textit{Be15b}.
This formula is coded in \textit{MembraneWorlds.nb}.
In the limit of an infinitely rigid core ($\hat\mu_m\rightarrow\infty$), Eq.~(\ref{hn0visco}) reduces to (\textit{Be15b}, Eq.~(C.1))
\begin{equation}
h_n^\circ = 1/\left(1- \xi_n \right) .
\label{hn0}
\end{equation}

%=========================================================================================
%=========================================================================================
\subsection*{Fluid-crust pressure Love number of degree~0}

Consider a 2-layer body made of an incompressible core and a compressible surface ocean.
If the density is uniform, the fluid-crust pressure Love number of degree~0 (see Section~\ref{Loads01}) can be expressed in terms of spherical Bessel functions of the first and second kind with indices 0 and 1 \citep{gilbert1968,takeuchi1972}.
In trigonometric form, the solution reads
\begin{equation}
h_0^{\circ P} = \frac{1}{12} \, \frac{\kappa (1-y) \cos (\kappa(1-y)) - (1 + \kappa^2 y ) \sin (\kappa(1-y))}{\kappa y \cos(\kappa(1-y)) + \sin(\kappa(1-y))} \, ,
\label{h0P2layers}
\end{equation}
where $\kappa^2=4/\hat K_o$.
In the limit $\kappa\ll1$, the solution is proportional to the ocean volume and inversely proportional to the ocean bulk modulus:
\begin{equation}
h_0^{\circ P} \approx - \frac{1-y^3}{9\hat K_o} \, .
\label{h0Papprox}
\end{equation}

%=========================================================================================
%=========================================================================================-
\subsection*{Homogeneous thick shell model}

In the homogeneous thick shell model, the floating shell is homogenous, incompressible, and has the same density as the top layer of the ocean.
Otherwise, the structure below the floating shell is arbitrary.
Extending the method of \textit{Be15a} (Appendix~A) from degree~2 to $n$, I obtain the tidal Love numbers in terms of the fluid-crust Love number $h_n^\circ$ and the properties of the shell (thickness, density, and shear modulus).
The results take the form of Eq.~(\ref{LoveThick}).
In simple incompressible models, $h_{n}^\circ$ can be expressed in analytic form (e.g.\ Eq.~(\ref{hn0visco}) or (\ref{hn0})), otherwise it must be computed by numerical integration (see Section~\ref{LaterallyUniformThickShell}).
The geometrical factors $(z_h,z_l)$ in Eq.~(\ref{LoveThick}) depend on the shell thickness (through $x$) and on the harmonic degree $n$:
\begin{eqnarray}
\left( z_h \, , z_l \right)&=& \frac{1}{2n+1} \left( -2\delta_n'\, \frac{\sum_j b_j \, x^j}{\sum_j a_j \, x^j} \,,  \frac{\sum_j c_j \, x^j}{\sum_j a_j \, x^j} \right) .
\label{zh}
\end{eqnarray}
The non-zero coefficients of the polynomials are given in Table~\ref{TablePoly} (the degree-2 formulas are coded in \textit{MembraneWorlds.nb}).

In the thin shell limit ($x\rightarrow1$),  $z_h\rightarrow0$ and $z_l\rightarrow3/(2n^2+2n-1)$, with an asymptotic behaviour given by Eqs.~(\ref{zhExp})-(\ref{zlExp}).
In the homogeneous limit, the shell fills the whole body: $x\rightarrow0$, $\rho\rightarrow\rho_b$, $h_n^\circ\rightarrow(2n+1)/(2(n-1))$ from Eq.~(\ref{hn0}), and
\begin{equation}
\left( z_h , z_l \right) \rightarrow \left( \frac{2\left(n-1\right)}{2n+1} \, \frac{2n^2+4n+3}{n} \, , \, \frac{3}{n \left( 2n+1 \right)} \right) ,
\label{zhHomog} 
\end{equation}
so that Eq.~(\ref{LoveThick}) tends to the Love numbers of a homogeneous incompressible body (\textit{Be16}, Appendix D).

%TABLE 6
\begin{table}[ht]\centering
\ra{1.3}
\scriptsize
\caption[]{\small
Degree-$n$ Love numbers of the homogeneous thick shell model: non-zero coefficients of the polynomials appearing in Eq.~(\ref{zh}).
The eigenvalues $\delta_n$ and $\delta_n'$ are defined by Eq.~(\ref{eigendelta}).
}
\begin{tabular}{@{}llr@{}}
\hline
 & \hspace{2mm} Arbitrary $n$ &  \hspace{1mm} $n=2$ \\
 \hline
$a_0$ & $\,\,\,\,\,\, n \left(n+2\right) \left(2n^2+1\right)$ & $3\times 24$ \\
$a_{2n-1}$ & $\,\,\,\,\, \left(2n+1\right) \delta_n \, \delta_n'$ & $3\times 40$ \\
$a_{2n+3}$ & $- \left(2n+1\right) \left(n^4+2n^3-n^2-2n+3\right)$ & $- \, 3\times 45$ \\
$a_{4n+2}$ & $- \left(n^2-1\right) \left(2n^2+4n+3\right)$ & $- \,3\times 19$ \\
\hline
$b_0$ & $\,\,\,\,\, \left(2n^2+1\right) \left(2n^2+4n+3\right)$ & $9 \times 19$\\
$b_{2n-1}$ & $- \left(2n+1\right)^2 \left(n^4+2n^3-n^2-2n+3\right)$ & $- \, 9\times 75$ \\
$b_{2n+1}$ & $\,\,\,\,\,\, 2 \left(2n-1\right) \left(2n+3\right) \delta_n \, \delta_n' $ & $9\times 112$\\
$b_{2n+3}$ & $=b_{2n-1}$ & $- \, 9\times 75$ \\
$b_{4n+2}$ & $=b_0$ & $\,\,\,\,\, 9 \times 19$ \\
\hline
$c_0$ & $\,\,\,\,\, 3 \left(n+2\right) \left(2n^2+1\right)$ & $\,\,\,\,\, 3 \times 36$\\
$c_{2n-1}$ &  $\,\,\,\,\, \left(2n+1\right)^2 \left(n^2+n-3\right) \delta_n'$ & $- \, 3\times 100$ \\
$c_{2n+1}$ & $- \left(2n-1\right) \left(2n+3\right) \left( 2n^2+2n-1\right) \delta_n' $ & $3\times 308$\\
$c_{2n+3}$ & $=b_{2n-1}$ & $- \, 3\times 225$  \\
$c_{4n+2}$ & $-3 \, \left( n-1 \right) \left(2n^2+4n+3\right)$ & $-\, 3 \times 19$ 
\\
\hline
\end{tabular}
\label{TablePoly}
\end{table}%

\FloatBarrier

\end{appendices}

\scriptsize
%%%%%%%%%%%%%%%%%%%%%%%%%%%%%%%%%%%%%%%%%%%%%%%%%%%%%%%%%%%
%%%%%%%%%%%%%%%%%%%%%%%%%%%%%%%%%%%%%%%%%%%%%%%%%%%%%%%%%%%
%%%%%%%%%%%%%%%%%%%%%%%%%%%%%%%%%%%%%%%%%%%%%%%%%%%%%%%%%%%
% REFERENCES
%%%%%%%%%%%%%%%%%%%%%%%%%%%%%%%%%%%%%%%%%%%%%%%%%%%%%%%%%%%
%%%%%%%%%%%%%%%%%%%%%%%%%%%%%%%%%%%%%%%%%%%%%%%%%%%%%%%%%%%
%%%%%%%%%%%%%%%%%%%%%%%%%%%%%%%%%%%%%%%%%%%%%%%%%%%%%%%%%%%

\end{document}